\documentclass[11pt]{article}              

\usepackage[margin=20mm]{geometry}
\usepackage{graphicx}
\usepackage{colortbl}
\usepackage[FIGBOTCAP]{subfigure}
 
\usepackage{fancyhdr}
\usepackage{amssymb,amsfonts,amsmath,amsthm}
\usepackage{natbib}

\usepackage{authblk}
	
\graphicspath{{./}}

\usepackage{bbm}
\usepackage{booktabs} % for toprule,...
\usepackage{multirow} % for multirow tabular

\usepackage{color}

\usepackage{lineno}

\newcommand{\ve}[1]{\boldsymbol{#1}}			% Vecteur
\newcommand{\Ve}[1]{\boldsymbol{#1}}			% Vecteur

\newcommand{\cm}{{\mathcal M}}

\newcommand{\ua}{\ve{\alpha}}
\newcommand{\ub}{\ve{\beta}}

\newcommand{\Var}[1]{{\rm Var}\left[ #1 \right]}

\begin{document}
\title{Global sensitivity analysis of natural convection in porous
    enclosure: Effect of thermal dispersion, anisotropic permeability
    and heterogeneity} 

\author[1]{N. Fajraoui} \author[2]{M. Fahs} \author[2,3,4]{A. Younes} \author[1]{B. Sudret} 

\affil[1]{Chair of Risk, Safety and Uncertainty Quantification,
    ETH Zurich, Stefano-Franscini-Platz 5, 8093 Zurich, Switzerland}
\affil[2]{LHyGeS, UMR-CNRS 7517, Universit\'e de Strasbourg/EOST, 1 rue Blessig, 67084 Strasbourg, France}
\affil[3]{IRD UMR LISAH, F-92761 Montpellier, France}
\affil[4]{LMHE, Ecole Nationale d'Ing\'enieurs de Tunis, Tunisie }

\date{}
\maketitle

\abstract{In this paper, global sensitivity analysis (GSA) and uncertainty quantification (UQ) have been applied to the problem of natural convection (NC) in a porous square cavity. This problem is widely used to provide physical insights into the processes of fluid flow and heat transfer in porous media. It introduces however several parameters whose values are usually uncertain. We herein explore the effect of the imperfect knowledge of the system parameters and their variability on the model quantities of interest (QoIs) characterizing the NC mechanisms. To this end, we use GSA in conjunction with the polynomial chaos expansion (PCE) methodology. In particular, GSA is performed using Sobol' sensitivity indices. Moreover, the probability distributions of the QoIs assessing the flow and heat transfer are obtained by performing UQ using PCE as a surrogate of the original computational model. The results demonstrate that the temperature distribution is mainly controlled by the longitudinal thermal dispersion coefficient. The variability of the average Nusselt number is controlled by the Rayleigh number and transverse dispersion coefficient. The velocity field is mainly sensitive to the Rayleigh number and permeability anisotropy ratio. The heterogeneity slightly affects the heat transfer in the cavity and has a major effect on the flow patterns. The methodology presented in this work allows performing in-depth analyses in order to provide relevant information for the interpretation of a NC problem in porous media at low computational costs. \\[1em]

        }

\maketitle

%%%%%%%%%%%%%%%%%%%%%%%%%%%%%%%%%%%%%%%%%%%%%%%%%%%%%%%%%%%%%%%%%%%%%%%%%%%%

\section{Introduction}

Natural convection (NC) in porous media can take place over a large range of scales that may go  from fraction of centimeters in fuel cells to several kilometers in geological strata \cite{nield2013convection}. This phenomenon is related to the dependence of the saturating fluid density on the temperature and/or compositional variations. A comprehensive bibliography about natural convection due to thermal causes can be found in the textbooks and handbooks by Nield and Bejan \cite{nield2013convection}, Ingham and Pop \cite{ingham2005transport}, Vafai \cite{vafai2005handbook} and Vadasz \cite{vadasz2008emerging}. Comprehensive reviews on NC due to compositional effects have been provided by Diersch and Kolditz \cite{diersch2002variable}, Simmons et al \cite{simmons2001variable}, Simmons \cite{simmons2005variable} and Simmons et al  \cite{simmons2010variable}. NC in porous media can be encountered in a multitude of technological and industrial applications such as building thermal insulation, heating and cooling processes in solid oxide fuel cells, fibrous insulation, grain storage, nuclear energy systems, catalytic reactors, solar power collectors, regenerative heat exchangers, thermal energy storage, among others \cite{nield2013convection,ingham2005transport,ingham2004emerging}. Important applications can be also found in hydro-geology and environmental fields such as in geothermal energy \cite{al2011computational, carotenuto2012new}, enhanced recovery of petroleum reservoirs \cite{almeida1995integral, riley1998compositional,  chen2007reservoir}, geologic carbon sequestration \cite{farajzadeh2007numerical, class2009benchmark,islam2013numerical, islam2014effects, vilarrasa2015geologic}, saltwater intrusion in coastal aquifers \cite{werner2013seawater} and infiltration of dense leachate from underground waste disposal \cite{zhang1995multispecies}. \\

Numerical simulation has emerged as a key approach to tackle the aforementioned applications in the last two decades. This is today a powerful and irreplaceable tool for understanding and predicting the behavior of complex physical systems. The literature concerning the numerical modeling and simulation of convective flow in porous media is abundant \cite{holzbecher1998modeling,pop2001convective,viera2012mathematical,miller2013numerical, su2015modeling,kolditz2015thermo}. The NC in porous media is usually described by the conservation equations of fluid mass, linear momentum and energy, respectively. Either Darcy or Brinkman models are used as linear momentum conservation laws. Darcy model is a simplification of the Brinkman model by neglecting the effect of viscosity. This simplification is valid for low permeable porous media. For high permeable porous media Brinkman model is more suitable because the effective viscosity is about 10 times the fluid viscosity \cite{givler1994determination, falsaperla2012double, shao2016high}.  In the traditional modeling analysis of NC in porous media, the governing equations are solved under the assumption that all the parameters are known. However, in real applications, the determination of the input parameters may be difficult or inaccurate.  For instance, in the simulation of geothermal reservoirs, the physical parameters (i.e. hydraulic conductivity and porosity) are subject to significant uncertainty because they are usually obtained by  model calibration procedures, that are often carried out with relatively insufficient historical data \cite{o2001state}. \\

The uncertainties affecting the model inputs may have major effects on
the model outputs. Typical examples about the significance of these
effects (that are not exhaustive) can be found in the design of clinical
devices or biomedical applications where small overheating can lead to
unexpected serious disasters \cite{davies1997sensitivity,
  ooi2011effects, wessapan2014aqueous}. Hence, the evaluation of how the
uncertainty in the model inputs propagates and leads to uncertainties in
the model outputs is an essential issue in numerical modeling. In this
context, uncertainty quantification (UQ) has become a must in all
branches of science and engineering \cite{brown2005assessing, SudretHDR,
  de2012modelling}. It provides a rigorous framework for dealing with
the parametric uncertainties. In addition, one wants to quantify how the
uncertainty in the model outputs is due to the variance of each model
input. This kind of studies is usually known as sensitivity analysis
(SA) \cite{Saltelli2002}. UQ aims at quantifying the variability of a
given response of interest as a function of uncertain input parameters,
whereas GSA allows to determine the key parameters responsible for this
variability. UQ and GSA are usually conducted by a multi-step analysis.
The first step consists on the identification of model inputs that are
uncertain and modelling them in a probabilistic context by means of
statistical methods using data from experiments, legacy data or expert
judgment. The second step consists in propagating the uncertainty in the
input through the model. Finally, sensitivity analysis is carried out by
ranking the input parameters according to their impact onto the
prediction uncertainty.
UQ and GSA have proven to be a powerful approach to assess the
applicability of a model, for fully understanding the complex processes,
designing, risk assessment and making decisions. They have been
extensively investigated in the literature
(e.g.
\cite{saltelli1999quantitative,sudret2008global,Xiu2003b,fesanghary2009design,blackwell2010technique,ghommem2011release,sarangi2014manifold,zhao2015effect,mamourian2016two,shirvan2017numerical,rajabi2014sampling}).
In the frame of flow and mass transfer in porous media, UQ and GSA have
been applied to problems dealing with saturated/unsaturated flow \cite{
  Younes2013}, solute transport \cite{Fajraoui2011b,Ciriello2013} and
density driven flow \cite{rajabi2014sampling,riva2015probabilistic}. 

A careful literature review shows that the investigation of sensitivity
analysis for NC in porous media has been limited to some special
applications \cite{shirvan2017numerical}. To the best of our knowledge,
these analyses have never been performed for a problem involving NC
within a porous enclosure. Yet, NC in porous enclosure has been largely
investigated for several purposes
\cite{oztop2009investigation,das2016studies} and several authors have
contributed important results for such a configuration
\cite{bejan1979boundary, prasad1984convective, beckermann1986numerical,
  gross1986application, moya1987numerical, lai1988natural,
  baytacs2000entropy, saeid2004transient, saeid2007conjugate,
  oztop2009investigation,sojoudi2014unsteady,chou2015effects,
  mansour2015numerical}.

Hence, keeping in view the various applications of NC in porous
enclosure and the importance of uncertainty analysis in numerical
modeling, a complete analysis involving GSA and UQ study is developed in
this work to address this gap. The considered problem deals with the
square porous cavity. Such a problem is widely used as a benchmark for
numerical code validation due to the simplicity of the boundary
conditions
\cite{walker1978convection,manole1993numerical,misra1995comparative,baytacs2000entropy,alves2000transient,fahs2015reference,
  shao2016high,shao2016new,zhu2017entropy}. It is also widely used to
provide physical insights and better understanding of NC processes in
porous media \cite{getachew1996natural, baytacs2000entropy,
  saeid2004transient, leong2004natural, mahmud2006mixed,
  choukairy2012numerical, malashetty2012linear} . As model inputs, we
consider the physical parameters characterizing the porous media and the
saturating fluid as the permeability, porosity, thermal diffusivity and
thermal expansion. All these parameters can be described by the Rayleigh
number which represents the ratio between the buoyancy and diffusion
effects.\\

 A common simplification for NC in porous media is to consider the saturated porous media with an equivalent thermal diffusivity (based on the porosity) and neglect the key process of heat mixing related to velocity dependent dispersion. Yet several studies have found that thermal dispersion plays an important role in NC systems \cite{howle1994natural, metzger2004optimal, pedras2008thermal, jeong2011estimation, yang2011analysis,kumar2009natural, sheremet2016effect, plumb1981non,cheng1981thermal,hong1987analysis,cheng1988transverse,amiri1994analysis,shih1992non} and applications related to transport in natural porous media \cite{abarca2007anisotropic, jamshidzadeh2013fluid, fahs2016henry}. Hence, main attention is given here to understand the impact of anisotropic thermal dispersion by including the longitudinal and transverse dispersion coefficients in the model inputs. Furthermore, anisotropy in the hydraulic conductivity is acknowledged as it is one of the properties of porous media which is a consequence of asymmetric geometry and preferential orientation of the solid grains \cite{shao2016new}. Finally heterogeneity of the porous media is considered as a source of uncertainty as it has a significant impact on NC in porous media \cite{simmons2001variable,  nield2007discussion,kuznetsov2010natural,zhu2017entropy}. As model outputs, we consider different quantities that are often used to assess the flow and the heat transfer processes in porous cavity as the temperature spatial distribution, the Nusselt number and the maximum velocity components.

In this work, we perform a global sensitivity analysis using a variance-based technique. In this particular context, the Sobol’ sensitivity indices \cite{Sobol1993,Homma1996,Sobol2001} are widely used as sensitivity metrics, because they do not rely on any assumption regarding the linearity or monotonous behavior of the physical model.  Various techniques have been proposed in the literature for computing the Sobol’ indices, see e.g. \cite{Archer1997,Sobol2001,Saltelli2002,Sobol2005,Saltelli2010}.  Monte Carlo (MC) is one of the most commonly used methods. However, it might become impractical, because of the large number of repeated simulations required to attain statistical convergence of the solution, especially for complex problem (e.g., \cite{sudret2008global,ballio2004convergence} and references therein).  In this context, new approaches based on advanced sampling strategies have been introduced to reduce the computational burden associated with Monte Carlo simulations. Among different alternatives, Polynomial Chaos Expansions (PCE) have been shown to be an efficient method for UQ and GSA \cite{BlatmanRESS2010, Blatman2010b,Blatman2011a}. In PCE, the key idea is to expand the model response in terms of a set of orthonormal multivariate polynomials orthogonal with respect to a suitable probability measure \cite{Ghanembook1991}. They allow one to uncover the relationship between different input parameters and how they affect the model outputs. Once a PCE representation is available, the Sobol' sensitivity indices are then obtained via a straightforward post-processing analysis without any additional computational cost \cite{sudret2008global}. It can also be used to perform an uncertainty quantification using Monte Carlo analysis at a significantly reduced computational cost (see, e.g., \cite{Fajraoui2011b} and references therein). \\

The structure of the present study is as follows. Section 2 is devoted to the description of the benchmark problem and the governing equations. Section 3 describes the numerical model. Section 4 describes the sensitivity analysis procedure using Sparse PCE. Section 5 discusses the GSA and UQ results for homogeneous and heterogeneous porous media. Finally, a summary and conclusions are given in Section 6.

\section{Problem statement and mathematical model}
The system under consideration is a square porous enclosure of length $H$ filled by a saturated heterogeneous porous medium. The properties of the fluid and the porous medium are assumed to be independent on the temperature. The porous medium and the saturating fluid are  locally in thermal equilibrium. We assume that the Darcy and Boussinesq approximations are valid and that the inertia and the viscous drag effects are negligible. Under these conditions, the fluid flow in anisotropic porous media can be described using the continuity equation and Darcy's law written in Cartesian coordinates as follows \cite{mahmud2006mixed}:

\begin{equation} 
	\label{eqn:1}
\frac{\partial u}{\partial x}+\frac{\partial v}{\partial y}=0,
\end{equation}

\begin{equation} 
\label{eqn:2}
u=-\frac{ k_x}{ \mu} \frac{\partial p}{\partial x},
\end{equation}

\begin{equation} 
\label{eqn:3}
v=-\frac{ k_y}{ \mu} \left(\frac{\partial p}{\partial y}+\left(\rho-\rho_c\right)g\right),
\end{equation}
where $u$ and $v\left[LT^{-1}\right]$ are the fluid velocity components in the $x$ and $y$ directions; $p \left[ML^{-1}T^{-2}\right]$ is the total pressure (fluid pressure and gravitational head); $k_x$ and $k_y\left[L^{2}\right]$ are the permeability components in the $x$ and $y$ directions; $\mu \left[ML^{-1}T^{-1}\right]$ is the dynamic viscosity; $\rho$ and $\rho_c \left[ML^{-1}\right]$ being respectively the density of the mixed fluid and density of the cold fluid; and $g \left[LT^{-2}\right]$ is the gravitational constant.\\

The heat transfer inside the cavity is modeled using the energy equation written as: 

\begin{equation} 
\label{eqn:4}
\frac{\partial T}{\partial t}+u\frac{\partial T}{\partial x}+v\frac{\partial T}{\partial y}=\alpha_m\left(\frac{\partial^2 T}{\partial x^2}+\frac{\partial^2 T}{\partial y^2}\right)+\frac{\partial }{\partial x}\left(\alpha^{xx}_{disp}\frac{\partial T}{\partial x}+\alpha^{xy}_{disp}\frac{\partial T}{\partial y}\right)+\frac{\partial }{\partial y} \left(\alpha^{xy}_{disp}\frac{\partial T}{\partial x}+\alpha^{yy}_{disp}\frac{\partial T}{\partial y}\right).
\end{equation}

\begin{figure}[!ht]
	\centering
	\includegraphics[width=.5\textwidth,clip = true, trim = 0 0 0 
	0]{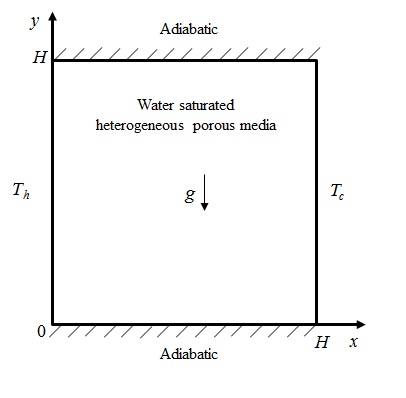}
	\caption{Schematic diagram of the heterogeneous porous-cavity problem}
	\label{fig:1}
\end{figure}

Here $T\left[\Theta\right]$ is the temperature; $\alpha_m\left[L^{2}T^{-1}\right]$ is the the effective thermal diffusivity and $\ve{\alpha_{disp}}\left[L^{2}T^{-1}\right]$ is the thermal dispersion tensor. In this work, we use the nonlinear model with anisotropic tensor as in \cite{howle1994natural}, that is defined as follows:

 \begin{equation} 
\label{eqn:5}
\alpha^{xx}_{disp}=\left(\alpha_L-\alpha_T\right)\frac{u^2}{\sqrt{u^2+v^2}}+\alpha_T\sqrt{u^2+v^2},
\end{equation}

\begin{equation} 
\label{eqn:6}
\alpha^{xy}_{disp}=\left(\alpha_L-\alpha_T\right)\frac{u\cdot v}{\sqrt{u^2+v^2}},
\end{equation}
\begin{equation} 
\label{eqn:7}
\alpha^{yy}_{disp}=\left(\alpha_L-\alpha_T\right)\frac{v^2}{\sqrt{u^2+v^2}}+\alpha_T\sqrt{u^2+v^2},
\end{equation}

where $\alpha_L$ and $\alpha_T\left[L\right]$ are respectively the longitudinal and transversal dispersivity coefficient, which are considered uniform in the system. No-flow boundary conditions are assumed across all boundaries. The left and right vertical walls are maintained at constant temperatures $ T_h$ and $T_c$ such that $(T_h>T_c)$, respectively. The horizontal surfaces are assumed to be adiabatic (Fig. \ref{fig:1}).\\

The system \eqref{eqn:1}-\eqref{eqn:7} is completed by specifying a constitutive relationship between fluid properties $\rho$ and the temperature $T$. The density of the mixed fluid is assumed to vary with temperature as a first-order polynomial, that is: 

\begin{equation} 
\label{eqn:8}
\rho=\rho_c[1-\beta(T-T_c)],
\end{equation}

where $\beta$ is the coefficient of thermal expansion. \\

Special attention is given in the literature to the NC in heterogeneous porous media because of the nonuniformity of the permeability and/or the thermal diffusivity affect significantly the overall rate of the heat transfer. The effect of heterogeneity is especially challenging in geothermal applications since hydraulic properties, such as permeability, can vary by several orders of magnitude over small spatial scales \cite{fahs2015reference}. In this context, the impact of heterogeneity has been studied for both external and internal natural convection using different heterogeneity configurations (stratified, horizontal, vertical, random, and periodic) \cite{kuznetsov2010natural} and references therein. In this work, the heterogeneity of the porous media is described via the exponential model as in \cite{FahsYounesMakradi2015,shao2016new,zhu2017entropy}. Based on the exponential model, the permeability in the $x-$ and $y-$ directions are given by:  

\begin{equation} 
\label{eqn:9}
k_x=k_{x0}e^{\sigma y},
\end{equation}

\begin{equation} 
\label{eqn:10}
k_y=k_{y0}e^{\sigma y},
\end{equation}

where $k_{x0}$ and $k_{y0}$ are respectively the permeability at $y=0$ and $x=0$;  $\sigma$ is the rate of change of $\ln (k)$ in the $y$ direction. \\

The system \eqref{eqn:1}-\eqref{eqn:3} can be reformulated in dimensionless form as: 

\begin{equation} 
\label{eqn:11}
\frac{\partial u^*}{\partial x^*}+\frac{\partial v^*}{\partial y^*}=0,
\end{equation}

\begin{equation} 
\label{eqn:12}
u^*=- e^{\sigma^*y^*}\frac{\partial p^*}{\partial x^*},
\end{equation}

\begin{equation} 
\label{eqn:13}
v^*=-r_k e^{\sigma^*y^*}\frac{\partial p^*}{\partial y^*}+Ra\cdot T^*,
\end{equation}

where $p^\ast=\dfrac{k_{x0}}{\mu \alpha_m}p$; $u^\ast=\dfrac{H}{\alpha_m}u$; $v^\ast=\dfrac{H}{\alpha_m}v$;  $T^\ast=\dfrac{T-T_c}{\Delta T}$; $\Delta T=T_h-T_c$ is the temperature difference between hot and cold walls; $x^\ast=\dfrac{x}{H}$; $y^\ast=\dfrac{y}{H}$;$\sigma^\ast=\sigma\cdot H; r_k= \dfrac{k_{y0}}{k_{x0}} $.
$Ra$ represents the Rayleigh number, that is given by:

  \begin{equation} 
  \label{eqn:14}
  Ra=\dfrac{k_{y}\cdot\rho_c\cdot\beta\cdot g\cdot\Delta T\cdot H}{\mu\alpha_m}.
  \end{equation}

 The steady-state energy equation can be written in dimensionless form as: 

\begin{equation} 
\label{eqn:15}
u^*\frac{\partial T^*}{\partial x^*}+v^*\frac{\partial T^*}{\partial y^*}=\left(\frac{\partial^2 T^*}{\partial x^*{^2}}+\frac{\partial^2 T^*}{\partial y^*{^2}}\right)+\frac{ \partial}{\partial x^*}\left(\alpha^{xx,*}_{disp}\frac{\partial T^*}{\partial x^*}+\alpha^{xy,*}_{disp}\frac{\partial T^*}{\partial y^*}\right)+\frac{\partial }{\partial y^*} \left(\alpha^{xy,*}_{disp}\frac{\partial T^*}{\partial x^*}+\alpha^{yy,*}_{disp}\frac{\partial T^*}{\partial y^*}\right),
\end{equation}

\begin{equation} 
\label{eqn:16}
\alpha^{xx,*}_{disp}=\left(\alpha_L^*-\alpha_T^*\right)\frac{\left(u^*\right)^2}{\sqrt{\left(u^*\right)^2+\left(v^*\right)^2}}+\alpha_T\sqrt{\left(u^*\right)^2+\left(v^*\right)^2},
\end{equation}

\begin{equation} 
\label{eqn:17}
\alpha^{xy,*}_{disp}=\left(\alpha_L^*-\alpha_T^*\right)\frac{u^*\cdot v^*}{\sqrt{\left(u^*\right)^2+\left(v^*\right)^2}},
\end{equation}

\begin{equation} 
\label{eqn:18}
\alpha^{yy,*}_{disp}=\left(\alpha_L^*-\alpha_T^*\right)\frac{\left(v^*\right)^2}{\sqrt{\left(u^*\right)^2+\left(v^*\right)^2}}+\alpha_T^*\sqrt{\left(u^*\right)^2+\left(v^*\right)^2},
\end{equation}

where $\alpha^\ast_L=\dfrac{\alpha_L}{H}$; $\alpha^\ast_T=\dfrac{\alpha_T}{H}$.\\

\section{The numerical model}
Numerical simulation of thermal-driven transfer problem is highly sensitive to discretization errors. Furthermore, hydraulic anisotropy, heterogeneity and anisotropic thermal dispersion render the numerical solution more challenging as they require specific numerical techniques.  Therefore, it is extremely important to select the appropriate numerical methods for solving the governing equations. In this work,  we use the advanced numerical model developed by Younes et al., \cite{younes2009solving} and Younes and Ackerer \cite{younes2008solving}. In this model, appropriate techniques for both time integration and spatial discretization are used to simulate coupled flow and heat transfer. For the spatial discretization, a specific method is used to achieve high accuracy for each type of equation. Thus, the Mixed Hybrid Finite method is used for the discretization of the flow equation. This method produces accurate and consistent velocity fields even for highly heterogeneous domains \cite{farthing2002mixed, durlofsky1994accuracy}. The heat transfer equation is discretized through a combination of a  discontinuous Galerkin (DG) and  Multipoint flux approximation (MPFA) methods. For the convective part, the DG method is used because it provides robust and accurate numerical solutions for problems involving steep fronts \cite{younes2008solving, tu2005slope}. For the diffusive part, the MPFA method is used because it allows for the handling of anisotropic heterogeneous domains and can be easily combined with the DG method \cite{younes2008solving}. The method of lines (MOL) is used for the time integration. This method improves the accuracy of the solution through the use of adaptive higher-order time integration schemes with formal error controls. The numerical model has been validated against lab experimental data for variable density flow \cite{konz2009variable}. It has been also validated by comparison against semi-analytical solutions for NC in porous square cavity \cite{fahs2014efficient,fahs2014new}, seawater intrusion in heterogeneous coastal aquifer \cite{younes2015extension} and seawater intrusion in anisotropic dispersive porous media \cite{fahs2016henry}.
We should note that the numerical model allows for transient simulations while in the GSA study we consider the steady state solutions. Hence transient solutions are performed until a long nondimensional
time to ensure steady conditions

\section{Polynomial Chaos Expansion for Sensitivity Analysis }

GSA is a useful tool that aims at quantifying which input parameters or combinations thereof contribute the most to the variability of the model responses, quantified in terms of its total variance. Variance-based sensitivity method have gained interest since the mid 90's, in this particular context. Here, we base our analysis on the the Sobol' indices, which are widely used as sensitivity metrics \cite{Sobol1993} and do not rely on any assumption regarding the linearity or monotonous behavior of the physical model.\\

In the sequel, we consider $Y=\mathcal{M}\left(\ve{X}\right)$, a mathematical model that describe a scalar output of the considered physical system, which depends on $M$-uncertain input parameters. $\mathcal{M}$ may, represent a scalar model response. In the case of vector-valued response, \textit{i.e.}; $\{Y \in \mathbb{R}^N , N > 1\}$, the following approach may be applied component-wise. We consider the uncertain parameters as independent random variables gathered into a random vector  $\ve{X}=\{X_1, . . . , X_M\}$ with joint probability density function (PDF) $f_{\Ve{X}}$ and marginal PDFs $ \{f_{X_i}(x_i), i = 1, . . . , M \} $. Within this context, we will introduce next the variance-based Sobol' indices. The interested reader is referred to \cite{sudret2008global,Gratietmetamodel2016} for a deeper insight into the details. 

\subsection{Anova-based sensitivity indices}
Provided that the function $\mathcal{M}$ is square-integrable with respect to the probability measure associated with $f_{\Ve{X}}$, it can be expanded in summands of increasing dimension as: 

\begin{equation}
\label{eqn:19}
\cm(\Ve{X})  =  \cm_0+\sum_{i=1}^{M} \cm_i(X_i)+\sum_{1\leq i<j\leq M} \cm_{ij}(X_i,X_j)+...+\cm_{12...M}(\Ve{X}),
\end{equation}

where $\cm_0$ is the expected value of $\cm(\Ve{X}) $, and the integrals of the summands $\cm_{i_1,i_2,...,i_s}$ with respect to their own variables is zero, that is:

\begin{equation}
\label{eqn:20}
\int_{\mathcal{D}_{X_{i_k}}}\cm_{i_1,i_2,...,i_s}(\Ve{X}_{i_1,i_2,...,i_s})f_{X_{i_k}}(X_{i_k}) =  0   \quad \text{~for~} 1\leq k\leq s,
\end{equation} 

where $\mathcal{D}_{X_{i_k}}$ and $f_{X_{i_k}}(X_{i_k})$
respectively denote the support and marginal PDF of $X_{i_k}$.\\
Eq \eqref{eqn:19} can be written equivalently as:  

\begin{equation}
\label{eqn:21}
\cm(\Ve{X})  =  \cm_0+\sum_{\Ve{u}\neq 0} \cm_{u}(\Ve{X}_{\Ve{u}}).
\end{equation}

Here $ \Ve{u}= \{i_1,i_2,...,i_M\} \subseteq \{1,2,...,M\}$ are index sets and $\Ve{X}_{\Ve{u}}$ are subvectors containing only those components of which the indices belong to ${\Ve{u}}$. This representation is called the Sobol' decomposition. It is unique under the orthogonality conditions between summand, namely: 

\begin{equation}
\label{eqn:22}
\mathbb{E}[\cm_{\Ve{u}}(\Ve{X}_{\Ve{u}})\cm_{\Ve{v}}(\Ve{X}_{\Ve{v}})]=0.
\end{equation}

Thanks to the uniqueness and orthogonality properties, it is straightforward
to decompose the total variance of $Y$, denoted  $D$ in a sum of partial variance $D_{u}$: 

\begin{equation}
\label{eqn:23}
D=\Var {\cm(\Ve{X})}=\sum_{\Ve{u}\neq 0} D_{u}= \sum_{\Ve{u}\neq 0}  \Var {\cm_{\Ve{u}}(\Ve{X}_{\Ve{u}})},
\end{equation}

where: 

\begin{equation}
\label{eqn:24}
D_{\Ve{u}}=\Var {\cm_{\Ve{u}}(\Ve{X}_{\Ve{u}})}=\mathbb{E}[\cm_{\Ve{u}}^2(\Ve{X}_{\Ve{u}})].
\end{equation}

This leads to a natural definition of the  Sobol' Indices $S_{\Ve{u}}$:

\begin{equation}
\label{eqn:25}
S_{\Ve{u}}=D_{\Ve{u}}/D,
\end{equation}

which measures the amount of the total variance due to the contribution of the subset $\Ve{X}_{\Ve{u}}$. In particular, the first-order sensitivity index is defined by: 

\begin{equation}
\label{eqn:26}
S_i=D_i/D.
\end{equation}

The first-order sensitivity indices $S_i$ measures the amount of variance of $Y$ that is due to the parameter $X_i$ considered separately. The overall contribution of a parameter $X_i$ to the response variance including its interactions with the other parameters is then given by the total sensitivity indices. They include the main effects $S_i$ and all the joint terms involving parameter $X_i$, \textit{i.e.} 

\begin{equation}
\label{eqn:27}
  S_i^T=\sum\limits_{\mathcal{I}_i} D_{\Ve{u}}/D, \quad \mathcal{I}_i=\{\Ve{u}\supset i\},
\end{equation}

In principle, one should rely upon the total sensitivity index to infer the relevance of the parameters \cite{saltelli2002relative}. The higher $S_i^T$, the more $X_i$ is an important parameter for the model response. In contrast, $X_i$ is termed unimportant (in terms of probabilistic modelling) if $S_i^T=0$.\\

The evaluation of Sobol' indices requires the computation of $2^M$ Monte Carlo integrals of the model response $\mathcal{M}\left(\ve{X}\right)$. This can be costly to manage, especially when dealing with time-consuming computational models. Fortunately, the Sobol' indices can easily be computed using the Polynomial Chaos Expansion (PCE) technique \cite{sudret2008global}. They are analytically obtained via a straightforward post-processing of the expansion. The PCE will be described in the next section. 

\subsection{Polynomial chaos expansion }
The model response can be casted into a set of orthonormal multivariate polynomial as: 

\begin{equation}
\label{eqn:28}
Y = \cm(\Ve{X})  =
\sum\limits_{\ua\in \mathcal{A}} y_{\ua}
{\Psi}_{\ua}(\Ve{X}),
\end{equation}

where $ \mathcal{A}$ is a multi-index $\ve{\ua}=\{{\ua_1,...,\ua_M}\}$, $\{ y_{\ua}, \ve{\ua} \in \mathcal{A} \} $ are the  expansion coefficients to be determined, $\{\Psi_{\ua} (\Ve{X}) , \ve{\ua} \in \mathcal{A} \}$ are multivariate polynomials which are orthonormal  with  respect to the joint pdf $f_{\Ve{X}}$ of $\Ve{X}$, i.e $\mathbb{E} [\Psi_{\ua} (\Ve{X})\Psi_{\ub} (\Ve{X})] = 1$ if $ \ua = \ub$ and 0 otherwise.\\

The multivariate polynomials  $\Psi_{\ua} $ are assembled as the tensor product of their appropriate univariate  polynomials, i.e 

\begin{equation}
\label{eqn:29}
\Psi_{\ua}(\ve{x}) = \prod_{i=1}^M \phi^{(i)}_{\alpha_i} (x_i),
\end{equation}

where $\phi^{(i)}_{\alpha_i}$ is a polynomial in the  $i $-th variable of degree ${\alpha_i}$. These bases are chosen according to the distributions associated with the input variables. For instance, if the input random variables are standard normal, a possible basis is the family of multivariate Hermite polynomials, which are orthogonal with respect to the Gaussian measure. Other common distributions can be used together with basis functions from the Askey scheme \cite{xiu2002}. A more general case can be treated through an isoprobabilistic transformation of the input random vector $\ve{X}$ into a standard random vector. The set of multi-indices $\mathcal{A}$ in Eq.~\eqref{eqn:28} is determined by an appropriate truncation scheme. In the present study, a hyperbolic truncation scheme \cite{Blatman2011a} is employed, which consists in selecting all polynomials satisfying the following criterion:

\begin{equation}
\label{eqn:30}
\lvert \ve{\ua} \rVert_q=\left(\sum_{i=1}^{M} \ua_i^q\right)^{1/q}\leq p,
\end{equation}

with $p$ being the highest total polynomial degree, $0 < q \leq 1$ being the parameter determining the hyperbolic truncation surface. This truncation scheme allows for retaining univariate polynomials of degree up to $p$, whereas limiting the interaction terms. \\

The next step is the computation of the polynomial chaos coefficients $\{y_{\ua}, \ua\in \mathcal{A}\}$. Several intrusive (\textit{e.g.} Galerkin scheme) or non-intrusive approaches (\textit{e.g.} stochastic collocation, projection, regression methods) \cite{sudret2008global,xiu2010numerical} are proposed in the literature. We herein focus our analysis on the regression methods also known as least-square approaches. A set of $N$ realization of the input vector, $\mathcal{X} =\{ \ve{x}^{(1)},..., \ve{x}^{(N)}\} $, is then needed, called experimental design (ED). The set of coefficient are then computed by means of the least-square minimization method, that is: 

\begin{equation}
\label{eqn:31}
\hat{\ve{y}}_{\ua}=\underset{\ve{y}_{\ua} \in \mathbb{R}^{\text{card}\mathcal{A}}} {\mathrm{argmin}} \frac{1}{N} \sum_{i=1}^{N} \left(\mathcal{M}\left(\Ve{x^{(i)}}\right)-\sum\limits_{\ua\in \mathcal{A}} y_{\ua} 
	{\Psi}_{\ua}(\Ve{x^{(i)}}) \right)^2.
\end{equation}

The number of terms in Eq. \eqref{eqn:28} may be unnecessarily large, thus a sparse PCE can be more efficient to capture the behavior of the model by disregarding insignificant terms from the set of regressors. We herein adopt the least angle regression (LAR) method proposed in \cite{Blatman2011a} which involves a sparse representation containing only a small number of regressors compared to the classical full representation. The reader is referred to \cite{Efron2004} for more details on the LARS technique and to \cite{Blatman2011a} for its implementation in the context of adaptive sparse PCE.\\

It can be worth noting that the constructed PCE can also be employed as a surrogate model of the target output in cases when evaluating a large number of model responses is not affordable. It is thus important to assess its quality. A good measure of the accuracy is the Leave-One-Out (LOO) error, which allows a fair error estimation at an affordable computational cost \cite{Blatman2010b}. The relative LOO error is defined as:
 
\begin{equation}
\label{eqn:32}
\epsilon_{LOO} = {\sum\limits_{i = 1}^N \left( 
	\frac{\cm(\Ve{x}^{(i)}) - 
		\cm^{PC}(\Ve{x}^{(i)})}{1-h_i}\right)^2}\bigg/{\sum\limits_{i = 1}^N 
	\left(\cm(\ve{x}^{(i)}) - \hat{\mu}_Y\right)^2},
\end{equation}

where $h_i$ is the $i^{th}$ diagonal term of matrix

 $\ve{\Psi}(\ve{\Psi}^{T}\ve{\Psi})^{-1}\ve{\Psi}^{T} $, where $\ve{\Psi}=\{\ve{\Psi}_{ij}=\Psi_j\left(\ve{X}^{i}\right)\}$ and $\hat{\mu}_Y=\frac{1}{N}\sum_{i=1}^{N} \cm(\Ve{x}^{(i)})$.\\

\subsection{Polynomial chaos expansions for sensitivity analysis}

Once the PCE is built, the mean $\mu$ and the total variance $D$ can be obtained using properties of the orthogonal polynomials \cite{sudret2008global}, such that: 

\begin{equation}
\label{eqn:33}
\mu= y_0,
\end{equation}

\begin{equation}
\label{eqn:34}
D=\sum\limits_{\ua\in \mathcal{A}\setminus{0}} y_{\ua}^2.
\end{equation}

As mentioned above, the Sobol' indices of any order can be computed in a straightforward manner. The first order and total Sobol' indices are then given by \cite{sudret2008global}: 

\begin{equation}
\label{eqn:35}
S_i=\sum\limits_{\ua\in \mathcal{A}_i} y_{\ua}^2/D, \quad \mathcal{A}_i=\{\ua\in \mathcal{A}: \alpha_i>0,\alpha_{j\neq i}=0\},
\end{equation}

and 

\begin{equation}
\label{eqn:36}
S_i^T=\sum\limits_{\ua\in \mathcal{A}_i^T} y_{\ua}^2/D, \quad \mathcal{A}_i^T=\{\ua\in \mathcal{A}: \alpha_i>0\}.
\end{equation}

Of particular interest is the marginal effect (also called univariate effect, see \cite{deman2016using}) of the parameters $X_i$, which enables investigation of the range of variation across which the model response is most sensitive to $X_i$. It corresponds to the sum of the mean values and first-order summands comprising univariate polynomials only, \textit{ i.e.}:  

\begin{equation}
\label{eqn:37}
\mathbb{E}[\mathcal{M}(\ve{X})\mid X_i=x_i]=\mathcal{M}_0+\sum_{\alpha\in \mathcal{A}_i} y_{\ve{\alpha}} \Psi_{\ve{\alpha}}(\ve{x}_i).
\end{equation}

\section{Results and discussions}

The PCEs presented in the previous section are used to perform UQ and GSA for the problem of natural convection in porous square cavity. The dimensionless form of the governing equations leads to define the model input parameters as follows:
\begin{itemize}
	\item The average Rayleigh number ($\overline{Ra}$): the Rayleigh number represents the ratio between the buoyancy and the diffusion effects. It depends on the porous media properties (porosity, thermal diffusivity and permeability), fluid properties (thermal diffusivity, viscosity, density and thermal expansion), the characteristic domain length and the temperature gradient. For isotropic porous media, the Rayleigh number is defined based on the scalar permeability of the porous media. For the general case of an anisotropic porous media, $Ra$ is defined based on the permeability in the vertical direction ($k_y$) \cite{bennacer2001double}. In this work, we are concerned with anisotropic heterogeneous porous media. Thus, we distinguish between local Rayleigh number based on the local permeability (see Eq. \eqref{eqn:14}) and the average Rayleigh number based on the overall average permeability \cite{fahs2015reference}. The local $Ra$ number can be  formulated as follows:
	
\begin{equation} 
\label{eqn:38}
{Ra}=Ra_0e^{\sigma ^\ast y^\ast},
\end{equation}

where $Ra_0$ is the local Rayleigh number at the bottom of the domain. The average Rayleigh number $\overline{Ra}$ is then obtained by integrating the local Ra overall the domain, that is given by:  

		\begin{equation} 
		\label{eqn:39}
		\overline{Ra}=Ra_0\int_{0}^{1}e^{\sigma^\ast y^\ast} \, dy^\ast=\dfrac{e^{\sigma^\ast  }-1}{\sigma^\ast}Ra_0.
		\end{equation}
	
The range of variability of the average Rayleigh number $\overline{Ra}$ is from 0 to 1000. This range of variation is physically plausible. 

\item  The permeability anisotropy ratio ($r_k$): this ratio is commonly used to describe the hydraulic anisotropy of the porous media \cite{abarca2007anisotropic,bennacer2001double}. In this work, the model used to describe the heterogeneity of the porous media leads to a constant anisotropy ratio, calculated based on the permeability on the $x$ and $y$ directions at the bottom of the domain ($k_{x_0}$ and $k_{y_0}$). As a common practice in porous media, the range of variability of $r_k$ is considered to be between 0 and 1 \cite{abarca2007anisotropic}.  

\item The non-dimensional dispersion coefficients ($\alpha_L^\ast$ and $\alpha_T^\ast$): these parameters correspond to the longitudinal and transverse thermal dispersion coefficients. They account for the enhancement of heat transfer due to hydrodynamic dispersion. The longitudinal dispersion $\alpha_L^\ast$ corresponds to the heat transfer along the local (Darcy) velocity vector while the transverse dispersion $\alpha_T^\ast$ acts normally to the local velocity. A detailed review about the physical understanding of longitudinal and transverse thermal dispersion is given in  Howle and Georgiadis \cite{howle1994natural}. According to Howle and Georgiadis \cite{howle1994natural}, Abarce et al., \cite{abarca2007anisotropic} and Fahs et al., \cite{fahs2016henry}, $\alpha_L^\ast$ was varied between $0.1$ and $1$ and $\alpha_T^\ast$ between $0.01$ and $0.1$. 
  
\item  The rate of heterogeneity variation ($\sigma_z^\ast$): this parameter is used to quantify the effect of the heterogeneity distribution on the model outputs. In fact, the geometrical distribution of the heterogeneity in the computational domain is not often well-defined. For instance, in hydrogeology, the heterogeneity distribution cannot be clearly described because hydraulic parameters do not correlate well with lithology. As in  Fahs et al., \cite{fahs2015reference} and shao et al., \cite{shao2016new}, the range of variability of ($\sigma_z^\ast$) is assumed to be from $0$ to $4$. 
\end{itemize}

Uncertainty in these parameters is related to our imperfect knowledge of the porous media properties (porosity, permeability tensor, heterogeneity distribution)  and thermo-physical parameters of both porous media grains and saturating fluid (thermal conductivity and dispersion). Without further information, and in view of drawing general conclusions, uniform distributions are selected for all parameters. Moreover, the parameters are assumed to be statistically independent. 

The results of the numerical model will be analyzed using several quantities of interest (QoI) which are controlled by the model inputs. To describe flow process we use the maximum dimensionless velocity components ($u^\ast_{max}$ and $v^\ast_{max}$). For the heat transfer process, the assessment is based on the spatial distribution of the dimensionless temperature ($T^\ast$). In addition, and as it is customary for the cavity problem, the heat processes are assessed using the wall average Nusselt number $\overline{Nu}$ given by: 

  \begin{equation} 
  \label{eqn:40}
  \overline{Nu}=\int_{0}^{1} Nu(y^\ast) \, dy^\ast,
  \end{equation}
  
 where $Nu$ is the local Nusselt number. 
 The local Nusselt number represents the net dimensionless heat transfer at a local point on the hot wall. It is defined as the ratio of the total convective heat flux to its value in the absence of convection. When thermal dispersion is considered, the local Nusselt number is defined as follows  \cite{howle1994natural,sheremet2016effect}: 
 
 \begin{equation}
 \label{eqn:41}
Nu=\left(1+\alpha_T^\ast \sqrt{\left(u^\ast\right)^2+\left(v^\ast\right)^2}  \right)\frac{\partial T^\ast}{\partial x^\ast}\bigg|_{x^\ast=0}.
 \end{equation}
  
\subsection{\textbf{Homogeneous case}}
\subsubsection{\textbf{Numerical details}}

Preliminary simulations were performed for different grid size in order to test the influence of grid discretization on the QoIs. They were performed under regular triangular mesh obtained by subdividing square elements into four equal triangles (by connecting the center of each square to its four nodes). Regular grids are used here to avoid instabilities and inaccuracies that can be caused by the change in mesh sizes within irregular grids. The most challenging configuration of the uncertain parameters is considered. This corresponds to the case with the highest Rayleigh number ($\overline{Ra}=1,000$) and anisotropy ratio ($r_k=1$) and lowest values of longitudinal and transverse thermal dispersion coefficients. For such a case, the heat transfer process is mainly dominated by the buoyancy effects which are at the origin of the rotating flow within the cavity. As a consequence, the steady state isotherms are sharply distributed and they have a spiral shape as they follow the flow structure due to the small thermal diffusivity. A relatively fine mesh should be used in this case to obtain a mesh independent solution. Several simulations are performed by increasing progressively the mesh refinement and by comparing the solution for two consecutive levels of grid. The tests revealed that the uniform grid formed by $40,000$ elements is adequate to render accurate results and capture adequately the flow and heat transfer processes. All simulations were run for $8$ minutes because this is the required time for the homogeneous problem to reach the steady state solution. These discretization parameters are kept fixed in subsequent simulations.\\

\begin{figure}[!ht]
	\centering
		\subfigure
		[Average Nusselt number $\overline{Nu}$]
		{
	\includegraphics[width=0.48\textwidth]{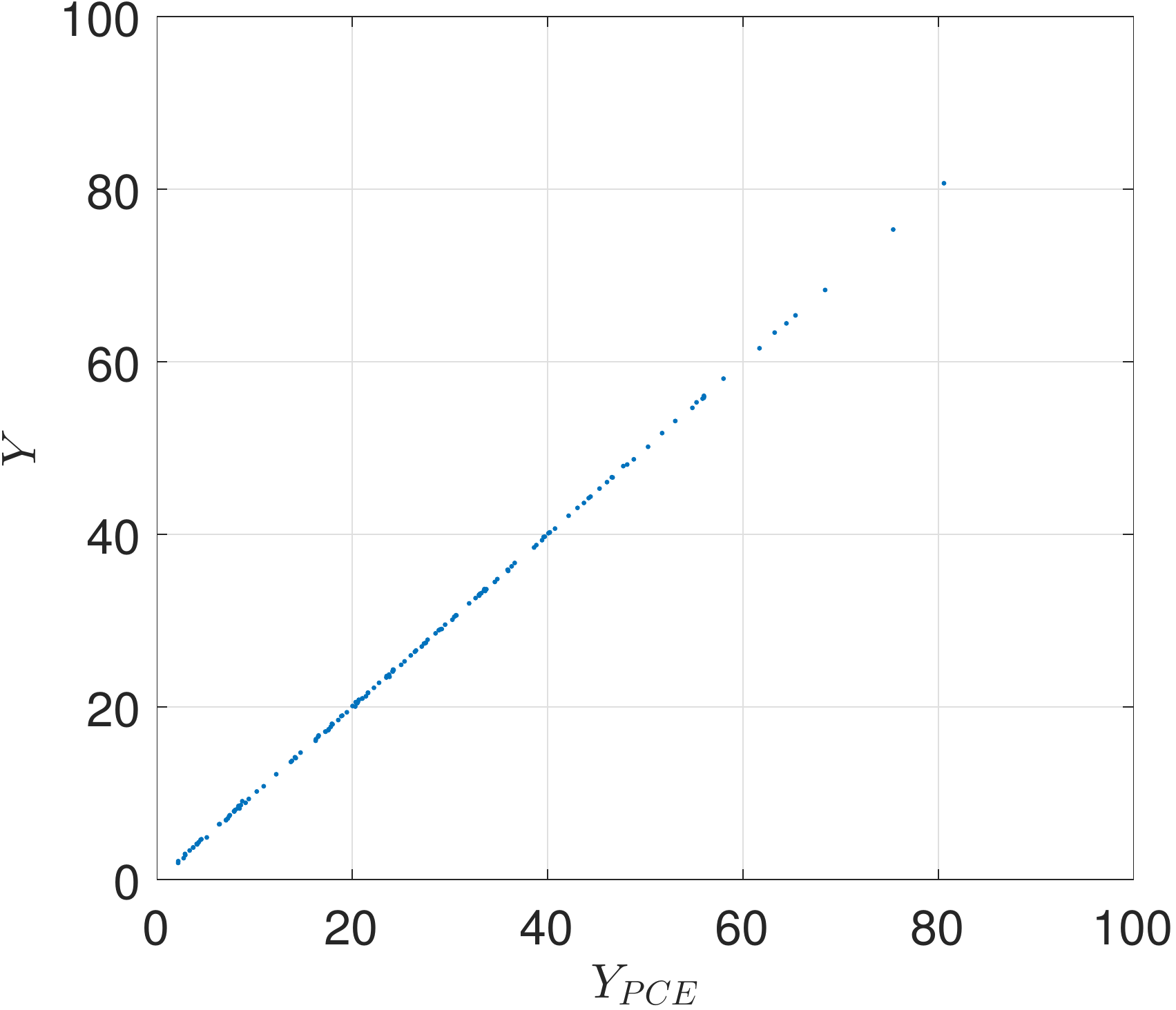}
}
		\subfigure
		[$u^\ast_{max}$]
		{
	\includegraphics[width=0.47\textwidth]{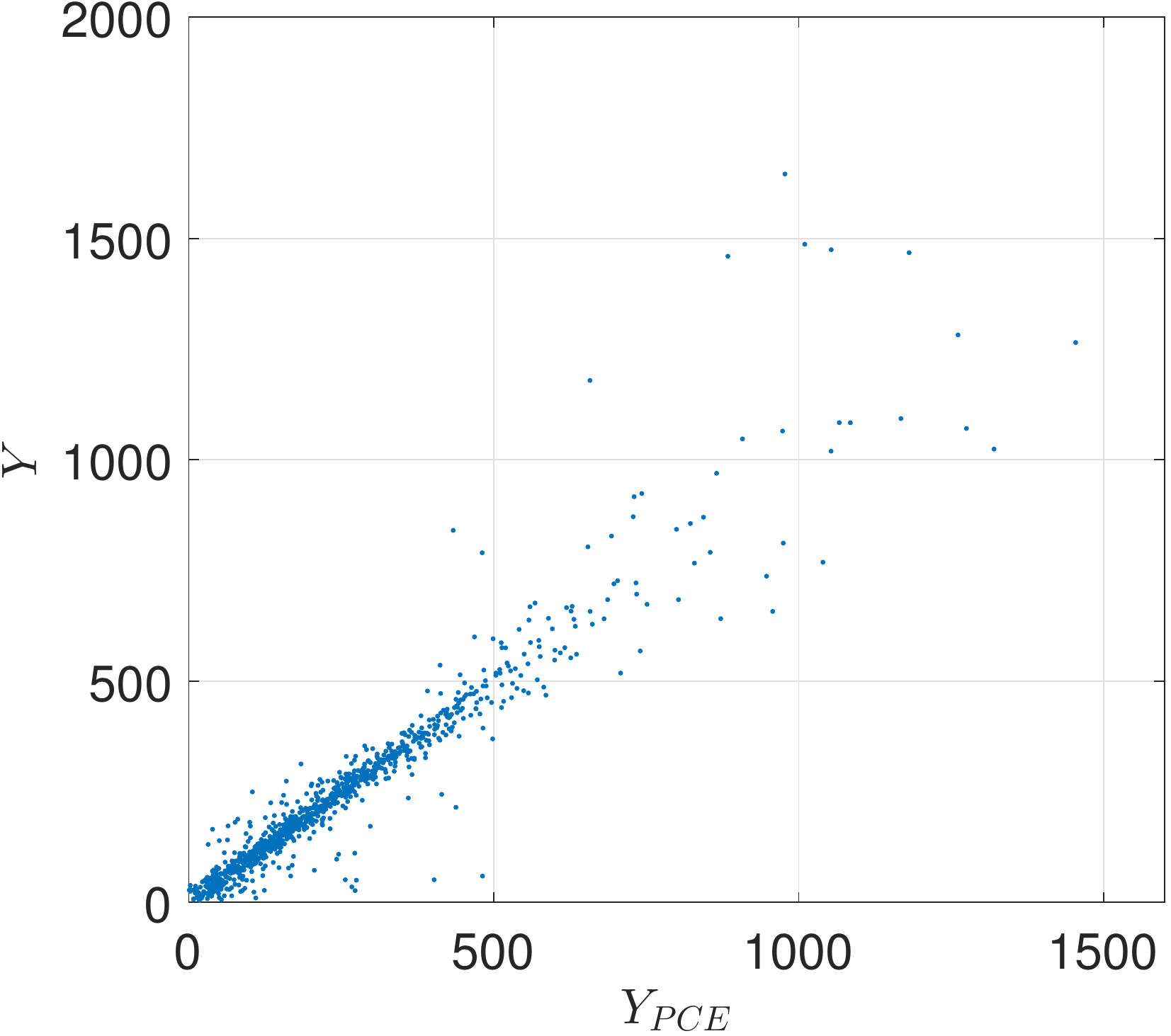}
}
		\subfigure
		[$v^\ast_{max}$]
		{
	\includegraphics[width=0.47\textwidth]{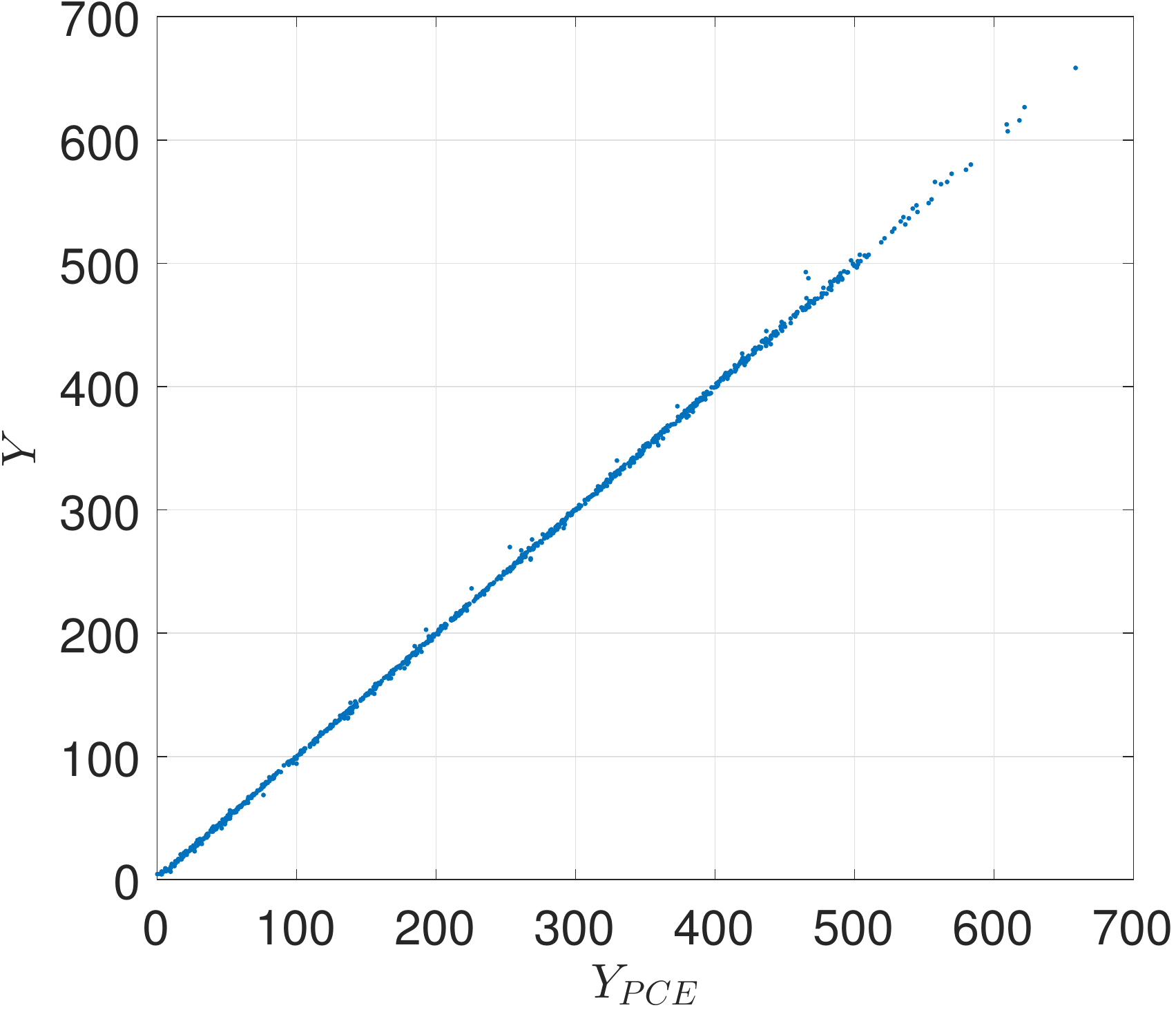}
}
	\caption{Homogeneous case: Comparison between PCE and the true model on $1,000$ validation runs for the model outputs. }
	\label{fig:2}
\end{figure}

In view of computing the PCE expansion of the model outputs in terms of the $4$ input random variables $\ve{X}=\{\overline{Ra},r_k,\alpha_L^\ast,\alpha_T^\ast\}$, several sets of parameter values sampled according to their respective pdf's are needed. For this purpose, we use an experimental design of size $N = 150$  drawn with Quasi Monte Carlo sampling (QMC). It is a well-known technique for obtaining deterministic experimental designs that covers at best the input space ensuring uniformity of each sample on the margin input variables. In particular, Sobol' sequences are used. PCE meta-models is constructed by applying the procedure described in Section $3$ for the considered QoI's. In the case of multivariate output (temperature), a PCE is constructed component-wise (\textit{i.e.;} for each points of the grid). The candidate basis is determined using a standard truncation scheme (see Eq.~\eqref{eqn:27}) with $q = 1$. The maximum degree $p$ is varied from $1$ to $20$ and the optimal sparse PCE is selected by means of the corrected relative $LOO$ error (see Eq.~\eqref{eqn:29}). The corresponding results (e.g. polynomial degree giving the best accuracy, relative $LOO$ error and number of retained polynomials) of the PCE are given in Table \ref{Tab:1} for the three scalar output $\overline{Nu}$, $u^\ast_{max}$ and $v^\ast_{max}$. For instance, when the average Nusselt number is considered, the optimal PCE is obtained for $p = 4$ and the corresponding $LOO$ error is $err_{LOO} = 8.6\times10^{-4}$. The sparse meta-model includes $63$ basis elements, whereas the size of full basis is $70$. In Fig.~\ref{fig:2}, we compare the values of the PCE  with the respective values of the physical model at a validation set consisting of $1,000$ MC simulations. We note that these simulations do not coincide with the ED used for the construction of the PCE. An excellent match is observed for both $\overline{Nu}$ and $v^\ast_{max}$; which is also illustrated by a small LOO error (less than $0.001$). Discrepancies between PCE and true model are observed for $u^\ast_{max}$; especially for larger values of $u^\ast_{max}$. The related LOO error is equal to $5.81\times10^{-2}$. 

\begin{table} [!ht]
	\centering
	\caption{Results of the utilized PCE}
	\label{Tab:1}
	\begin{tabular}{c c c c}
		\hline   & $\overline{Nu}$ & $u^\ast_{max}$ & $v^\ast_{max}$  \\
		\hline $p_{opt}$ & $5$ & $6$  & $4$ \\
		$err_{LOO}$ & $8.6\times10^{-4}$ & $5.81\times10^{-2}$  & $3.93\times10^{-4}$  \\
		Size of the Sparse Basis & $63$ & $47$  & $41$   \\
		\hline       
	\end{tabular}
\end{table}

\subsubsection{\textbf{Global sensitivity analysis}}
This section is devoted to GSA in order to identify the most influential parameters and to understand the marginal effect of the parameters onto the model outputs. Depending on the output QoI's, a different behaviour of the parameters is observed. The first and total Sobol' indices are computed, as well as second-order one, based on the obtained PCEs of the various  QoI's. Referring to the results in Table~\ref{Tab:1}, the relative LOO error varies from $0.04$\% to $5.8$\%. It is important to emphasize that excellent GSA results are obtained by PCE as soon as $err_{LOO}<10^{-3}$. Moreover, the results obtained for $u^\ast_{max}$  are also deemed acceptable. 
\begin{itemize}
	\item \textit{GSA of the temperature distribution}
\end{itemize}

\begin{figure}[!ht]
	\centering
	\subfigure
	[spatial distribution of the mean value of the temperature]
	{
	\includegraphics[width=.4\textwidth,clip = true, trim = 0 0 0 
	0]{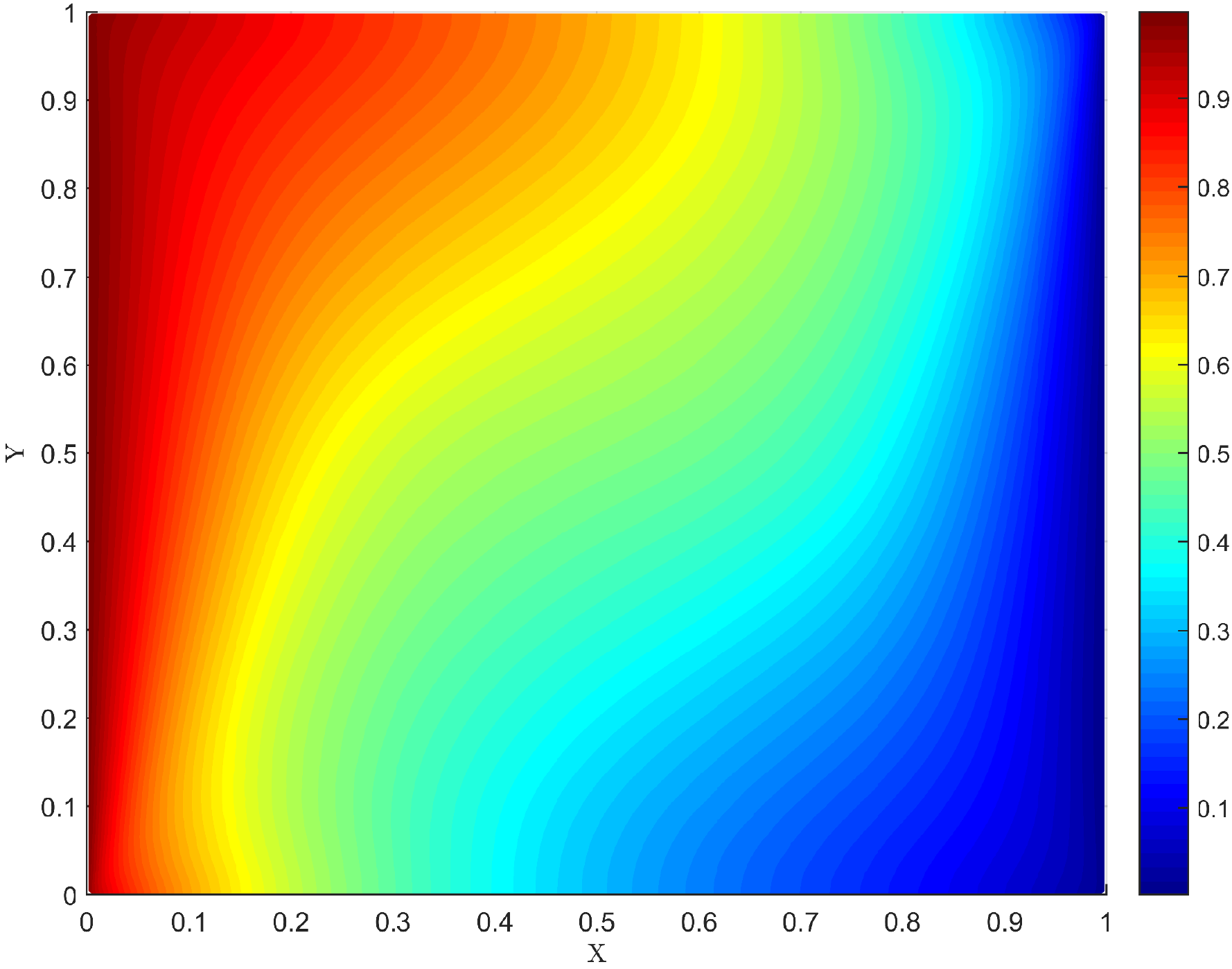}
}
	\subfigure
	[ spatial distribution of the variance]
	{
	\includegraphics[width=.4\textwidth,clip = true, trim = 0 0 0 
	0]{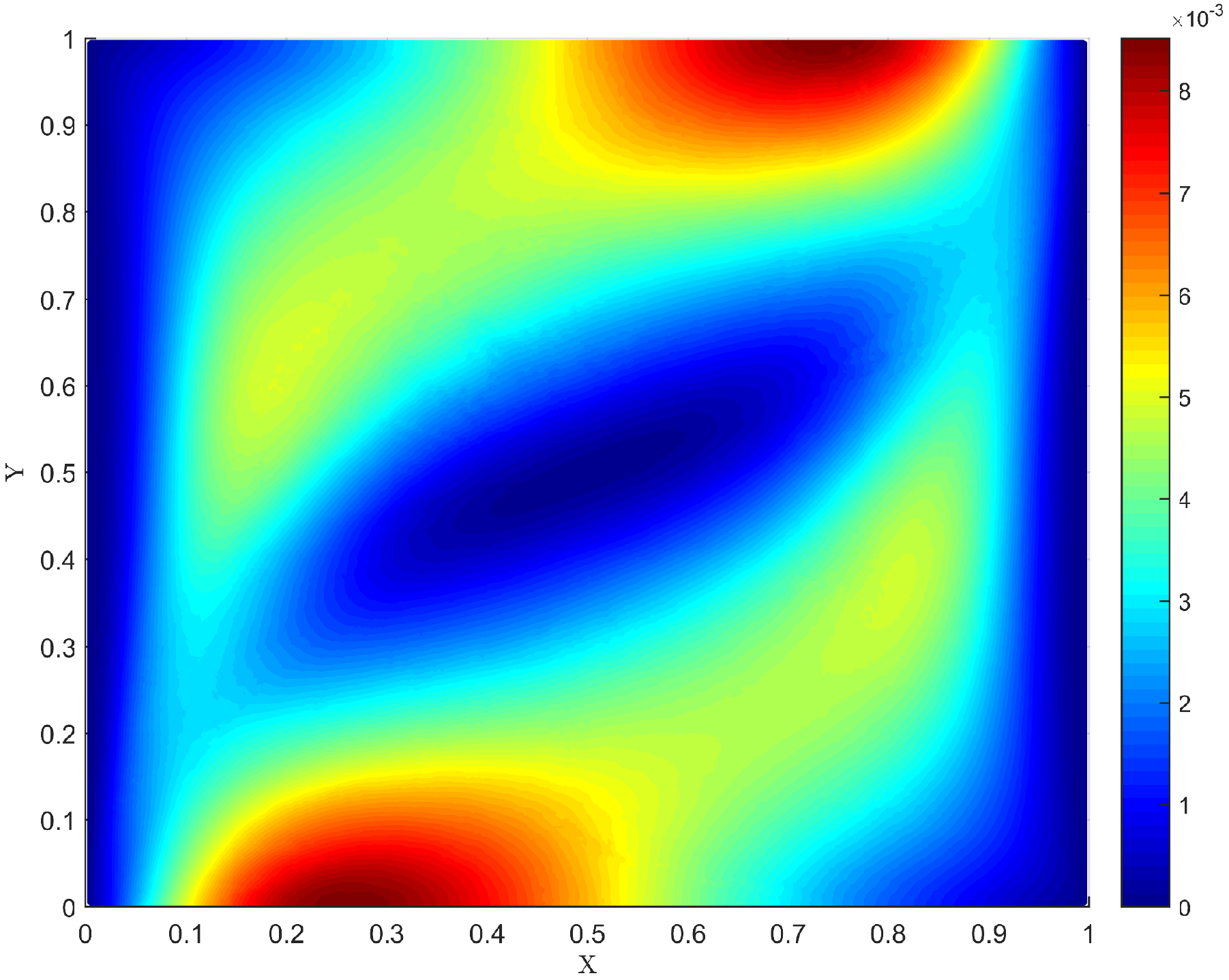}
}
	\caption{Homogeneous case - spatial distribution of the temperature statistical moments.}
	\label{fig:3}
\end{figure}

Fig. \ref{fig:3}a illustrates the spatial distribution of the mean of the temperature based on PCE. In this case, the presented approach is applied component-wise. Indeed, the PCE of a numerical model with many outputs is carried-out by metamodelling independently each model output. Fig. \ref{fig:3}a shows that the distribution of the mean temperature reflects the general behavior of the heat transfer in the case of NC in square porous cavity. The isotherms are not vertical as they are affected by the circulation of the fluid saturating the porous media. In order to evaluate how far the temperatures are spread out from their mean, we plot in Fig. \ref{fig:3}b the distribution of the temperature variance. As a general comment, a symmetrical behavior of the variance around the center point is observed. The temperature variance is negligible in the thermal boundary layers of the hot and cold walls (deterministic boundary conditions) and in the relatively slow-motion rotating region at the core of the square. It becomes significant at the horizontal top and bottom surfaces of the porous cavity. The largest variance values are located toward the cold wall at the top surface and the hot wall at the bottom surface. In these zones the flow is nearly horizontal. The fluid is cooled down (resp. heated) at the top (resp. bottom) by the effect of the cold (resp. hot) wall.

\begin{figure}[!ht]
	\centering
	\subfigure
	[ Spatial distribution of total Sobol' index of $\overline{Ra}$;]
	{
		\includegraphics[width=.47\textwidth,clip = true, trim = 0 0 0 
		0]{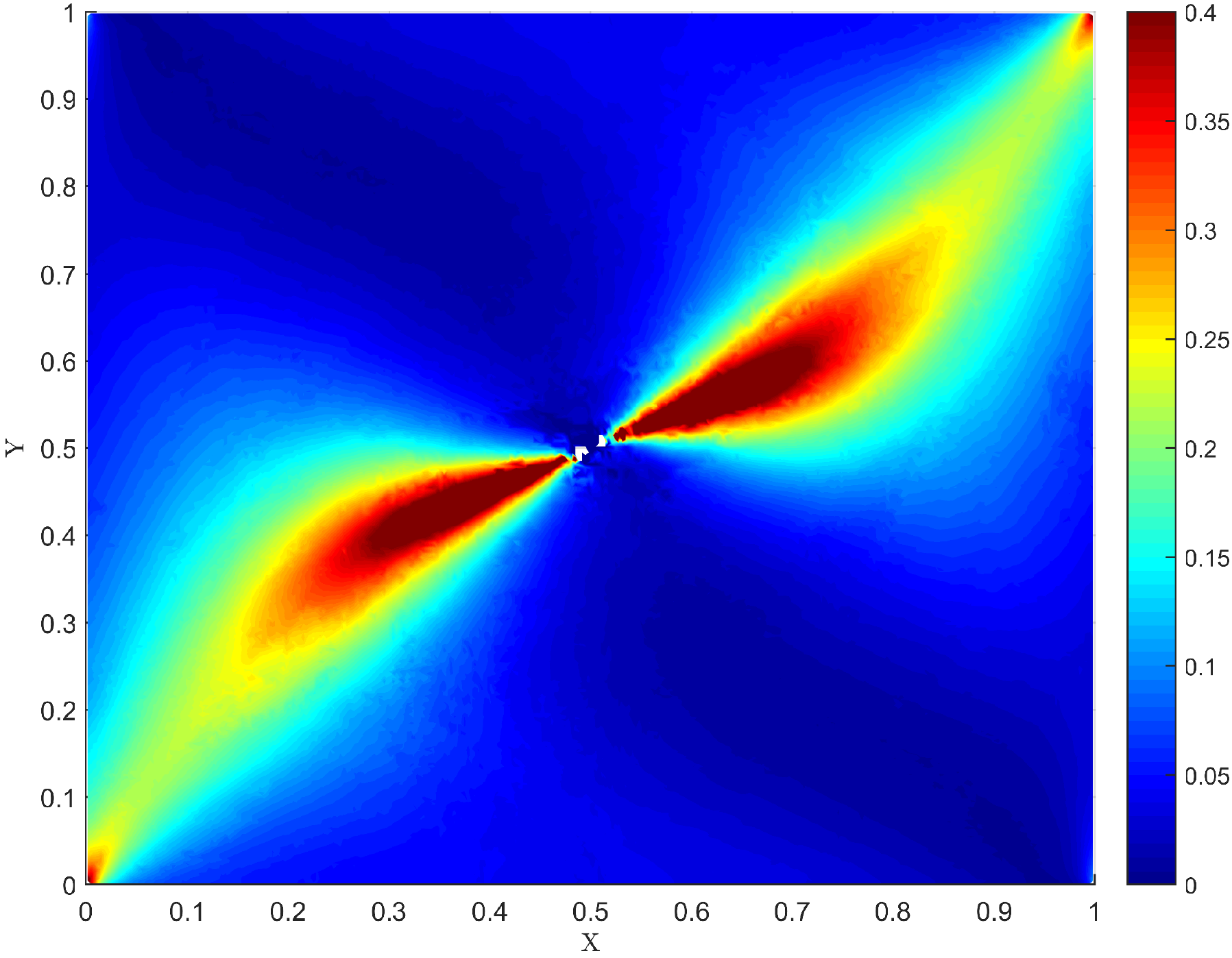}
	}
	\subfigure
	[Spatial distribution of total Sobol' index of $r_k$;]
	{
		\includegraphics[width=.47\textwidth,clip = true, trim = 0 0 0 
		0]{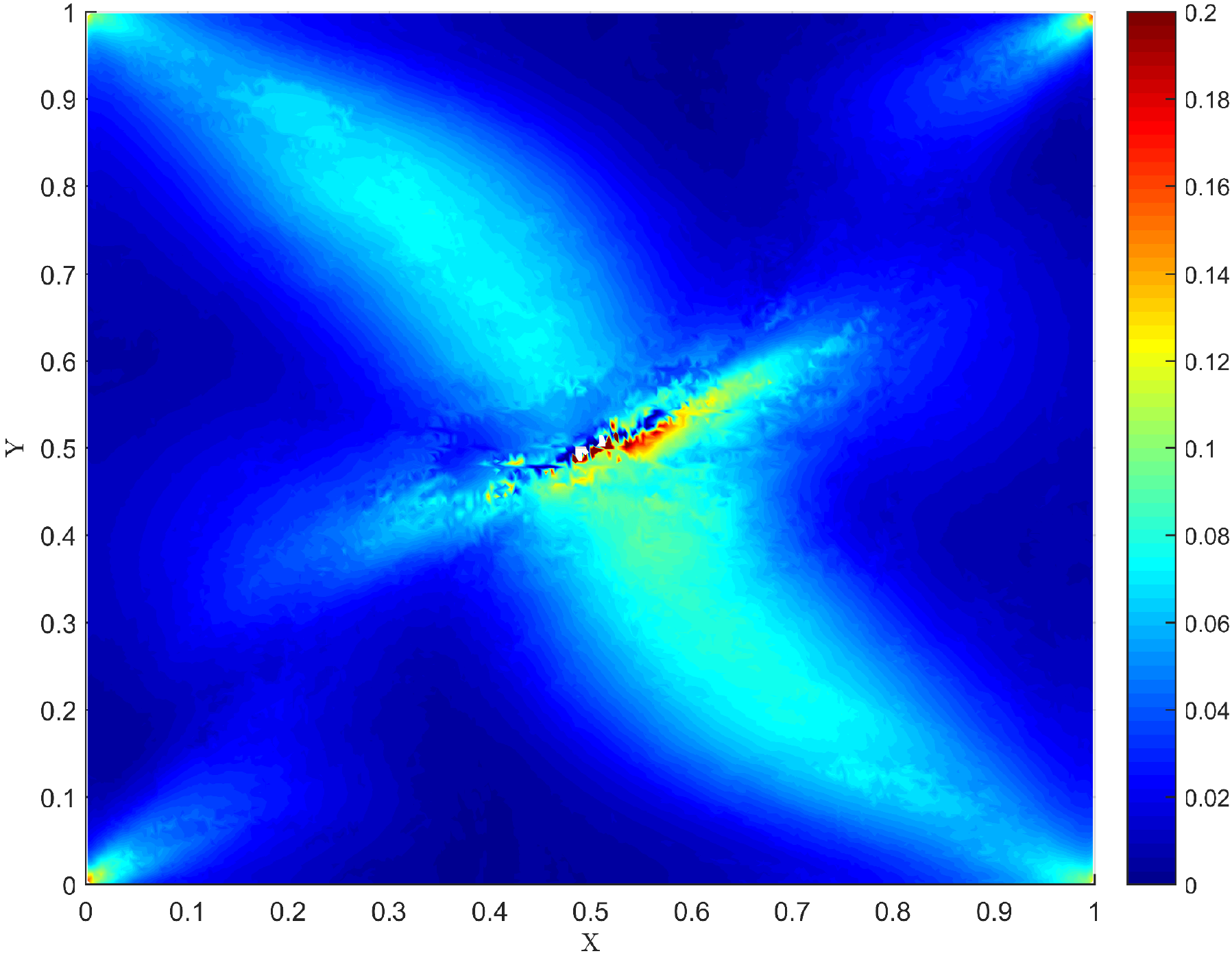}
	}
	\subfigure
	[ Spatial distribution of total Sobol' index of $\alpha_L^\ast$;]
	{
			\includegraphics[width=.47\textwidth,clip = true, trim = 0 0 0 
			0]{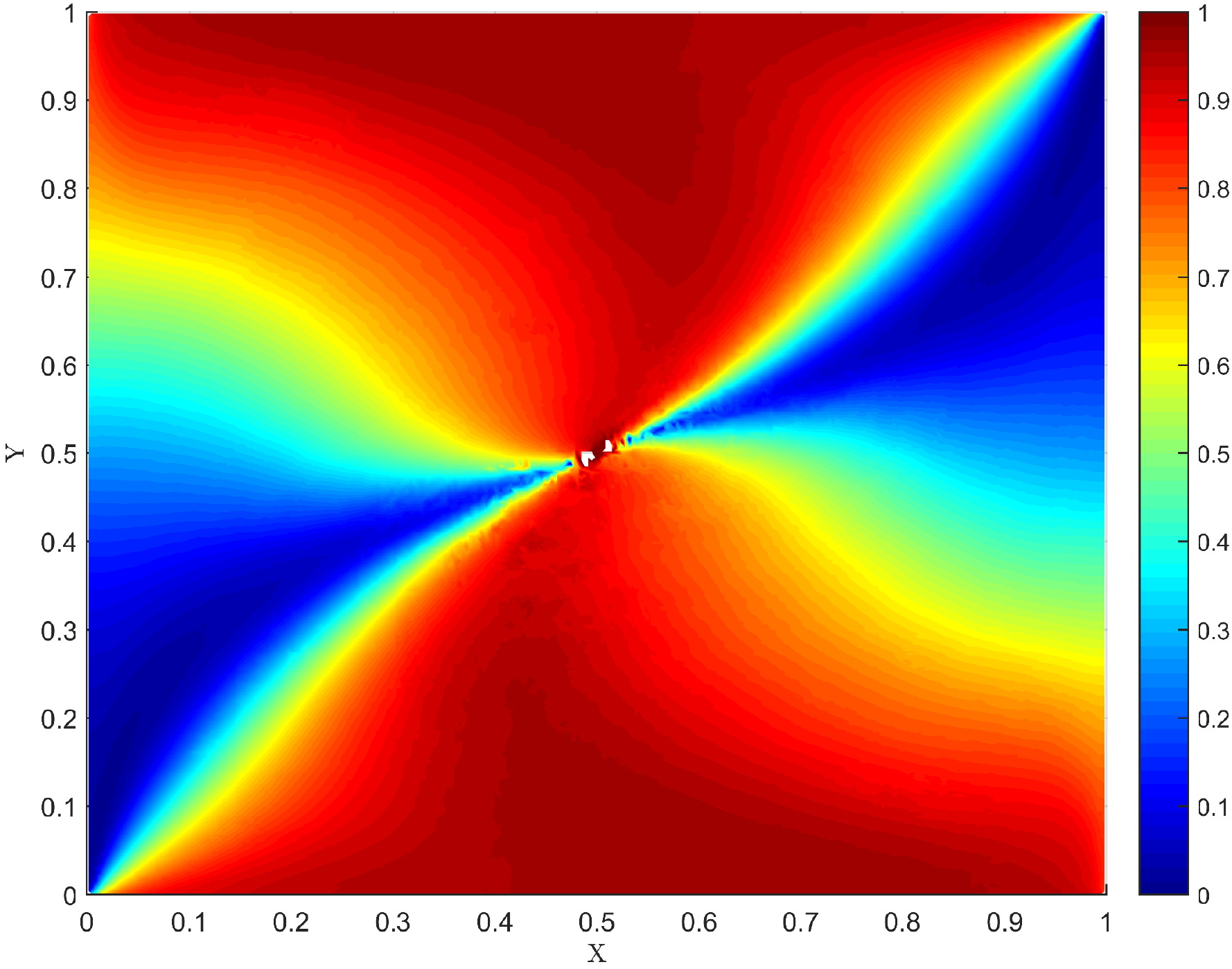}
	}
	\subfigure
	[Spatial distribution of total Sobol' index of $\alpha_T^\ast$;]
	{
		\includegraphics[width=.47\textwidth,clip = true, trim = 0 0 0 
		0]{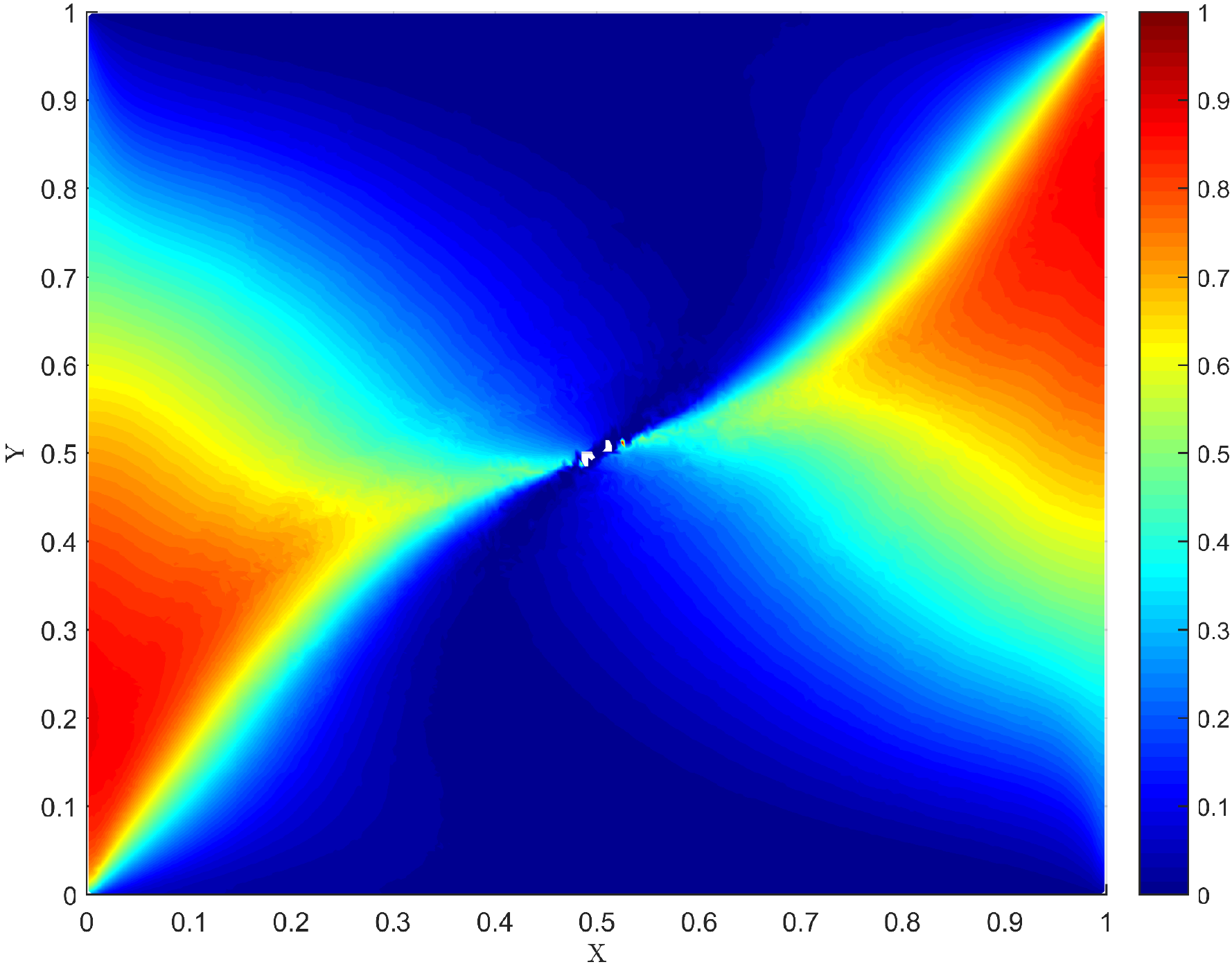}
	}

	\caption{Homogeneous case - Spatial distribution of the total Sobol' indices for the temperature}
	\label{fig:4}
\end{figure}

The sensitivity of the temperature field to the variability of the random parameters can be assessed by means of spatial maps of the Sobol' indices. Fig. \ref{fig:4} shows the spatial distribution of total Sobol' indices due to uncertainty in $\overline{Ra}$, $r_k$, $\alpha_L^\ast$ and $\alpha_T^\ast$. We recall that the total Sobol' indices involve the total effect of a parameter including nonlinearities as well as interactions. Thus, they allow us to rank the parameters according to their importance. Focusing on Fig. \ref{fig:4}, we can see that the most influential parameters are $\alpha_L^\ast$ and $\alpha_T^\ast$. A complementary effect between these parameters is observed. The effect of $\alpha_L^\ast$ is more pronounced than that for $\alpha_T^\ast$ as its zone of influence is located in the region where the temperature variance is maximum. It is worth nothing that complementary effect between the influence of $\alpha_L^\ast$ and $\alpha_T^\ast$ can be explained by reformulating the dispersive heat flux in terms of the dot product of the velocity and temperature gradient vectors. Considering Eqs. \eqref{eqn:5}-\eqref{eqn:7} the thermal dispersive flux $\ve{q_{disp}}$ can be rearranged as follows: 

\begin{equation} 
\label{eqn:42}
\left [ \begin{array}{l}
q_{disp}^x\\
q_{disp}^z
\end{array} \right ]
=\left  [
\begin{array}{l}
\alpha_T \left( \mid\ve{V}\mid \dfrac{\partial \ve{T}}{\partial x} -\dfrac{u}{\mid \ve{V} \mid} \left(\ve{V} \cdot \nabla \ve{T}\right) \right)+ \dfrac{u \alpha_L}{ \mid \ve{V} \mid} \left(\ve{V} \cdot \nabla \ve{T}\right)\\
\alpha_T \left( \mid \ve{V} \mid \dfrac{\partial \ve{T}}{\partial y} -\dfrac{v}{\mid \ve{V} \mid} \left(\ve{V} \cdot \nabla \ve{T}\right) \right)+ \dfrac{v \alpha_L}{ \mid \ve{V} \mid} \left(\ve{V} \cdot \nabla \ve{T}\right)
\end{array} 
\right ] .
\end{equation}
where $\ve{V}$ is the velocity vector. This equation reveals that the longitudinal (resp. transverse) dispersion is an increasing (resp. decreasing) function of $\ve{V}\cdot \nabla T$. Around the top and bottom surfaces of the cavity, the velocity is almost horizontal and parallel to the thermal gradient. Hence, $\ve{V}\cdot \nabla T$  exhibits its maximum value and by consequence temperature distribution in these zones is mainly controlled by $\alpha_L^\ast$  (see Fig. \ref{fig:4}). The velocity near the vertical walls is vertical and relatively perpendicular to the thermal gradient. $\ve{V}\cdot \nabla T$ tends towards zero (minimum value). Hence, the temperature distribution in these regions is mainly controlled by $\alpha_T^\ast$ (see Fig. \ref{fig:4}d).\\

The sensitivity of the temperature field due to $\overline{Ra}$ is less important than that of $\alpha_L^\ast$ and $\alpha_T^\ast$ (Fig. \ref{fig:4}a). Its zone of influence expands along the cavity diagonal bisector with increasing magnitude toward the cavity center and near the corners. The effect of $\overline{Ra}$ on the temperature distribution is related to the heat mixing by thermal diffusion and/or the convection due to the buoyancy effects. Around the top right and bottom left corners the fluid velocity is very small as it should be zero right on the corners to satisfy the boundary conditions. Hence heat mixing between hot and cold fluids by thermal dispersion is almost negligible. In addition, around these corners the thermal boundary layers are very thin. Thus, the temperature gradient is very important so that mixing by thermal diffusion is dominating and by consequence temperature distribution is sensitive to $\overline{Ra}$. In the center of the cavity the thermal dispersion tensor is also negligible because the rotating flow is relatively slow. Hence mixing is mainly related to diffusion and by consequence sensitivity to the $\overline{Ra}$ is relatively important. 
Fig. \ref{fig:4}b indicates that the temperature distribution is slightly sensitive to the permeability ratio. The zone of influence of $r_k$ matches well with the region in which the flow is strongly bidirectional. This is physically understandable, since $r_k$ expresses the ability of a porous media to transmit fluid in a direction perpendicular to the main flow. Hence $r_k$ is a non-influent parameter in the zones where the flow is almost unidirectional.

\begin{itemize}
		\item \textit{GSA of the scalar QoIs}
\end{itemize}

\begin{figure}[!ht]
	\centering
	\includegraphics[width=.6\textwidth,clip = true, trim = 0 0 0
        0]{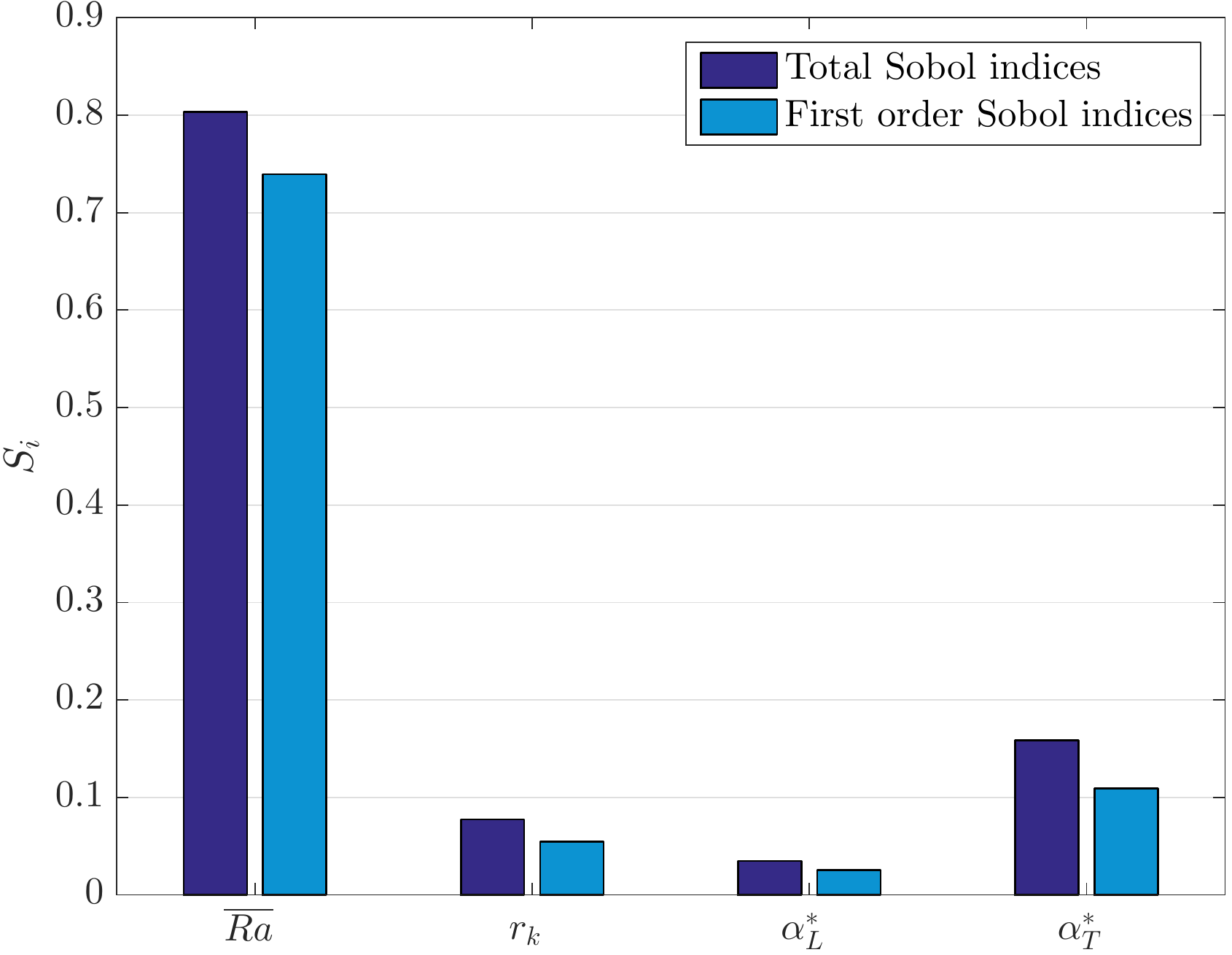}
	\caption{Homogeneous case - Total Sobol' indices for the average of the average of Nusselt number $\overline{Nu}$;}
	\label{fig:5}
\end{figure}

Fig. \ref{fig:5} shows bar-plots of the first order and total Sobol’ indices of the average Nusselt number $\overline{Nu}$. Inspection of the sensitivity indices showed that the variability of $\overline{Nu}$ is mainly due to the principal effects of $\overline{Ra}$ and $\alpha_T^\ast$. The most influential parameter is $\overline{Ra}$ with $S^{T}_{\overline{Ra}}=0.8$. A small influence of $r_k$ and $\alpha_L^\ast$ is also observed. Interactions between the random parameters are not significant (not shown). The maximum value obtained is $S_{\overline{Ra},\alpha_T^\ast}=0.035$.  The results are summarized in Table \ref{Tab:2}.\\

\begin{table} [!ht]
	\centering
	\caption{Homogeneous case - Sobol' indices for $\overline{Nu}$}
	\label{Tab:2}
	\begin{tabular}{c c c c c}
		\hline    & $\overline{Ra}$& $r_k$ &$\alpha_L^\ast$ &    $\alpha_T^\ast$  \\
		\hline $S_{i}$ & $0.739$ & $0.0547$ & $0.0252$& $0.1093$  \\
		$S^{T}$ & $0.803$ & $0.077$ & $0.0235$& $0.158$   \\
				\hline       
	\end{tabular}
\end{table}

\begin{figure}[ht]
	\centering
	\subfigure
	[Effect of $\overline{Ra}$]
	{
		\includegraphics[width=.4\textwidth,clip = true, trim = 0 0 0 
		0]{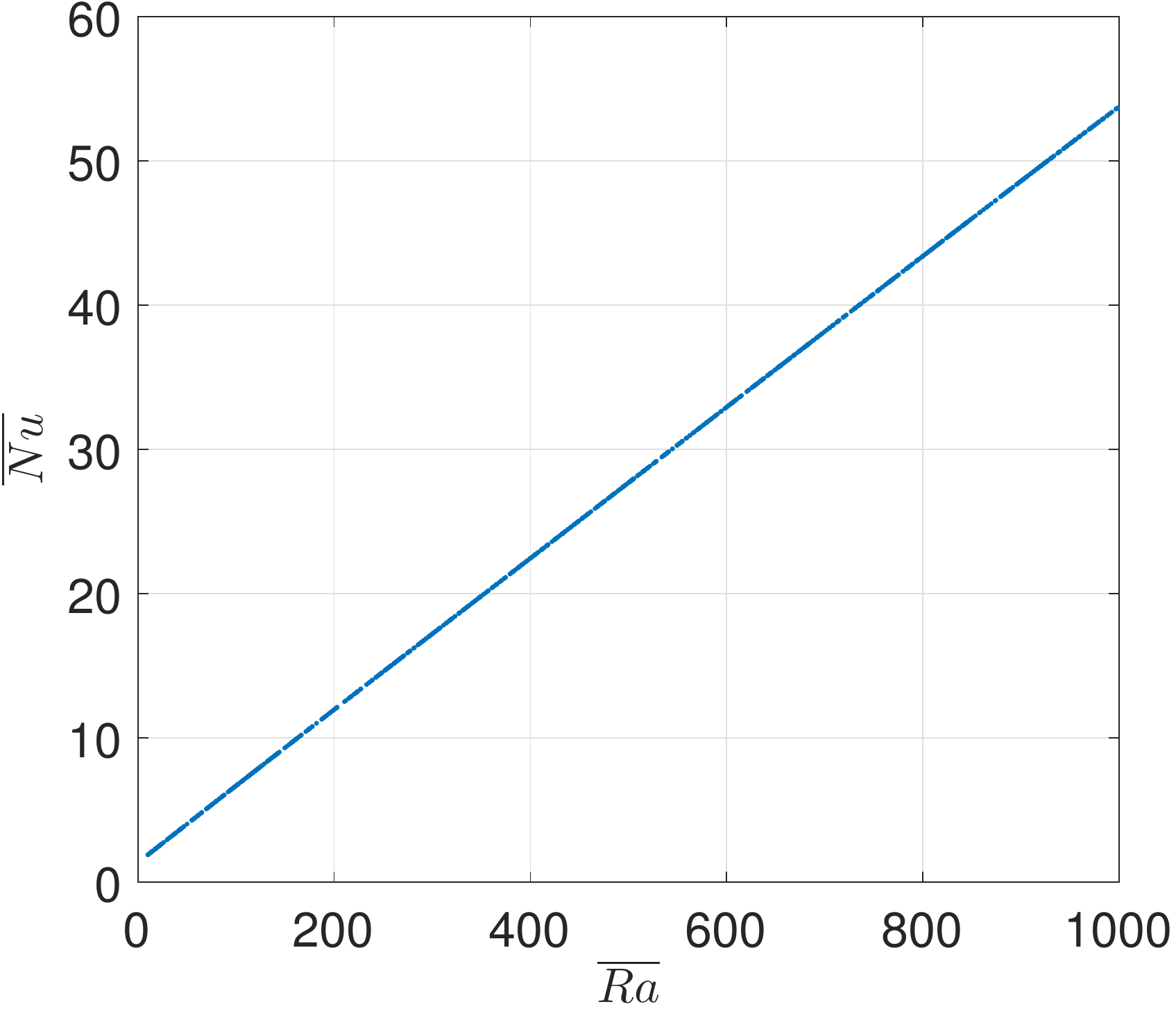}
	}
	\subfigure
	[Effect of $r_k$]
	{
		\includegraphics[width=.4\textwidth,clip = true, trim = 0 0 0 
		0]{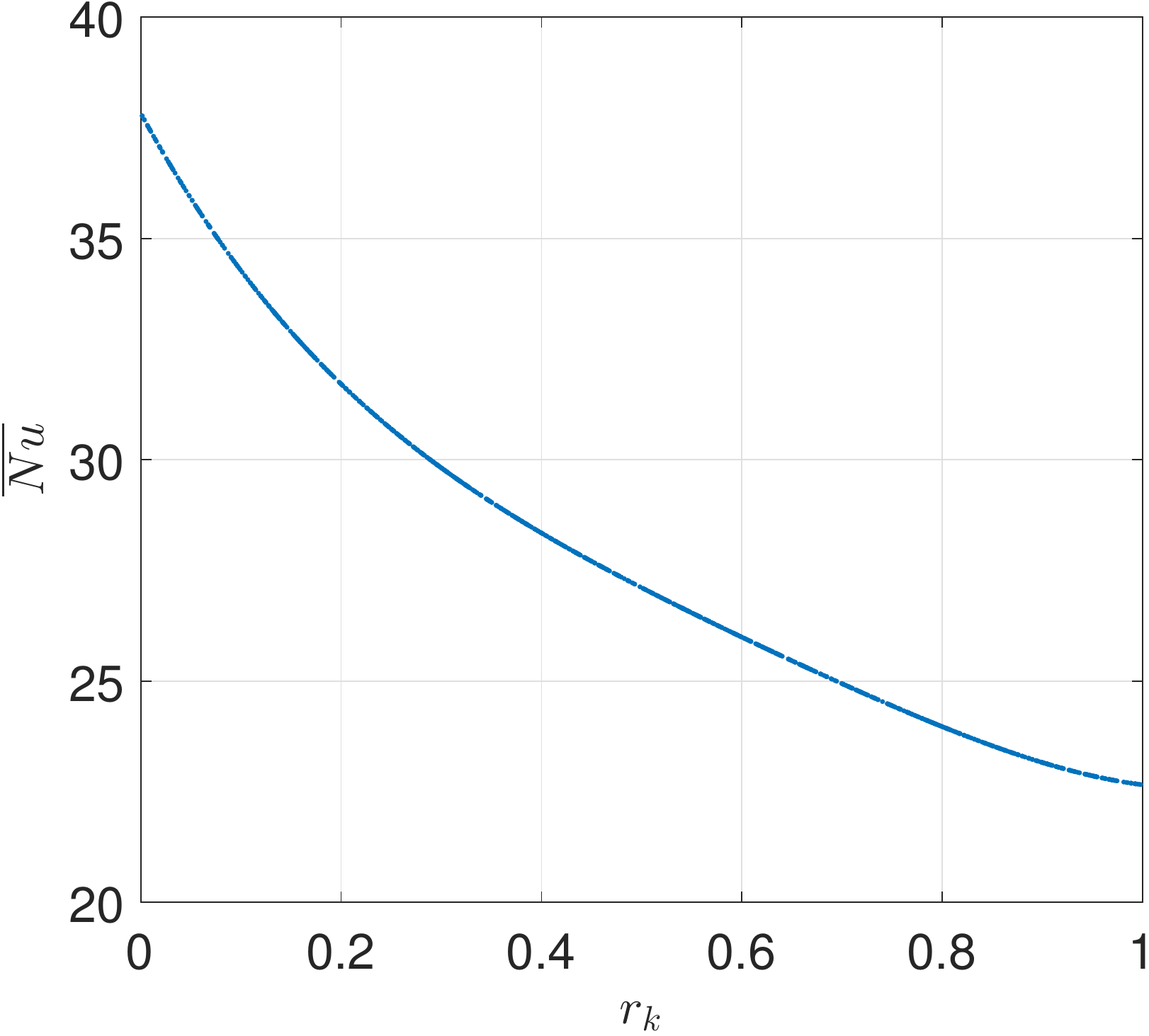}
	}
	\subfigure
	[Effect of $\alpha_L^\ast$]
	{
		\includegraphics[width=.4\textwidth,clip = true, trim = 0 0 0 
		0]{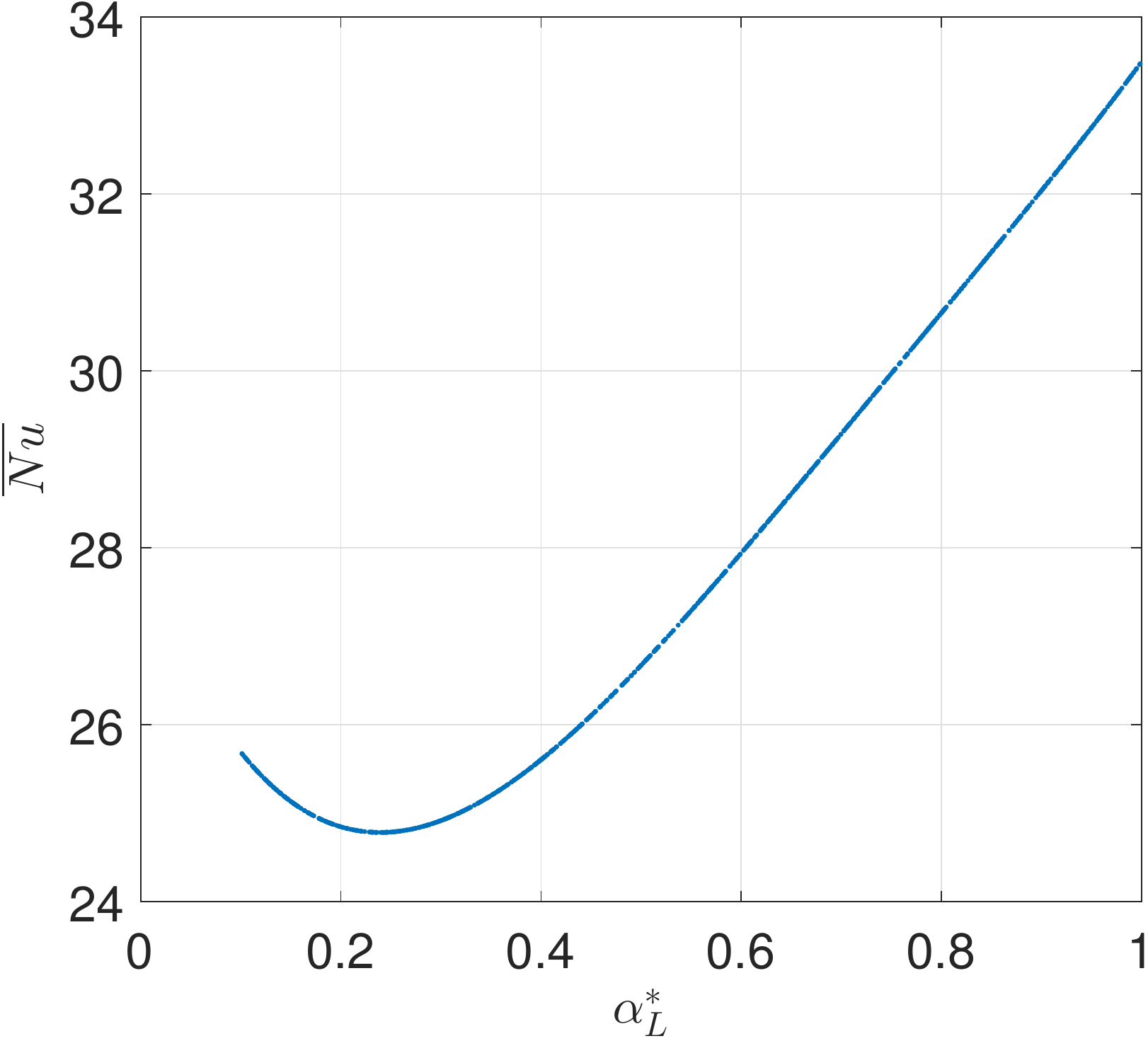}
	}
	\subfigure 
	[Effect of $\alpha_T^\ast$]
	{
		\includegraphics[width=.4\textwidth,clip = true, trim = 0 0 0 
		0]{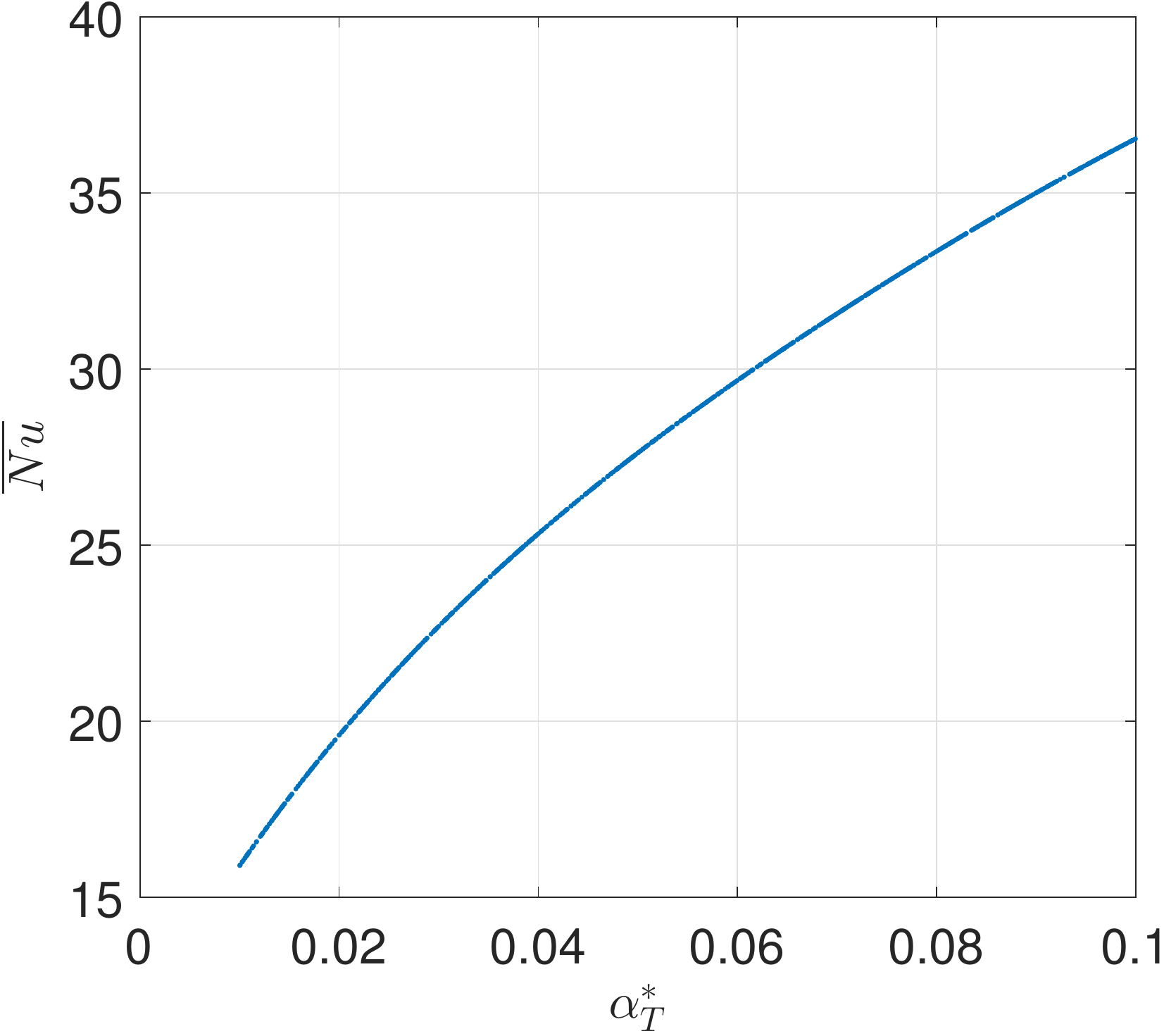}
	}
	\caption{Homogeneous case - Univariate effects of the input parameters on the average Nusselt number $\overline{Nu}$}
	\label{fig:6}
\end{figure}

To further elaborate our investigation, we examine the marginal effect of the uncertain parameters on the model response. This effect corresponds to the evolution of the model output with respect to a single parameter averaged on the other parameters (Eq. \eqref{eqn:37}). Indeed, if a parameter is sensitive, significant variations (positive or negative slopes) are expected whereas a weakly sensitive parameter results in small variations of the model responses. Results for the marginal effect of the parameters on $\overline{Nu}$ are shown in Fig. \ref{fig:6}. Different scales are observed indicating their level of influence. In general, the marginal effects are in agreement with the global sensitivity analysis.\\

Fig. \ref{fig:6}a demonstrates that $\overline{Nu}$ increases with $\overline{Ra}$. Indeed, the increase of $\overline{Ra}$ enhances the buoyancy effects and reduces the thickness of the thermal boundary layer in the lower part of the hot wall. This leads to higher values of $\overline{Nu}$ as the temperature distribution becomes steep near the hot wall, especially around the bottom corner. Same behavior of $\overline{Nu}$ is observed against $\alpha_T^\ast$ (Fig. \ref{fig:6}d). This parameter affects slightly the velocity field (as it will be shown later in this paper). It slightly affects the temperature distribution at the hot wall as we can see in the Figures \ref{fig:3} and \ref{fig:4}d. Hence considering the expression of $\overline{Nu}$ (Eq.~\eqref{eqn:41} ) it is logical that $\overline{Nu}$ increases with $\alpha_T^\ast$. Similar results have been reported in \cite{hong1987analysis} where authors showed that when the transverse dispersion effect dominates, the heat transfer is greatly increased. This is also in agreement with the results reported by Sheremet et al., \cite{sheremet2016effect} for natural convection in a porous cavity filled with a nanofluid.\\

 Fig. \ref{fig:6}b shows that $\overline{Nu}$ decreases with the increase of the permeability anisotropy ratio ($r_k$). This is consistent with the results obtained in Bennacer et al., \cite{bennacer2001double} and Ni and Beckermann \cite{ni1991natural} for natural convection (without thermal dispersion) and for equivalent ranges of parameters. This behavior can be explained by the fact that at a constant value of the Rayleigh number (i.e. $k_y$ is constant), the increase of $r_k$ can be interpreted as a decrease of the permeability in the horizontal direction $k_x$. This entails a weaker convective flow with more expanded thermal boundary layers ( in the lower part of the hot wall) and by consequence smaller $\overline{Nu}$. The marginal effect of $\alpha_L^\ast$ (Fig. \ref{fig:6}c) indicates the existence of two regimes for the evolution of the Nusselt number. Thus, we have a decreasing  $\overline{Nu}$ for $\alpha_L^\ast<0.25$ and an inverse behavior for $\alpha_L^\ast>0.25$. Indeed, when $\alpha_L^\ast$ is increased, the mixing zone between the hot and cold fluids expands in the direction of the flow. This will push the highest and lowest isotherms towards the vertical walls and increase by consequence the thermal gradient in the vertical boundary layers. On the other hand, this redistribution of the thermal gradient leads to an attenuation of the rotating flow within the cavity.  Thus, referring to the expression of the Nusselt number (Eq.~\eqref{eqn:41}), we can deduce that, for small values of $\alpha_L^\ast$ $(<0.25)$, $\overline{Nu}$ decreases as the velocity variation is predominating. For large values of $\alpha_L^\ast$ $(>0.25)$, $\overline{Nu}$ increases because the effect of the thermal gradient becomes significant and more important than the velocity.\\

\begin{figure}[!ht]
	\centering
	\includegraphics[width=.6\textwidth,clip = true, trim = 0 0 0 
	0]{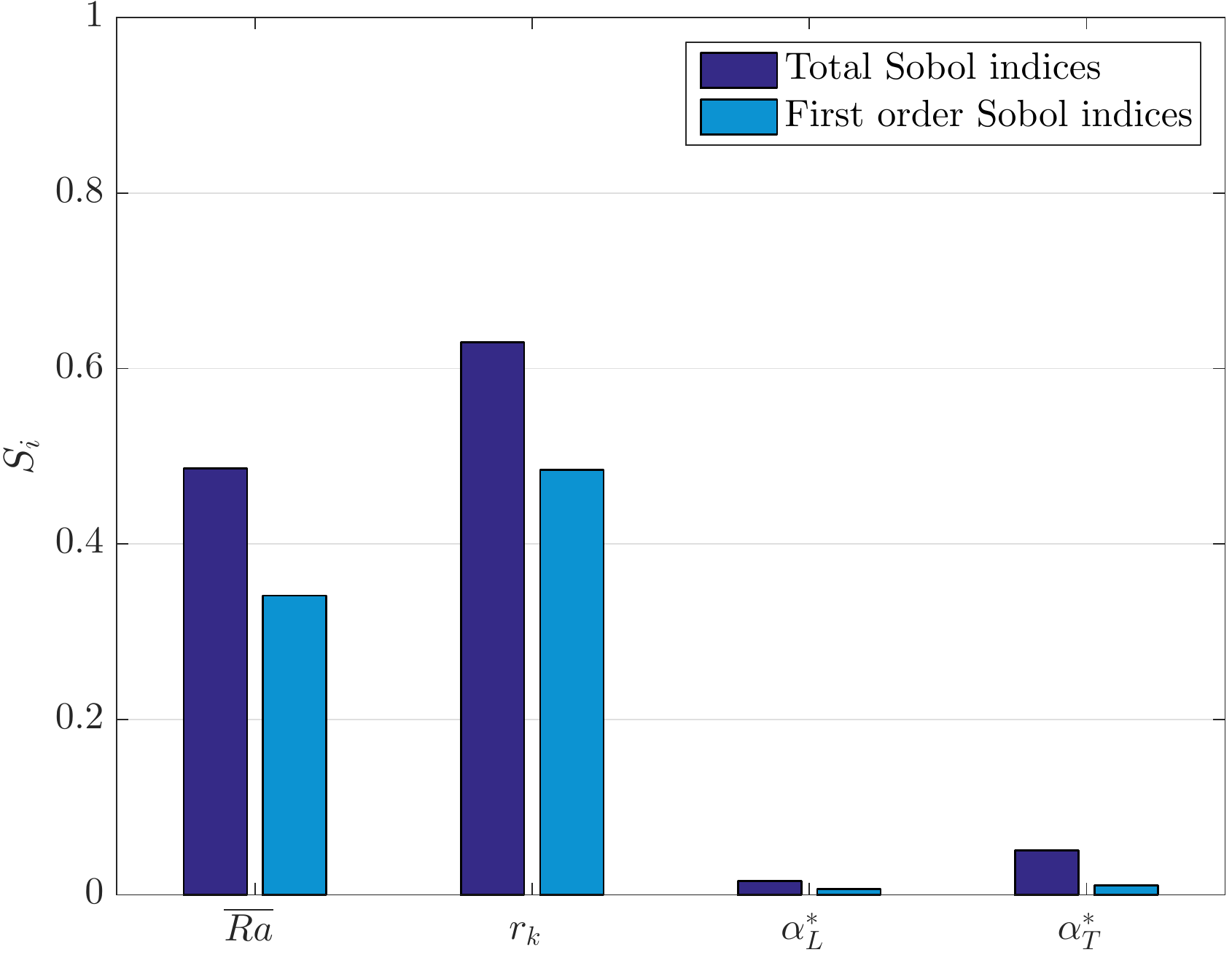}
	\caption{Homogeneous case - Total Sobol' indices for  $u_{max}^\ast$;}
	\label{fig:7}
\end{figure}

\begin{figure}[!ht]
	\centering
	\subfigure
	[	Effect of $\overline{Ra}$]
	{
	\includegraphics[width=.4\textwidth,clip = true, trim = 0 0 0 
	0]{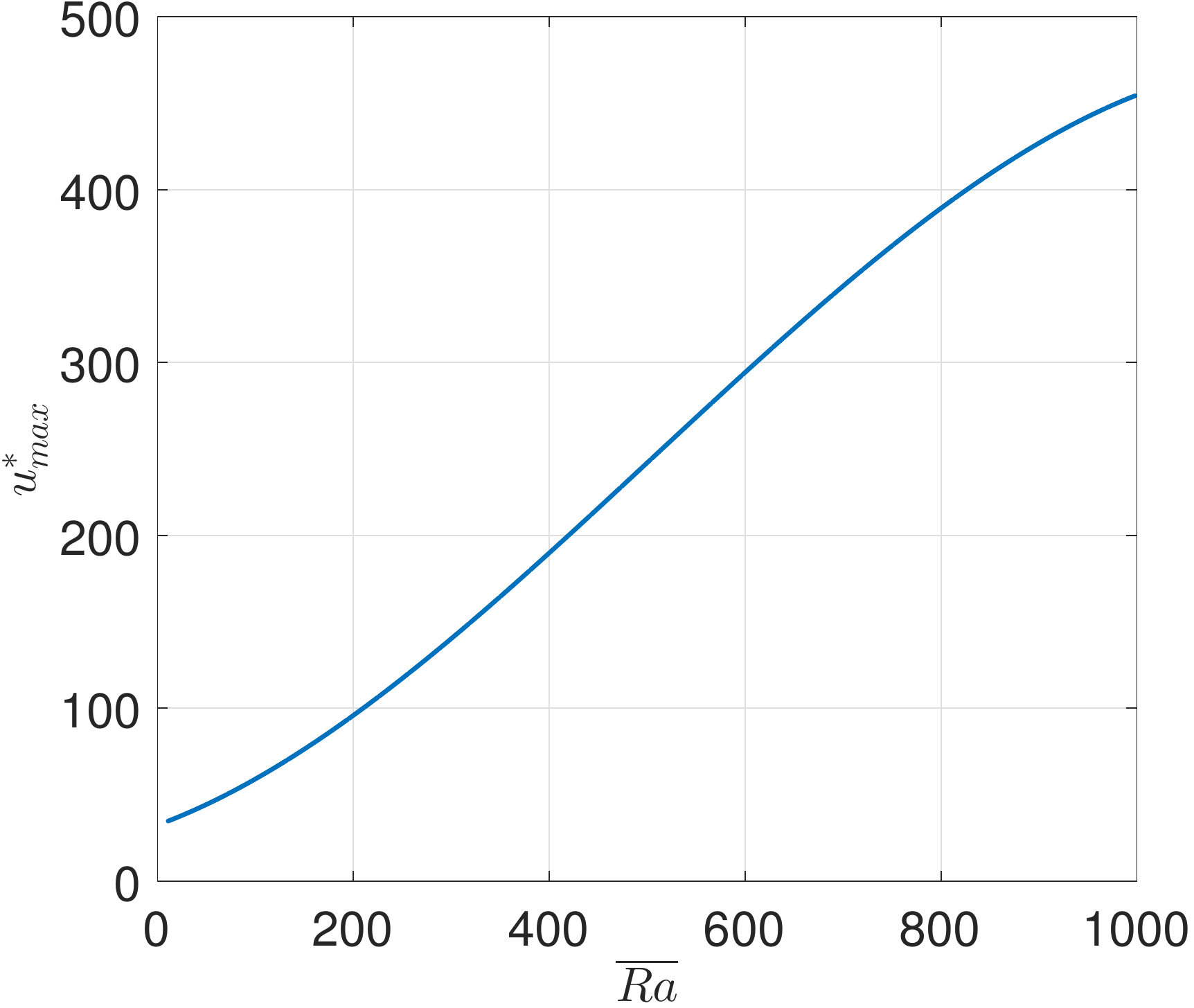}
}
\subfigure
[	Effect of $r_k$]
{
	\includegraphics[width=.4\textwidth,clip = true, trim = 0 0 0 
	0]{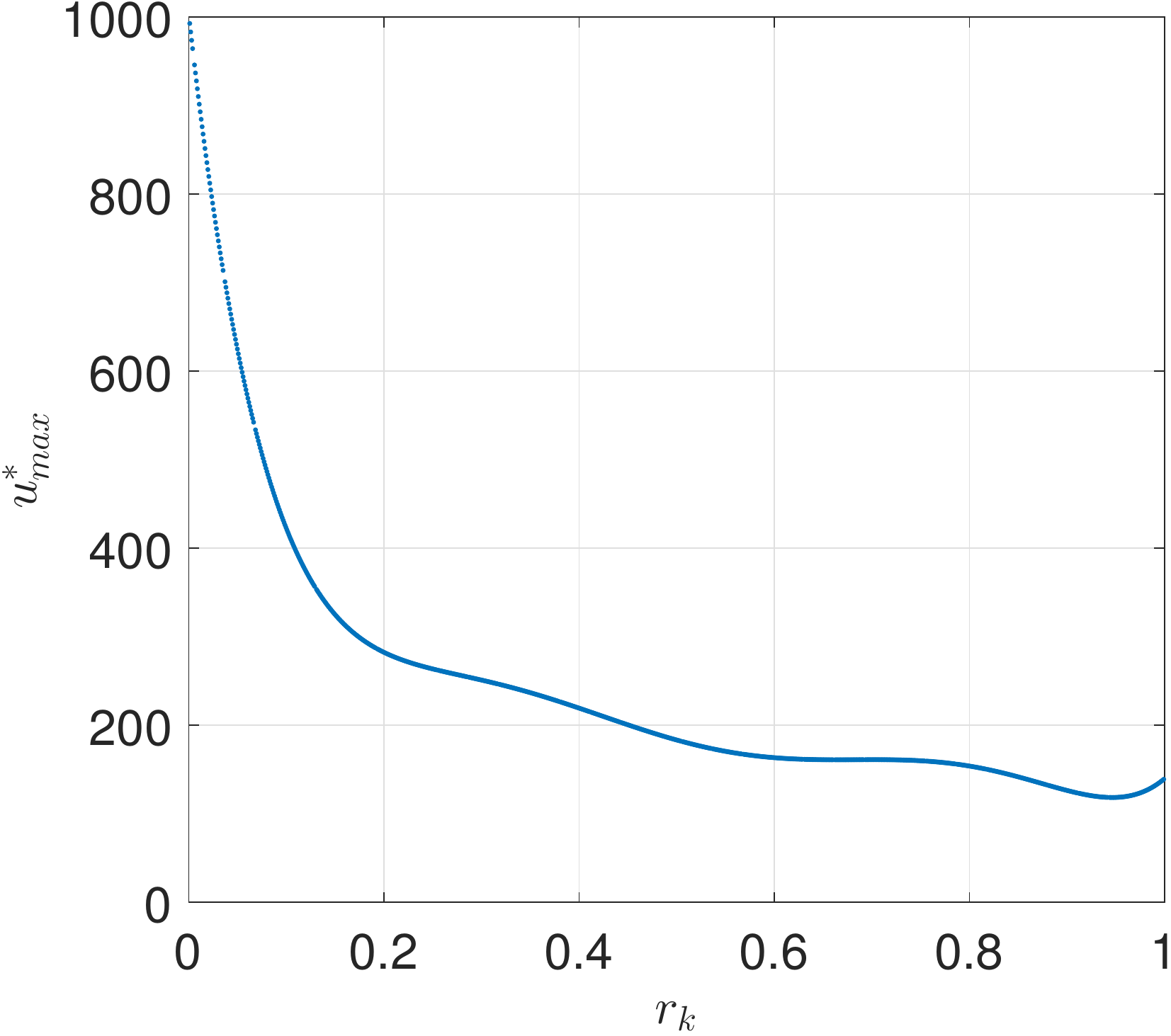}
}
\subfigure
[	Effect of $\alpha_L^\ast$]
{
	\includegraphics[width=.4\textwidth,clip = true, trim = 0 0 0 
	0]{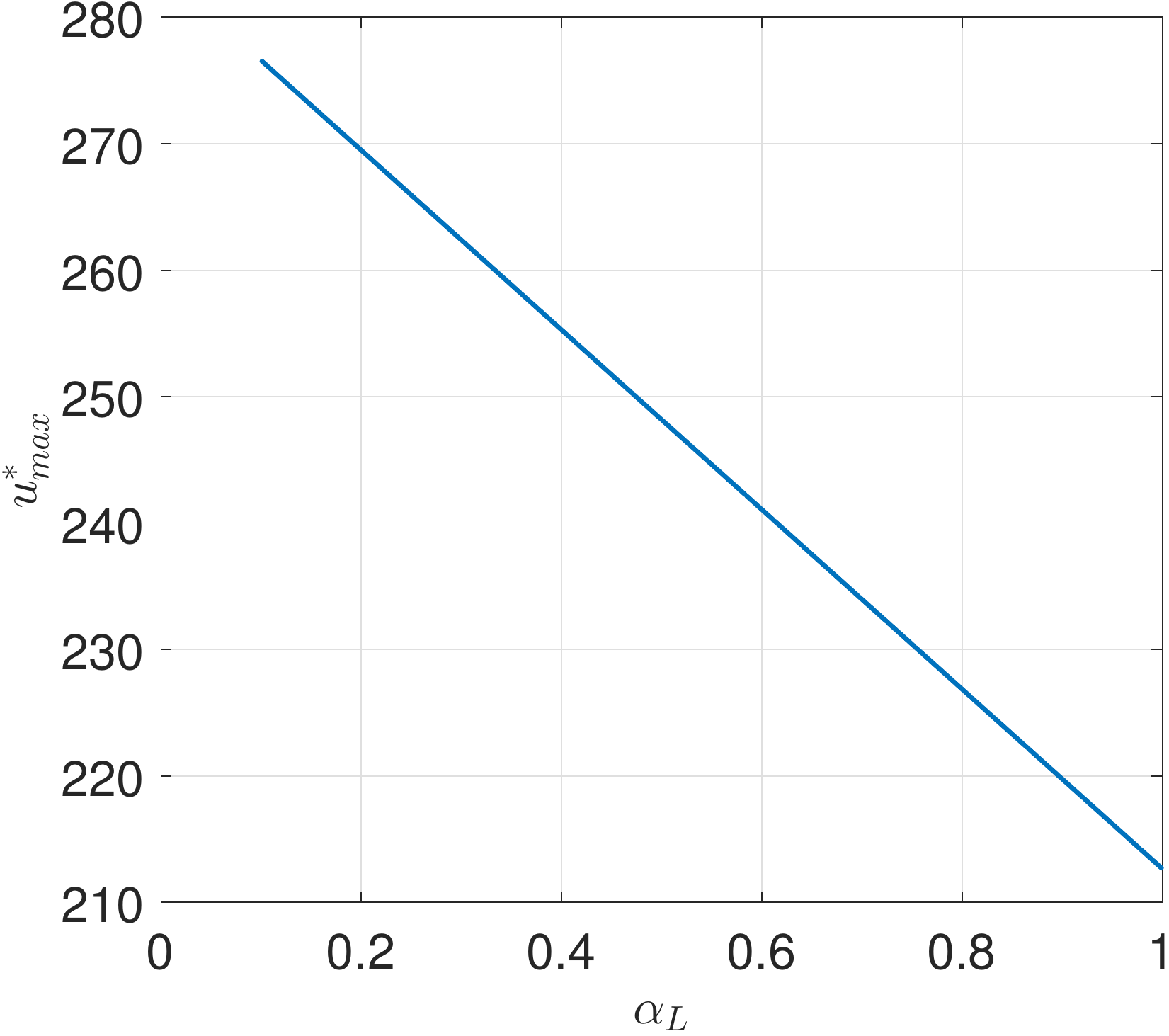}
}
\subfigure
[	Effect of $\alpha_T^\ast$]
{
	\includegraphics[width=.4\textwidth,clip = true, trim = 0 0 0 
	0]{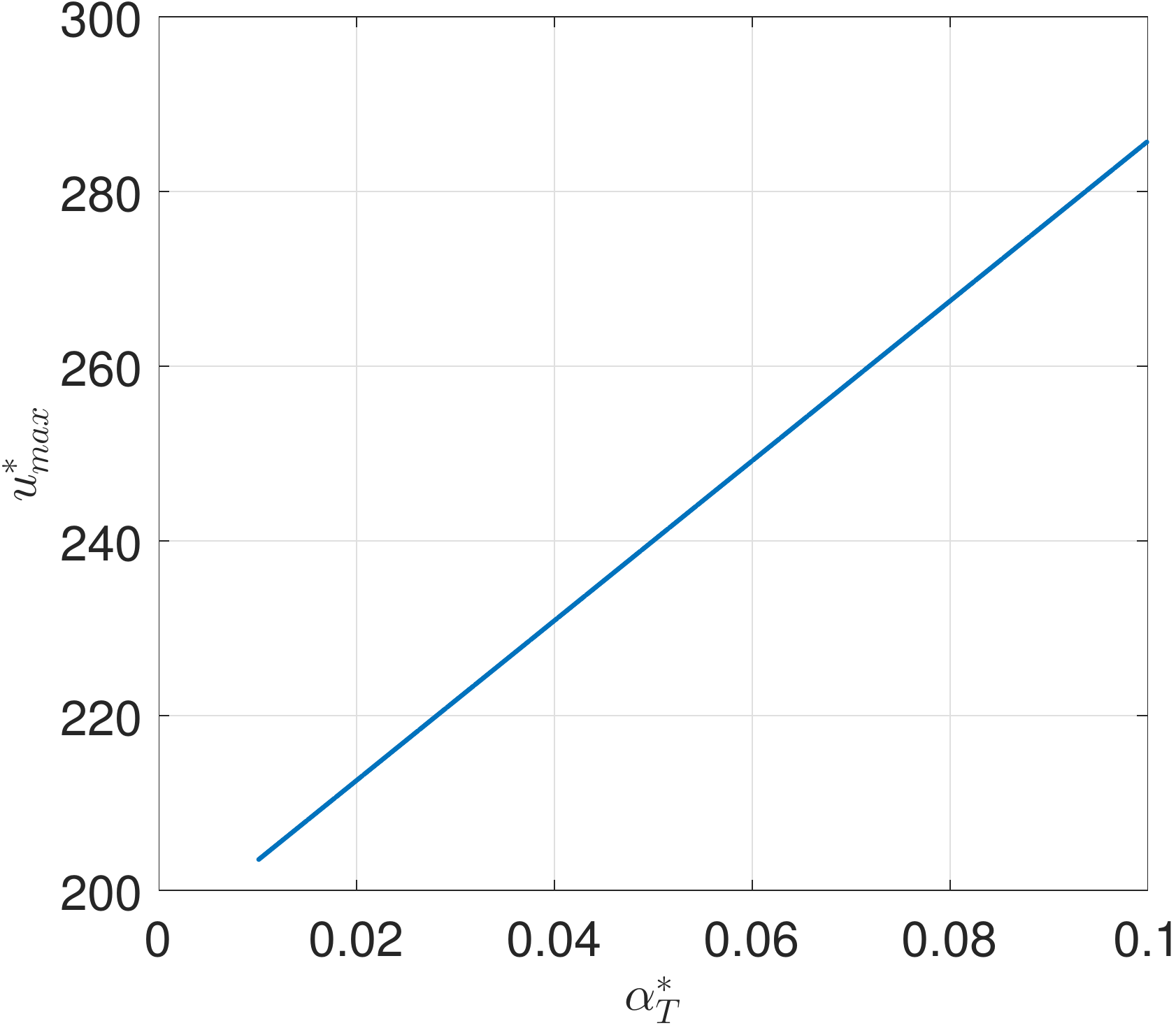}
}

	\caption{Homogeneous case - Univariate effects of the input parameters on $u_{max}^\ast$}

	\label{fig:8}
\end{figure}

Fig. \ref{fig:7} shows bar-plots of the first order and total Sobol'
indices of the maximum velocity $u_{max}^\ast$. Results indicate that
the variability of $u_{max}^\ast$ is mainly controlled by
$\overline{Ra}$ and $r_k$. Interactions between $\overline{Ra}$ and
$r_k$ are also observed. They explain $14.5 \%$ of the total variance of
$u_{max}^\ast$. The total effect of $\alpha_T^\ast$ accounts for
approximately $1.0 \%$. In Fig.~\ref{fig:8}, we display the marginal
effect of the uncertain parameters on $u_{max}^\ast$. One can observe
that $u_{max}^\ast$ increases with the increase of Rayleigh number
$\overline{Ra}$, indicating that the buoyancy-induced flow along the
horizontal surfaces becomes much stronger as $\overline{Ra}$ is
increased. On the contrary, we note that $u_{max}^\ast$ decreases with
the increase of $r_k$, as the latter corresponds to the decrease of the
permeability in the horizontal direction. We can also note that small
variations of $u_{max}^\ast$ are observed when $r_k>0.2$. \\

Figs. \ref{fig:8}c-d confirm that $u_{max}^\ast$ is slightly sensitive
to $\alpha_L^\ast$ and $\alpha_T^\ast$. One can observe that an increase
of $\alpha_L^\ast$ is associated with a decrease of $u_{max}^\ast$. The
effect of $\alpha_L^\ast$ on the velocity can be understood with the
help of the stream function form of the flow equation. This form can be
obtained by applying the curl operator on the Darcy's law. It is a
Poisson equation with the horizontal component of the temperature
gradient as source term. This equation is subject to zero stream
function as boundary conditions. The corresponding solution is
concentric streamlines. The shape of these streamlines (center,
orientation, spacing and density) depends on the source function. For
the problem of natural convection in square cavity, the maximum
horizontal temperature gradient is located at the right bottom and left
top corners. The resulting streamlines have a concentric ellipsoidal
shape with focal axis oriented in the direction of the line connecting
the maximum gradient points (the cavity first bisector). The increase of
$\alpha_L^\ast$ leads to the enhancement of the heat mixing by
longitudinal dispersion in the zones where the velocity is parallel to
the temperature gradient (outside the boundary layers of the hot and
cold walls). By consequence, the horizontal temperature gradient
decreases in this zone and increases outsides. This implies that, on the
one hand, the reorientation of the focal axis of the ellipsoidal shaped
streamlines in the horizontal direction and, on the other hand, an
attenuation of the stream function maximum value. Hence, the rotating
flow decelerates, $u_{max}^\ast$ decreases and the point of maximum
velocity moves toward the center of the cavity surfaces. Reverse
behavior can be observed when $\alpha_T^\ast$ is increased. In such a
case the transverse heat mixing is enhanced in the zone where the
velocity is orthogonal to the temperature gradient (within the boundary
layers). Transverse heat flux is horizontal in this case. Hence, the
horizontal temperature gradient decreases within the boundary layers and
increases outside. The direction of the focal axis moves toward the
first bisector and the maximum value of the stream function
increases. The streamlines becomes more spaced at the vertical walls and
more closed at the top surfaces. The value of $u_{max}^\ast$ increases
and its location at the top (resp. bottom) surface moves toward the cold
(resp. hot) wall.\\ 

\begin{figure}[!ht]
	\centering
	\includegraphics[width=.6\textwidth,clip = true, trim = 0 0 0 
	0]{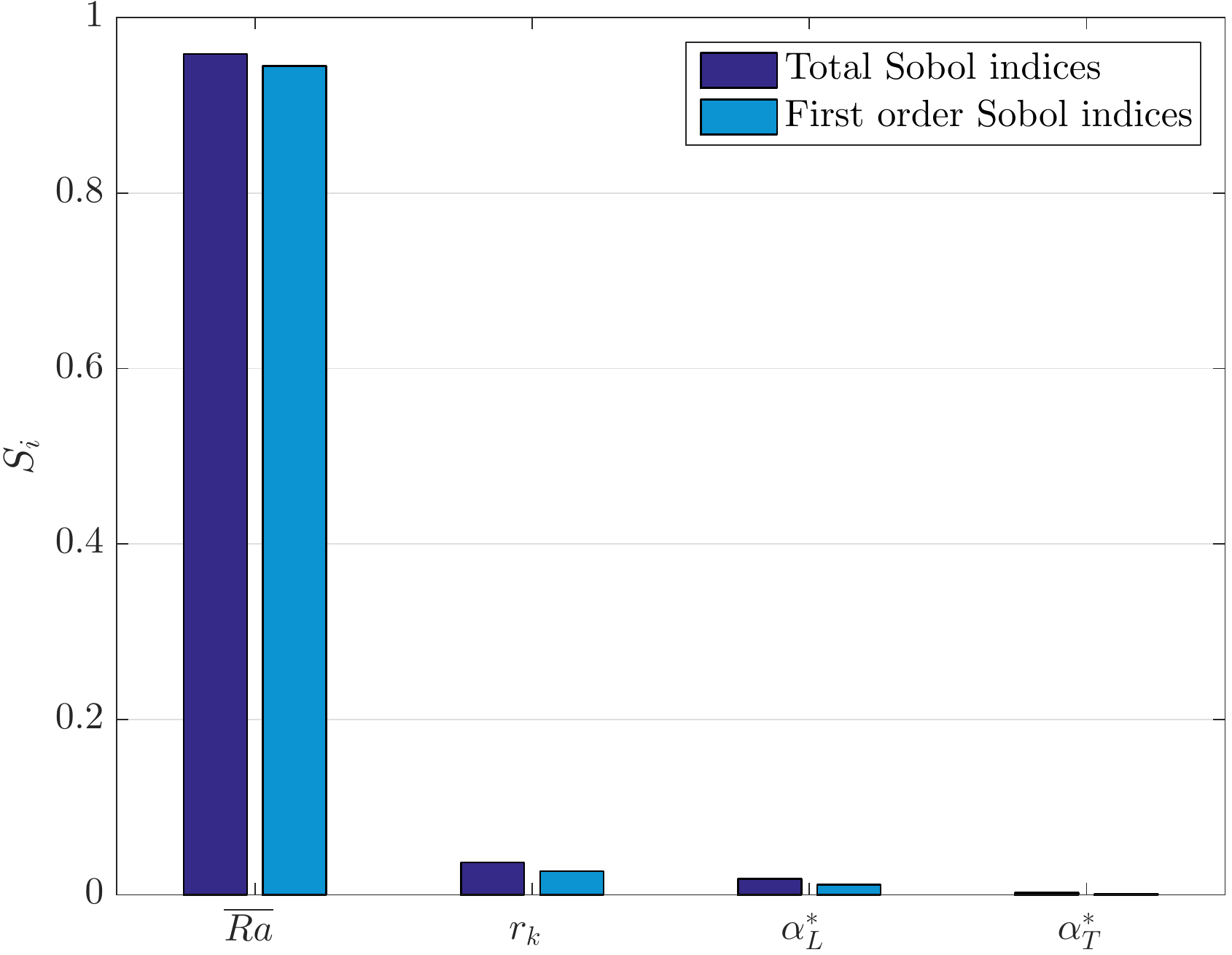}
	\caption{Homogeneous case - Total Sobol' indices for  $v_{max}^\ast$;}
	\label{fig:9}
\end{figure}

Fig. \ref{fig:9} shows the bar-plots of the first order and total
Sobol’ indices of the maximum velocity $v_{max}^\ast$. Results shows
that the variability of $v_{max}^\ast$ is mainly due to the principal
effects of $\overline{Ra}$. Unlike $u_{max}^\ast$, $v_{max}^\ast$ is
only slightly sensitivity to $r_k$. 

The differences between the first-order and total indices are
negligible, indicating insignificant interactions between the
parameters.

\begin{figure}[!ht]
	\centering
	\subfigure
	[	Effect of $\overline{Ra}$]
	{
		\includegraphics[width=.4\textwidth,clip = true, trim = 0 0 0 
		0]{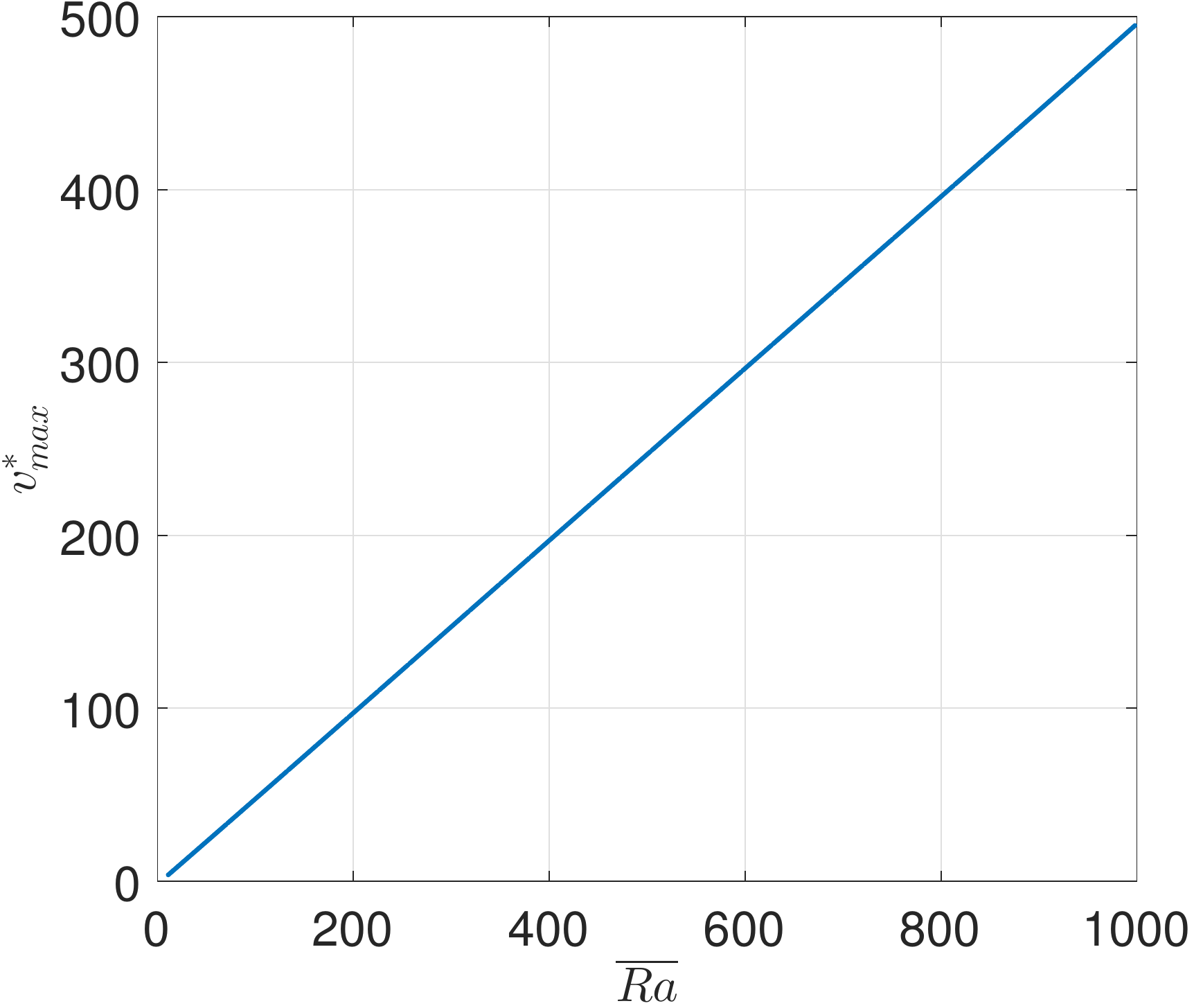}
	}
	\subfigure
	[	Effect of $r_k$]
	{
		\includegraphics[width=.4\textwidth,clip = true, trim = 0 0 0 
		0]{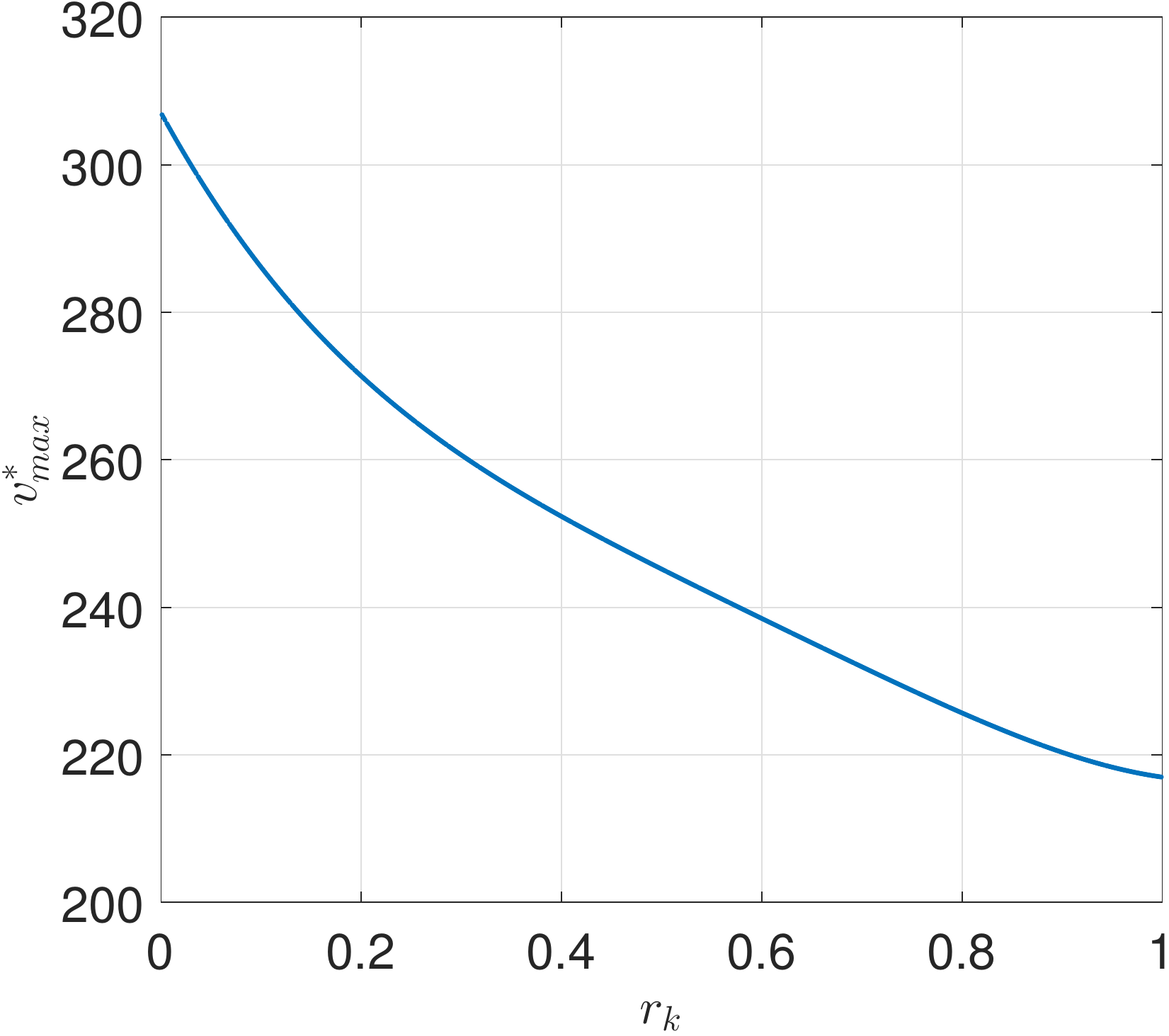}
	}
	\subfigure
	[	Effect of $\alpha_L^\ast$]
	{
		\includegraphics[width=.4\textwidth,clip = true, trim = 0 0 0 
		0]{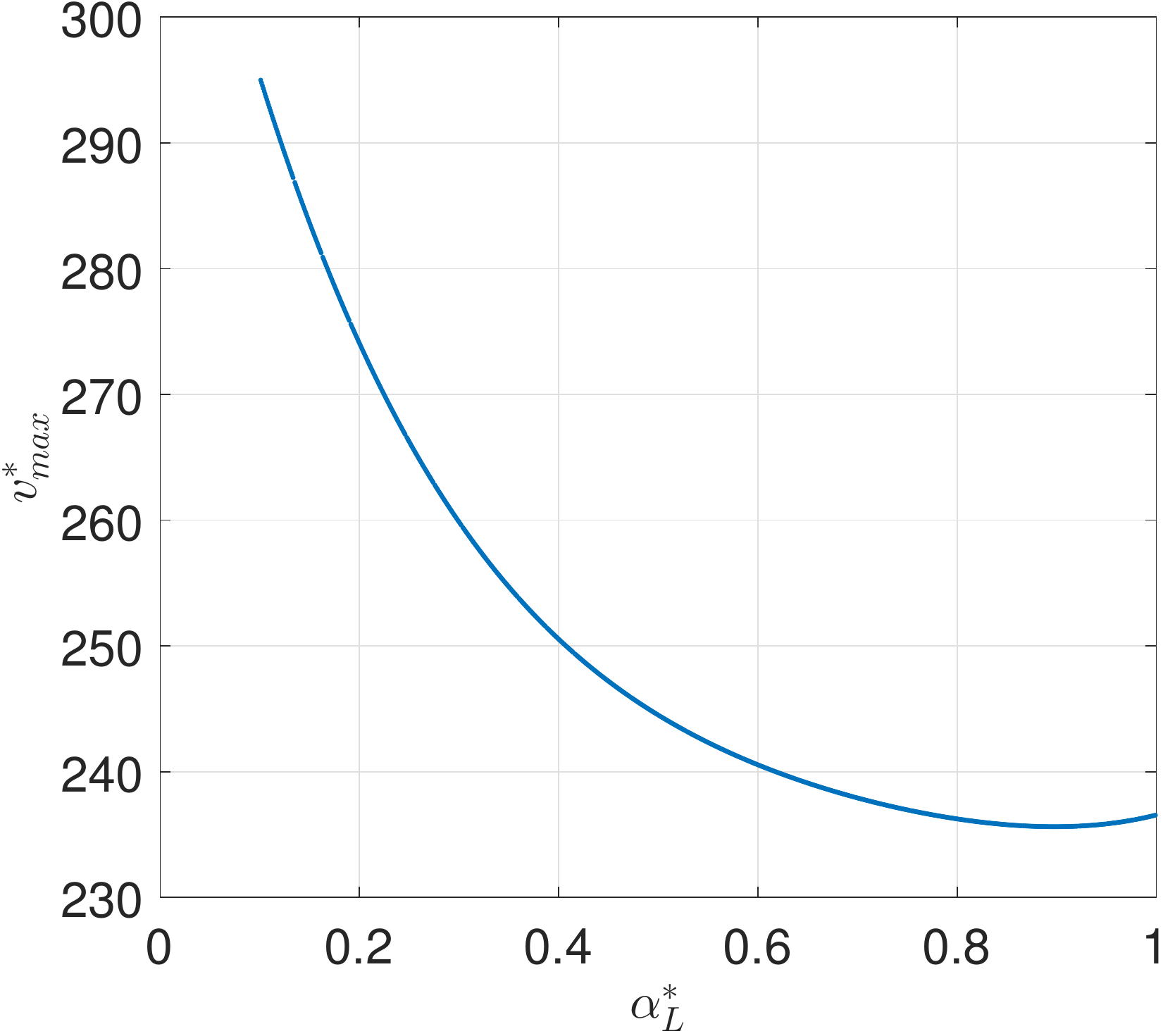}
	}
	\subfigure
	[	Effect of $\alpha_T^\ast$]
	{
		\includegraphics[width=.4\textwidth,clip = true, trim = 0 0 0 
		0]{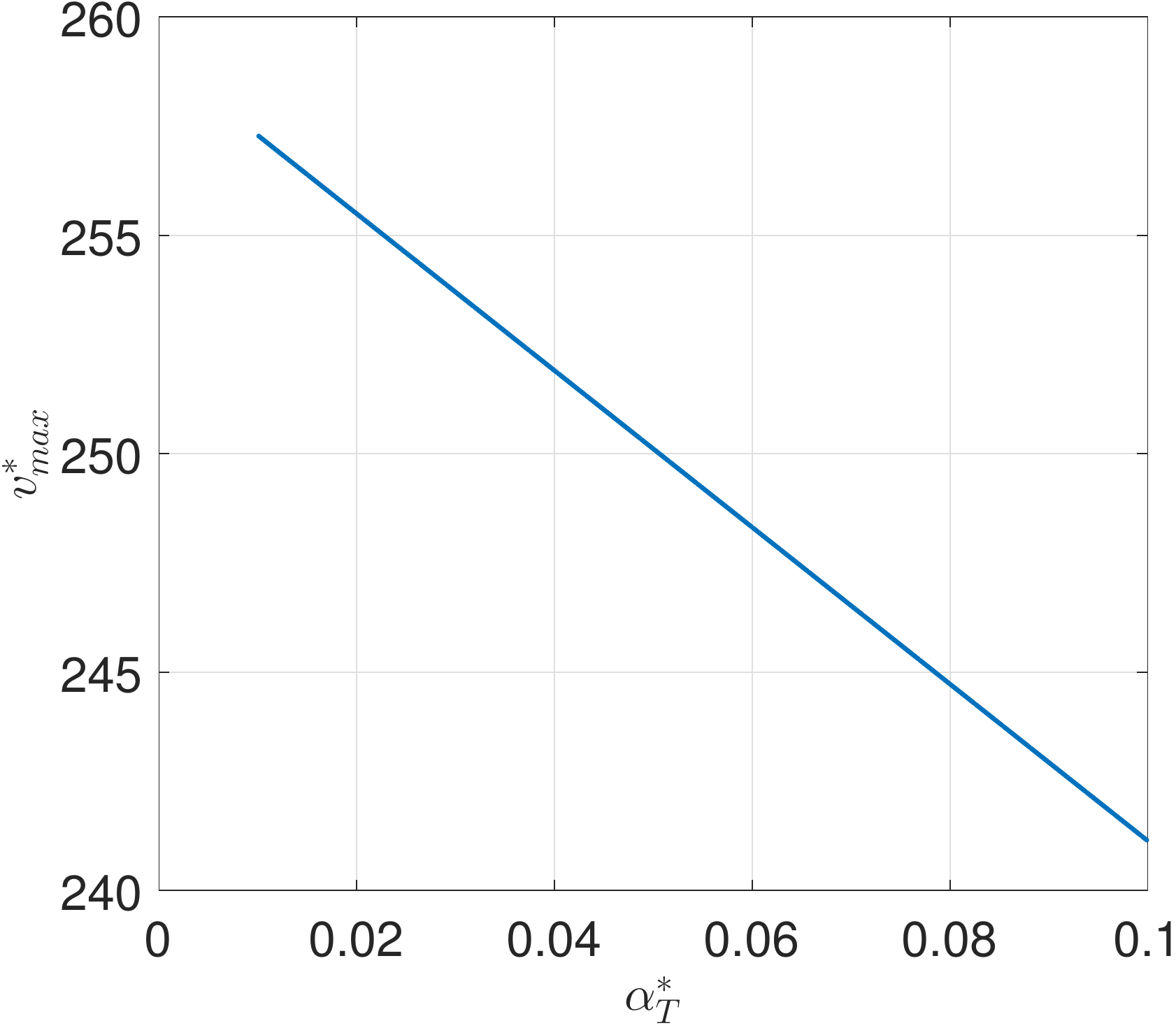}
	}
		\caption{Homogeneous case - Univariate effects of the input parameters on $v_{max}^\ast$}
	\label{fig:10}
\end{figure}

Fig. \ref{fig:10} represents the marginal effect of the
different parameters on the $v_{max}^\ast$. As expected, a different
magnitude of variations is obtained indicating the level of influence of
the parameters. The largest variation is obtained with the Rayleigh
number $\overline{Ra}$. Fig. \ref{fig:10}a shows that $v_{max}^\ast$
increases with the increase of $\overline{Ra}$ as this latter intensify
the rotating flow within the cavity. The marginal effect of
$v_{max}^\ast$ to the permeability ratio $r_k$ is slightly flatter (see
Fig. \ref{fig:10}b) and confirm the weak sensitivity of $v_{max}^\ast$
to $r_k$. The obtained negative slope is the consequence of the
horizontal velocity reduction caused by the decrease of $k_x$.  This
finding is consistent with the results obtained in Bennacer et al.,
\cite{bennacer2001double}. As expected, the marginal effects of
$v_{max}^\ast$ to $\alpha_L^\ast$ and $\alpha_T^\ast$ are nearly flat
(Fig. \ref{fig:10}(c-d)). A rather negative slope of $v_{max}^\ast$
versus $\alpha_L^\ast$ is observed indicating that the enhancement of
the longitudinal dispersive mixing between the hot and cold fluid leads
to an attenuation of the convective flow. Fig. \ref{fig:10}d shows that
$v_{max}^\ast$ decreases with the increase of $\alpha_T^\ast$ due to the
redistribution of the temperature gradient.

\clearpage

\subsubsection{\textbf{Uncertainty quantification}}  

\begin{figure}[!ht]
	\centering
	\includegraphics{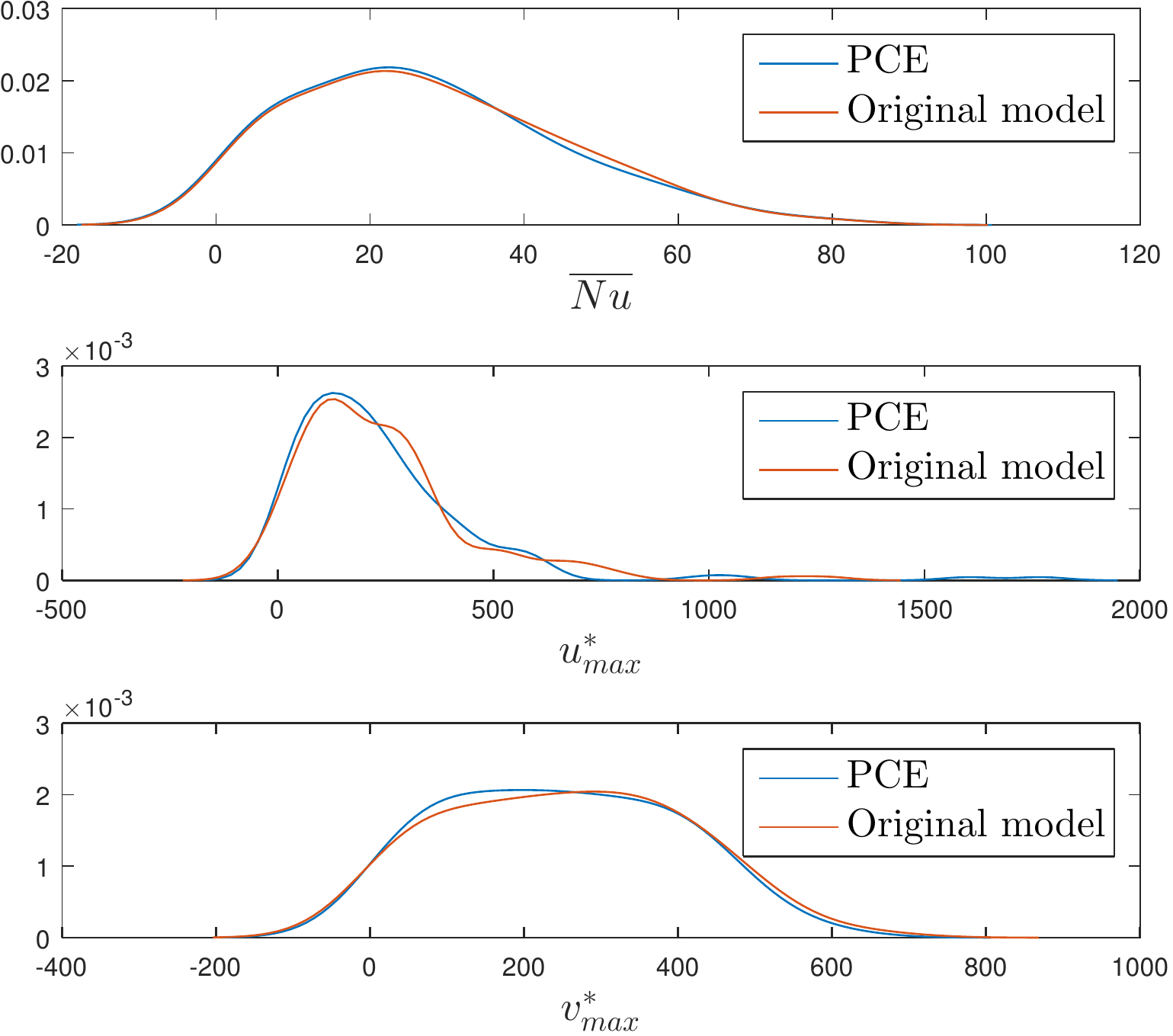}
	\caption{Homogeneous case - Marginal pdf obtained by relying on PCE (red lines) and $1,000$ MC Simulations (blue lines).}
	\label{fig:11}
\end{figure}

The constructed PCE can be employed as a surrogate model of the target output. Statistical analysis is then affordable by performing a large Monte Carlo simulations on the PCE approximations at a very low additional computational cost upon sampling the random input parameter space. We depict in Fig.~\ref{fig:11} the probability density function (PDF) of the scalar QoIs. The PDFs computed by relying on the PCE using $10^5$ Monte Carlo simulations are compared with the PDFs computed with the $1,000$ Monte Carlo (MC) simulations of the physical model. Note that we limit our comparison to $1,000$ MCs of the complete model because of the computational burden. A number of conclusions can be drawn with respect to Fig.~\ref{fig:11}. The marginal PDFs resulting from the PCEs compare very well with the PDF obtained by relying on numerical MC simulations at a total cost of $150$ simulations though. Positively skewed distributions, with longer tails toward larger values are observed for both output $\overline{Nu}$ and $u_{max}^\ast$ . The long tail of the PDF of the $\overline{Nu}$ is associated with setting characterized by low values of $\alpha_T^\ast$, whereas the long tail of $u_{max}^\ast$ is associated with setting characterized by low values of $r_k$. A flat distribution with short tails is obtained for $v_{max}^\ast$ . This is due to the fact that $v_{max}^\ast$ is mainly sensitive to the Rayleigh number $\overline{Ra}$ and that they are linearly related. \\

In the following, we investigate the level of correlation between QoI's pairs by determining the correlation coefficient, defined as the covariance of the two output variables divided by the product of their standard deviation. Table \ref{Tab:3}(a) lists the correlation coefficient evaluated using $1,000$ MC simulations. To further elaborate on the accuracy of the PCE, the correlation coefficient results obtained by relying on PCE are also shown (see Table \ref{Tab:3}(b)). Again, the agreement between the full model and the PCE is quite remarkable. A strong positive correlation between $u_{max}^\ast$ and $v_{max}^\ast$ is observed. This is related to the fact that $u_{max}^\ast$ and $v_{max}^\ast$ are both affected by the Rayleigh number $\overline{Ra}$ and they both increase with its increase. 

\begin{table} [!ht]
	\centering
	\caption{Correlation coefficient of the QoIs evaluated with $1,000$ MC simulations (Black fonts) and PCE.}
	\label{Tab:3}
		\begin{subtable}[Correlation coefficient of the QoIs based on $1,000$ MC simulations]
			{\begin{tabular}{c c c c}
				\hline   & $\overline{Nu}$ & $u_{max}^\ast$ & $v_{max}^\ast$ \\
				\hline $\overline{Nu}$ & $1$ & ${0.078}$  & ${0.209}$ \\
				$u_{max}^\ast$ & ${0.078}$ & $1$  & ${0.717}$  \\
				$v_{max}^\ast$ & ${0.209}$ & ${0.717}$  & $1$   \\
				\hline    
			\end{tabular}}   
		\end{subtable}
\hspace{1cm}
	\begin{subtable}[Correlation coefficient of the QoIs based on PCE]
		{\begin{tabular}{c c c c}
			\hline   & $\overline{Nu}$ & $u_{max}^\ast$ & $v_{max}^\ast$ \\
			\hline $\overline{Nu}$ & $1$ & $0.070$  & $0.226$ \\
			$u_{max}^\ast$ & ${0.07}$ & $1$  & $0.712$  \\
			$v_{max}^\ast$ & ${0.226}$ & ${0.712}$  & $1$   \\
			\hline   
		\end{tabular}}    
		\end{subtable}
\end{table}

\subsection{\textbf{Effect of heterogeneity}}

In this section, an heterogeneous porous media will be considered. The heterogeneity is assumed to follow the exponential model given in Eq.~\eqref{eqn:9} and~\eqref{eqn:10}. The effect of heterogeneity is expressed in terms of the rate of change of the permeability $\sigma^\ast$. Consequently, five independent input random variables $\ve{X}=\{\overline{Ra},r_k,\alpha_L^\ast,\alpha_T^\ast,\sigma^\ast \}$ are now considered uncertain with uniform marginal distributions. \\

The spatial discretization required to reach the converged numerical solution is highly dependent on the degree of heterogeneity. Indeed, an increase in the heterogeneity degree results in an increase of the local Rayleigh number, leading to a locally steeper and rougher solution than for the homogeneous case \cite{shao2016high}. Consequently, finer meshes are required to obtain a mesh independent solution. As for the homogeneous case, the most challenging configurations of parameters are thoroughly tested. A special irregular mesh is used to obtain the converged finite element solution. This mesh involves local refinement on the high-permeability zones where the buoyancy effects are more significant. A non uniform grid of $64,000$ nodes is used. All simulations were performed for $8 [t]$ in order to be sure that the steady state solution is reached. These discretization parameters are kept fixed in subsequent simulations.\\

In view of computing the PCE expansion of the model outputs, an experimental design drawn with QMC of size $N=150$ is considered. As in the previous case, the candidate basis is determined using a standard truncation scheme with $q =1$ for all the outputs. The corresponding results (e.g. polynomial degree giving the best accuracy, relative $LOO$ error and number of retained polynomials) of the PCE are given in Table \ref{Tab:4} for the three scalar output $\overline{Nu}$, $u_{max}^\ast$ and $v_{max}^\ast$. An accurate PCE is obtained for both $\overline{Nu}$ and $v_{max}^\ast$, where $LOO$ error is about $1$\%. A less accurate PCE is obtained for $u_{max}^\ast$, where $LOO$ error is larger than $0.1$.

\begin{table} [!ht]
	\centering
	\caption{Results of the utilized PCE}
	\label{Tab:4}
	\begin{tabular}{c c c c}
		\hline   & $\overline{Nu}$ & $u_{max}^\ast$ & $v_{max}^\ast$ \\
		\hline $p_{opt}$ & $6$ & $5$  & $6$ \\
		$err_{LOO}$ & $1.1\times10^{-2}$ & $2.88\times10^{-1}$  & $9.68\times10^{-3}$  \\
		size of the sparse basis & $46$ & $26$  & $63$   \\
		\hline       
	\end{tabular}
\end{table}

\begin{itemize}
	\item \textit{GSA of the temperature field}
\end{itemize}

\begin{figure}[!ht]
	\centering
	\subfigure
	[Spatial distribution of the mean value of the temperature]
	{
	\includegraphics[width=.4\textwidth,clip = true, trim = 0 0 0 
	0]{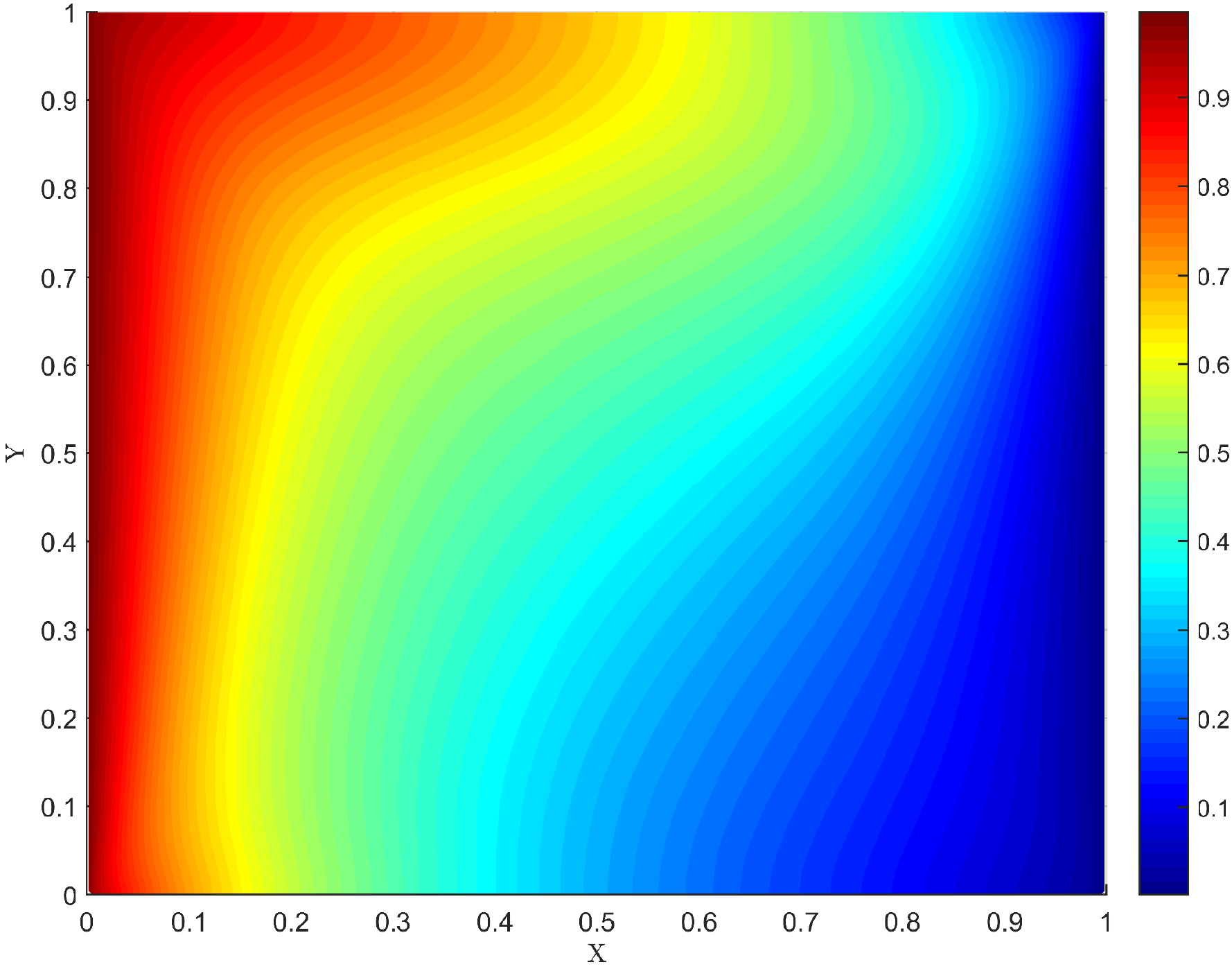}
}
\subfigure
[Spatial distribution of the variance]
{
	\includegraphics[width=.4\textwidth,clip = true, trim = 0 0 0 
	0]{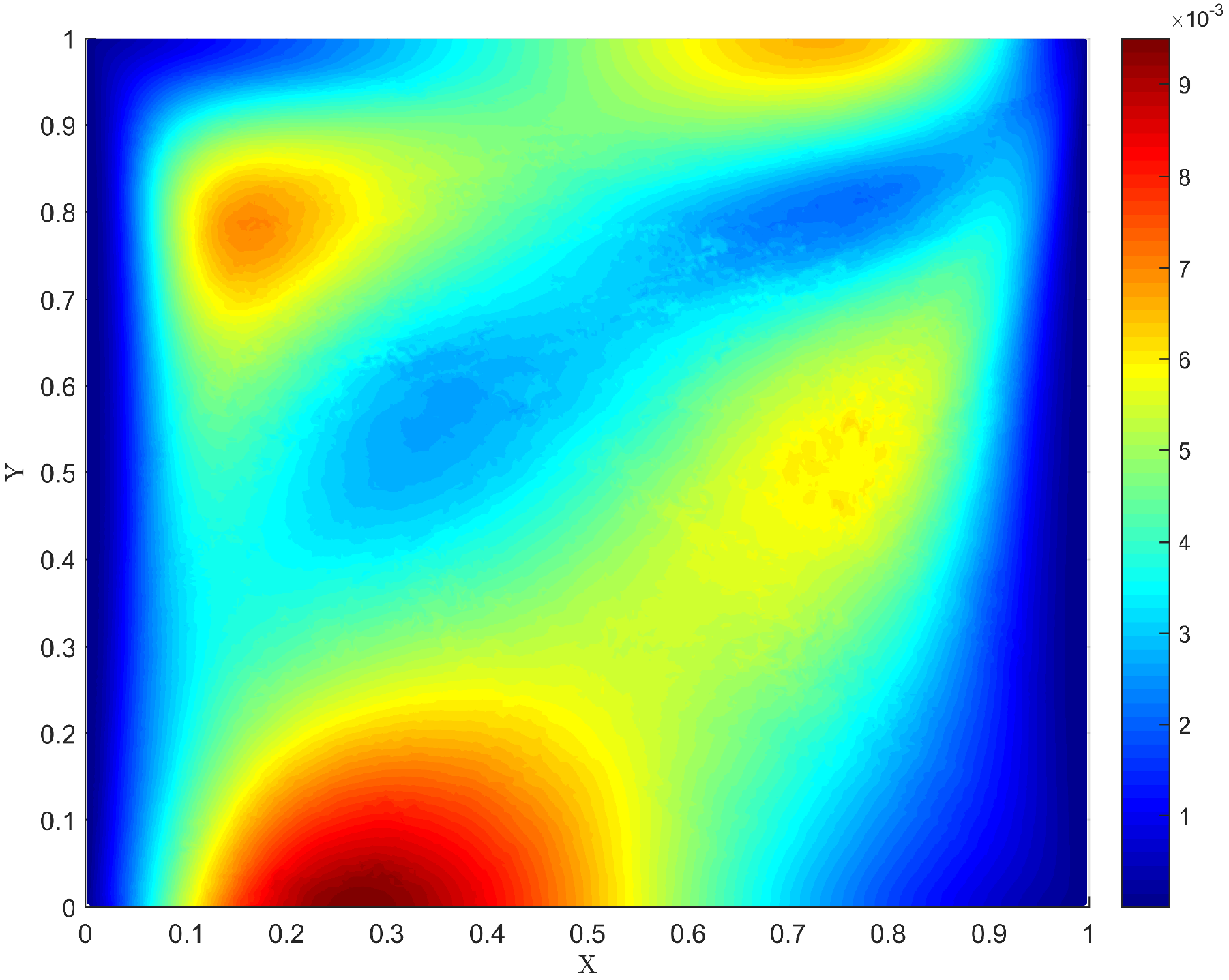}
}
	\caption{Heterogeneous case - spatial distribution of the temperature statistical moments.}
	\label{fig:12}
\end{figure}

Fig. \ref{fig:12}a illustrates the spatial distribution of the mean
temperature by relying on the PCE. It shows that the isotherms are more
affected by the circulating flow than that of the homogeneous case;
especially in the high permeable zones (near the top surface of the
cavity). Indeed, the local Rayleigh number exceeds the average value
$\overline{Ra}$, leading eventually to a more intense convective
flow. 

The corresponding depictions of the temperature variance is shown in
Fig. \ref{fig:12}b. First, we observe is that the spatial distribution
of the variance is not anymore symmetric as in the homogeneous case. The
heterogeneity results in a different distribution of the variance while
maintaining the same level of variability. The smallest variation zone
is located at the boundary layer of the vertical walls and in the
slow-motion region, as for the homogeneous case. It should be noted here
that, due to the effect of increased permeability, the slow-motion
region expands horizontally and moves up toward the right corner
\cite{fahs2015reference}. The zone of low temperature variance exhibits
similar behavior. The largest variance zone moves to the low permeable
layers and it is shifted toward the hot wall. The high permeability at
the top layers leads to a reduction of the variance of
the temperature. \\

\begin{figure}[!ht]
	\centering
	\subfigure
		[ Spatial distribution of total Sobol' index of $\overline{Ra}$;]
	{
	\includegraphics[width=.4\textwidth,clip = true, trim = 0 0 0 
	0]{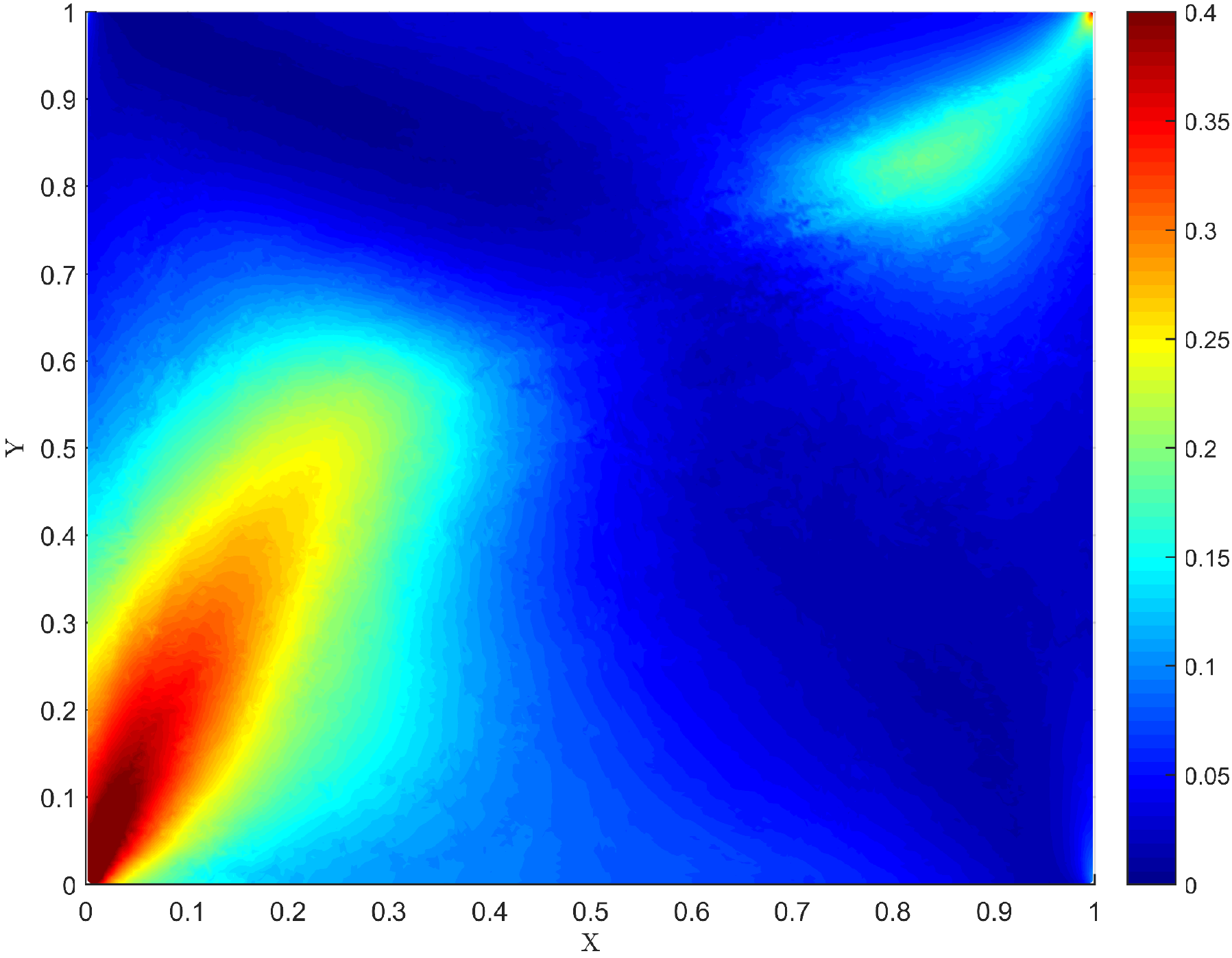}
}
\subfigure
	[ Spatial distribution of total Sobol' index of $r_k$;]
{
	\includegraphics[width=.4\textwidth,clip = true, trim = 0 0 0 
	0]{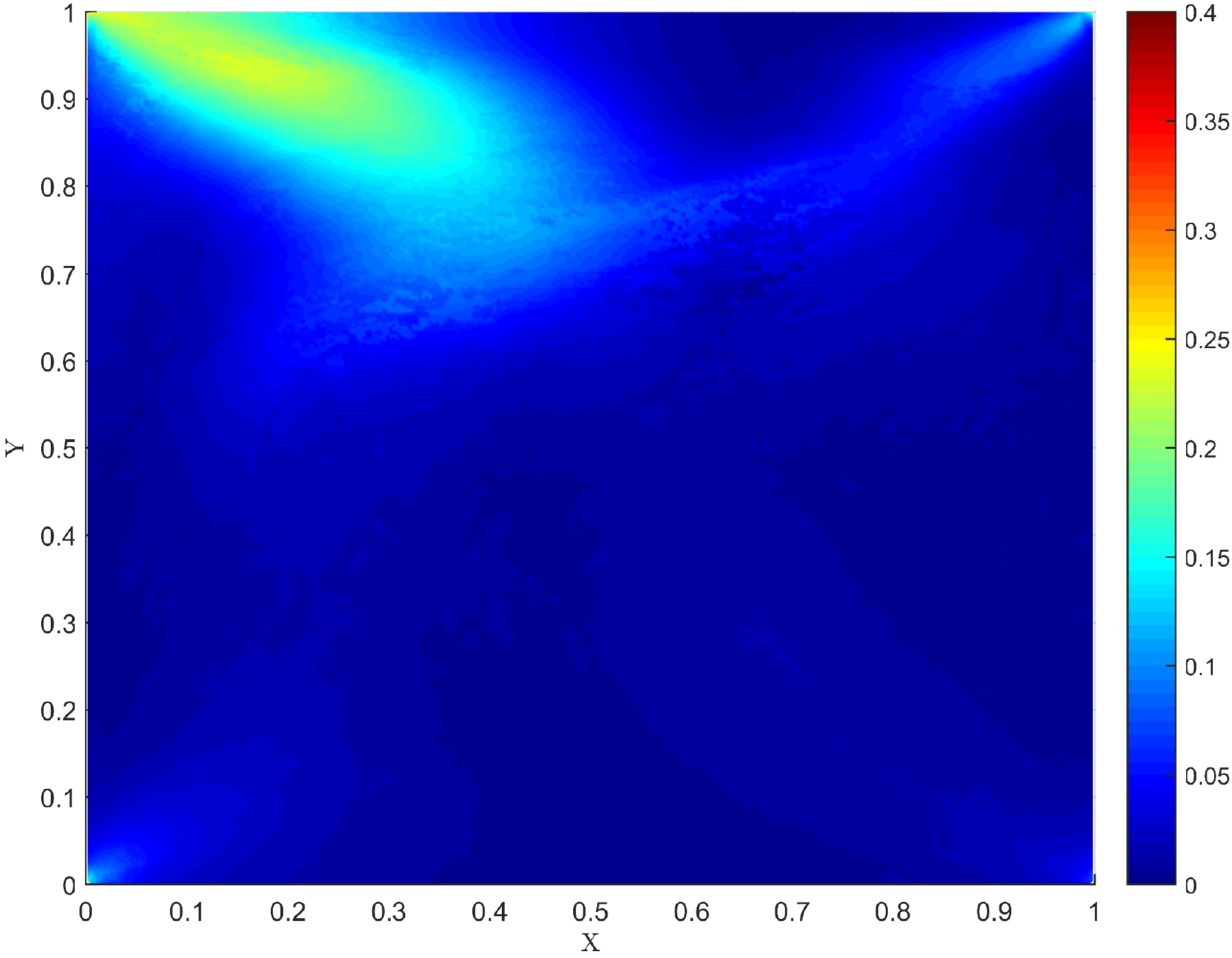}
}
\subfigure
	[ Spatial distribution of total Sobol' index of $\alpha_L^\ast$;]
{
	\includegraphics[width=.4\textwidth,clip = true, trim = 0 0 0 
	0]{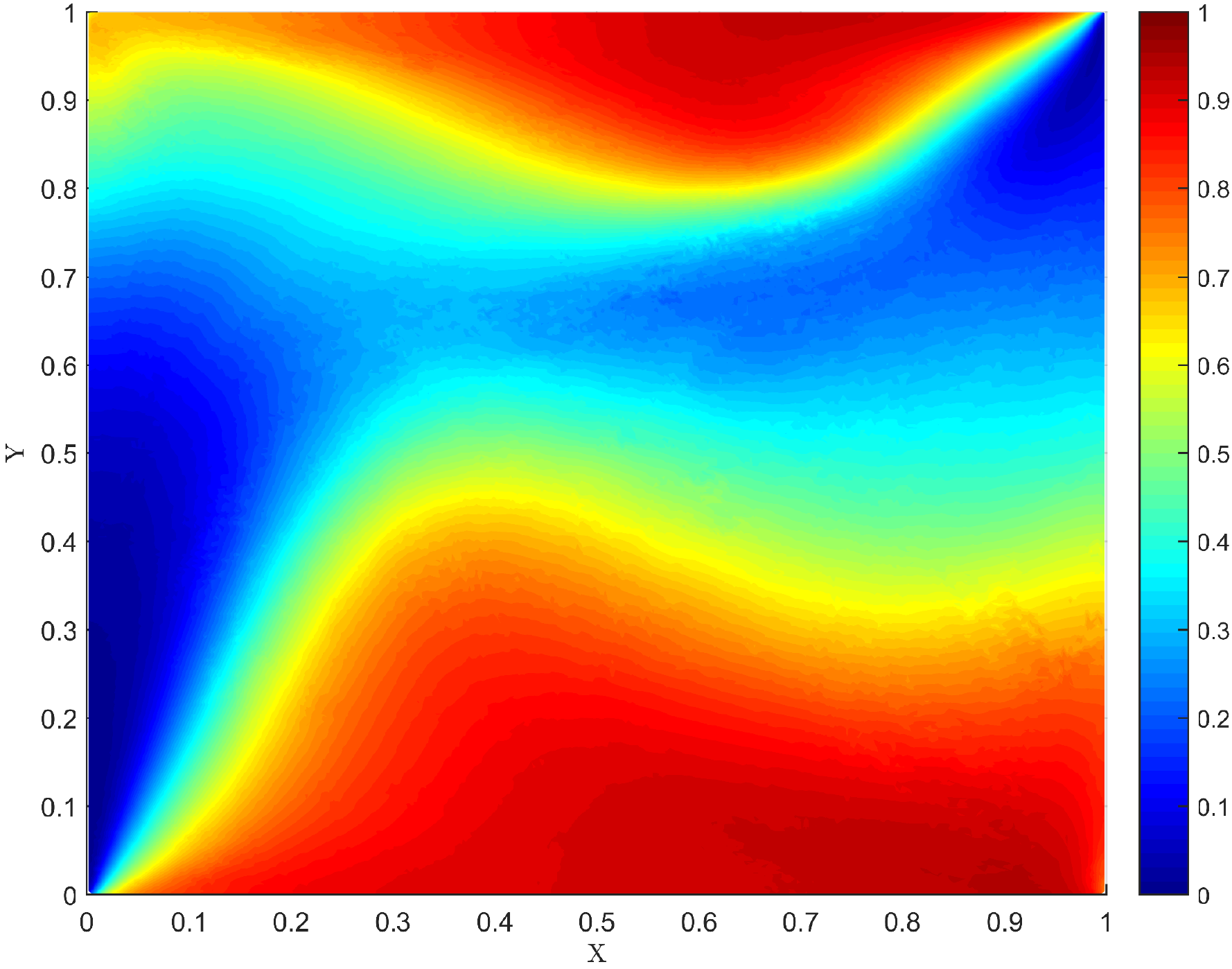}
}
\subfigure
	[ Spatial distribution of total Sobol' index of $\alpha_T^\ast$;]
{
	\includegraphics[width=.4\textwidth,clip = true, trim = 0 0 0 
	0]{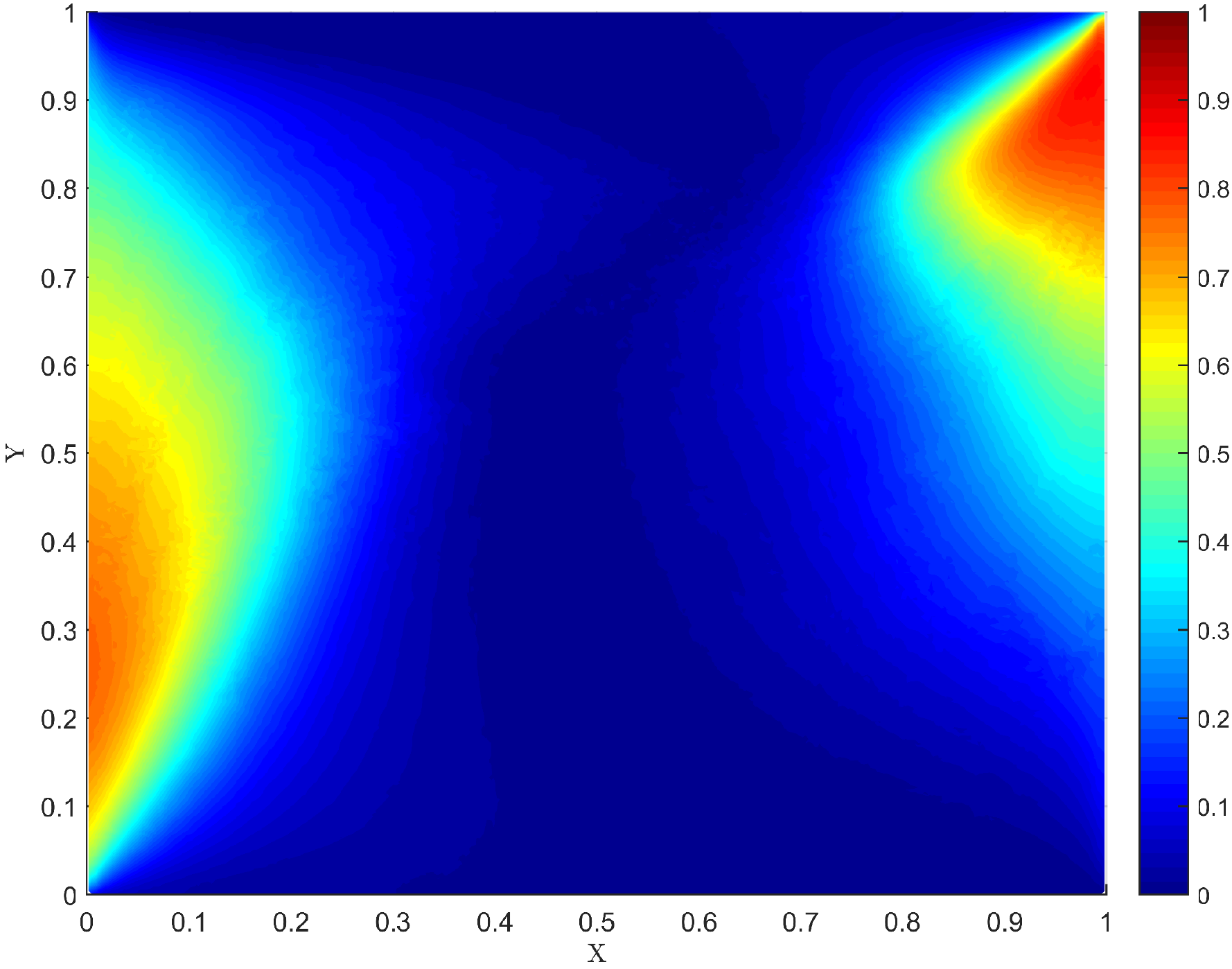}
}
\subfigure
	[ Spatial distribution of total Sobol' index of $\sigma^\ast$;]
{
	\includegraphics[width=.4\textwidth,clip = true, trim = 0 0 0 
	0]{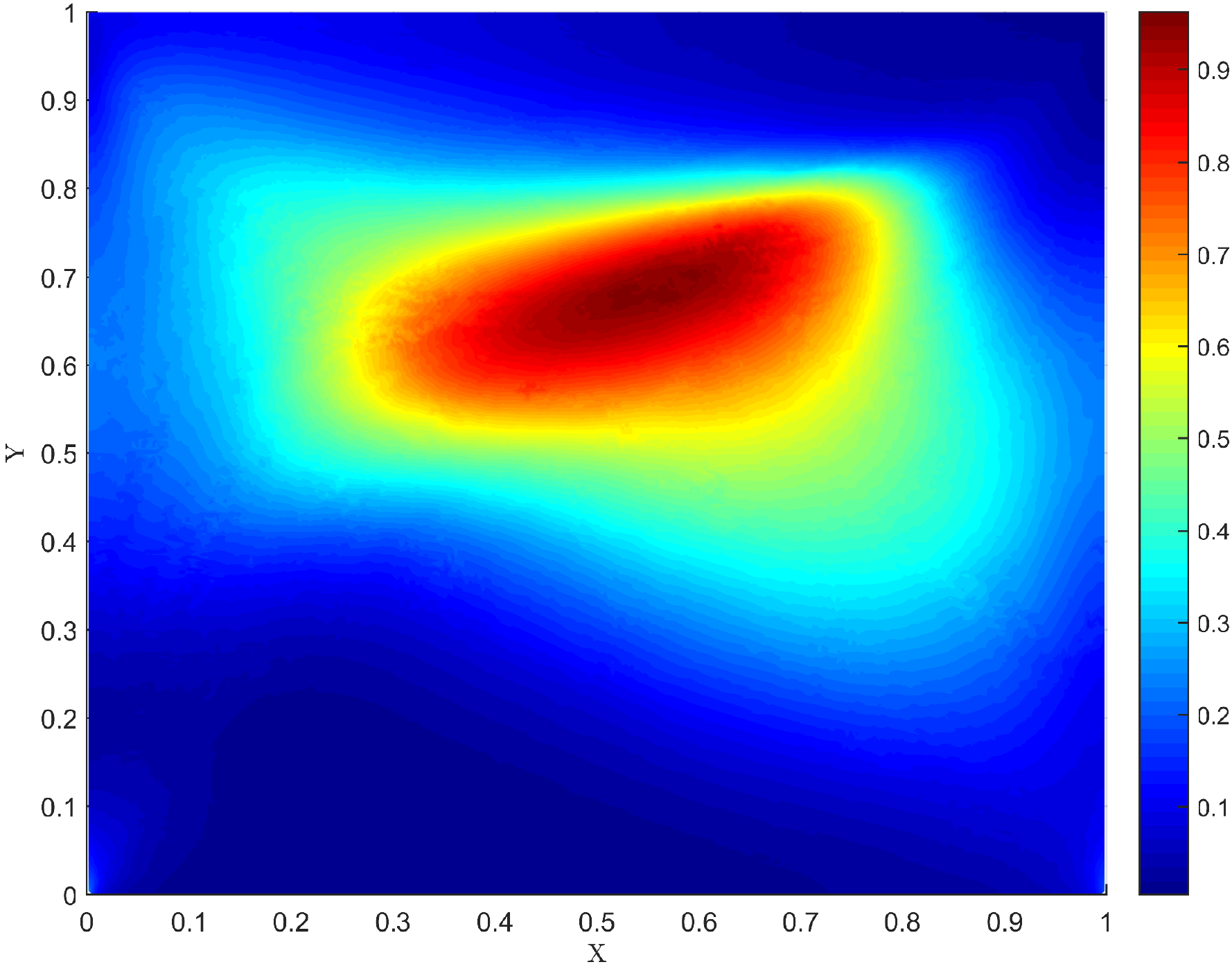}
}
	\caption{Heterogeneous case - Spatial distribution of the total Sobol' indices for the temperature}
	\label{fig:13}
\end{figure}

The spatial maps of the total Sobol' indices are depicted in
Fig.~\ref{fig:13}. It demonstrates that, due to the heterogeneity, the
symmetry of the Sobol' indices spatial distribution around the center of
the cavity is completely destroyed. Fig.~\ref{fig:13} indicates that
$\alpha_L^\ast$, $\alpha_T^\ast$ and $\sigma^\ast$ have significant
influence on the temperature distribution. 

A closer look to the
temperature variance confirms that, as for the homogeneous case,
$\alpha_L^\ast$ is the most sensitive parameter since its zone of
influence intersects well with the zone of maximum temperature
variance. Comparing to the homogeneous case, the zone of influence of
$\alpha_L^\ast$ expands in the zones of low permeability (toward the
bottom surface of the cavity) and contracts in the high permeable zones
(at the top surface). Indeed, when the slow-motion region moves up
toward the right corner, the zones in which the velocity vector is
horizontal (parallel to the gradient) shrinks near the top surface and
grows in the lower part of the cavity. The reverse is true for the zone
of influence of $\alpha_T^\ast$. This zone expands vertically near the
hot wall and contracts near the cold wall. This behavior is also
attributable to the shifting of the slow-motion region toward the top
right corner due to heterogeneity. The sensitivity of the temperature
distribution to the rate of change of heterogeneity $\sigma^\ast$ is
mainly important around the zone of slow rotating motion as it can be
seen in Fig. \ref{fig:13}e. The zone of influence of $\overline{Ra}$
expands around the right bottom corner (Fig. \ref{fig:13}a). This
behavior is related to the expansion of the zone in which the velocity
is relatively weak as a consequence of the low permeability near the
cavity bottom surface. Conversely, around the top left corner the high
permeability induces faster convective flow associated with higher
dispersion tensor. Hence, mixing by thermal dispersion dominates and
temperature distribution becomes less sensitive to
$\overline{Ra}$. Fig. \ref{fig:13}b shows that in the case of
heterogeneous porous media, $r_k$ becomes more influential on the
temperature distribution than for the homogeneous case. Its zone of
influence expands around the top left corner.

\begin{itemize}
	\item \textit{GSA of the scalar QoIs}
\end{itemize}

\begin{figure}[!ht]
	\centering
	\subfigure
		[Total Sobol' indices for the average of Nusselt number $\overline{Nu}$;]
		{
	\includegraphics[width=.4\textwidth,clip = true, trim = 0 0 0 
	0]{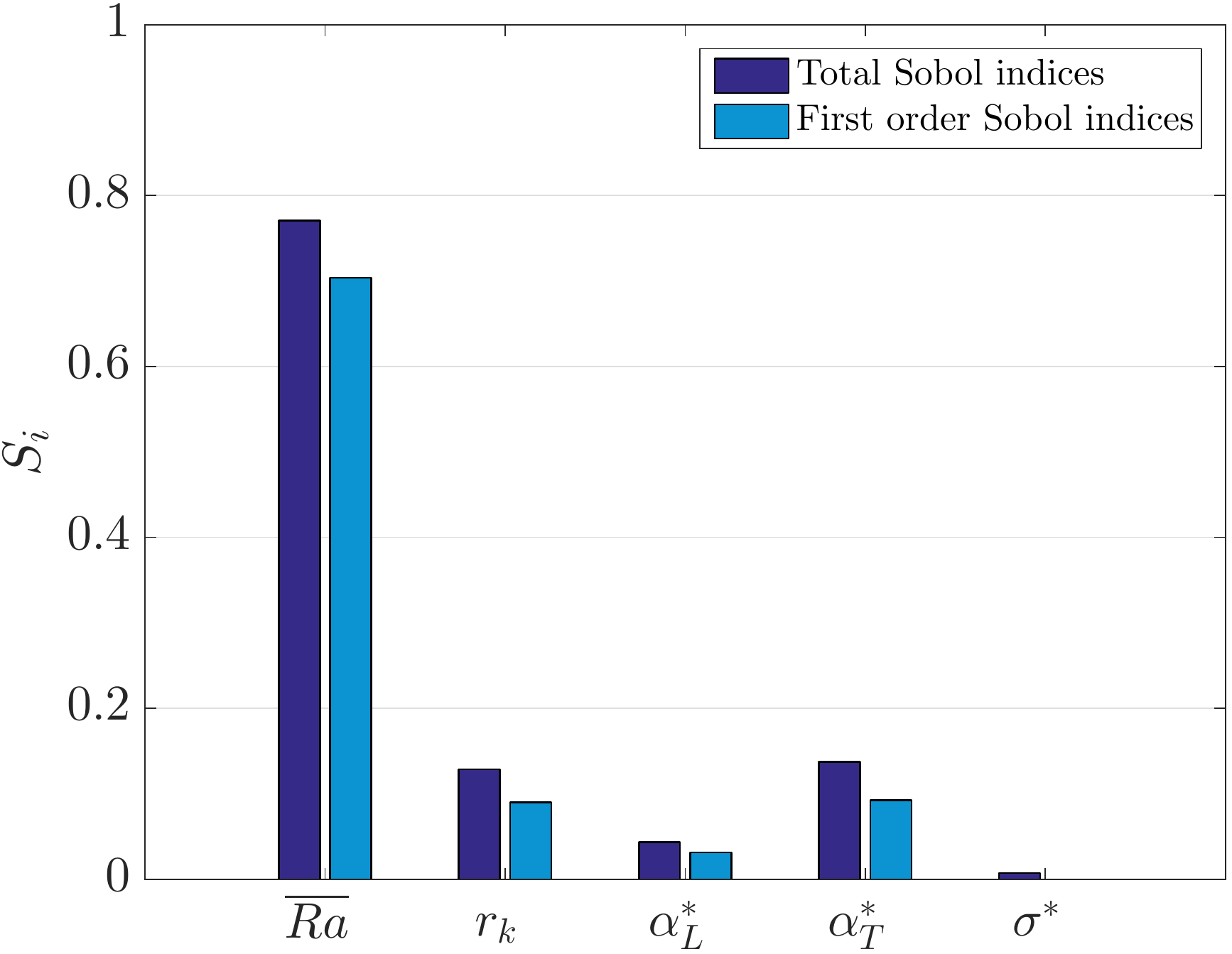}
}
	\subfigure
	[Total Sobol' indices for $u_{max}^\ast$ ;]
	{\includegraphics[width=.4\textwidth,clip = true, trim = 0 0 0 
		0]{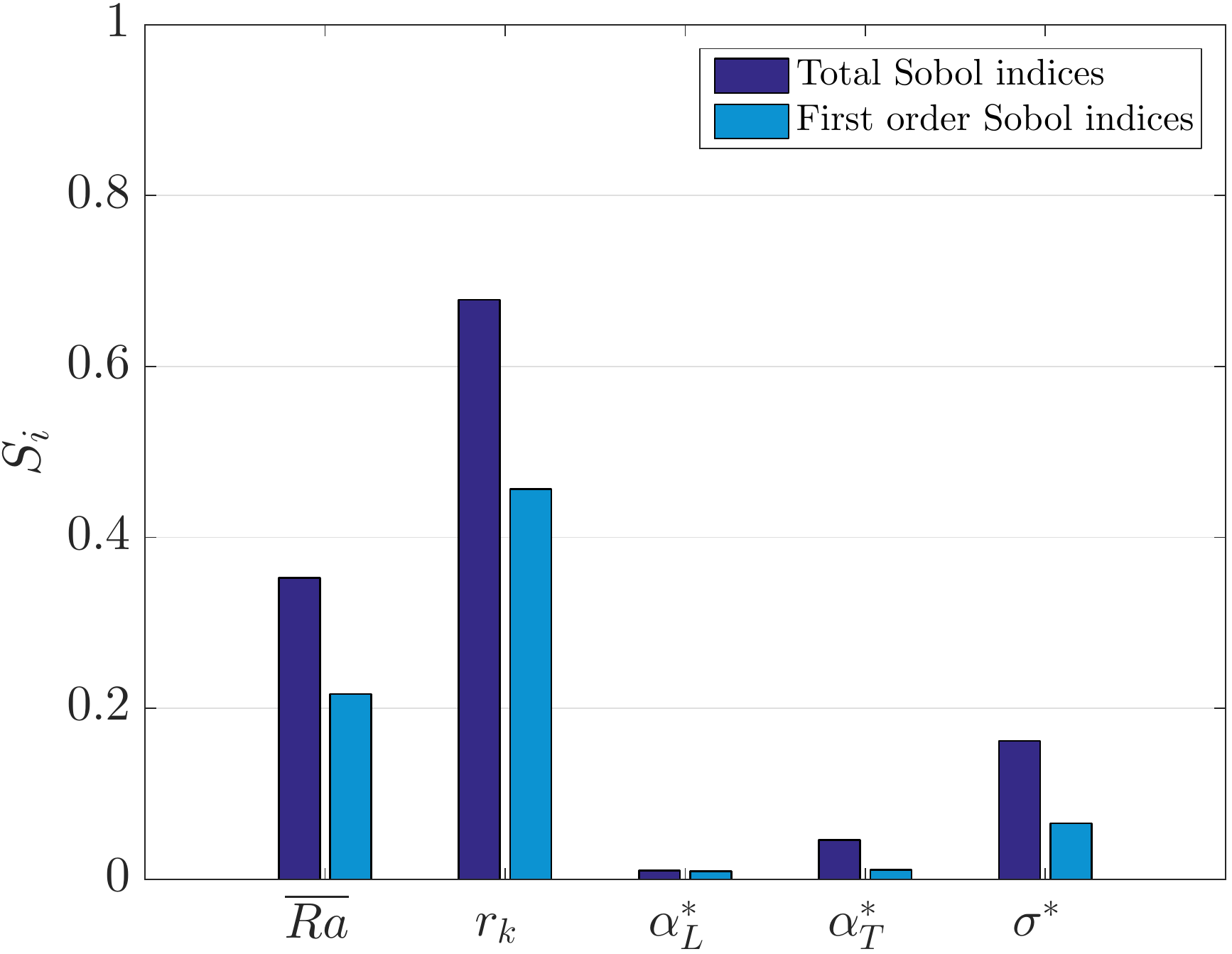}
	}
	
		\subfigure
		[Total Sobol' indices for $v_{max}^\ast$;]
		{\includegraphics[width=.4\textwidth,clip = true, trim = 0 0 0 
			0]{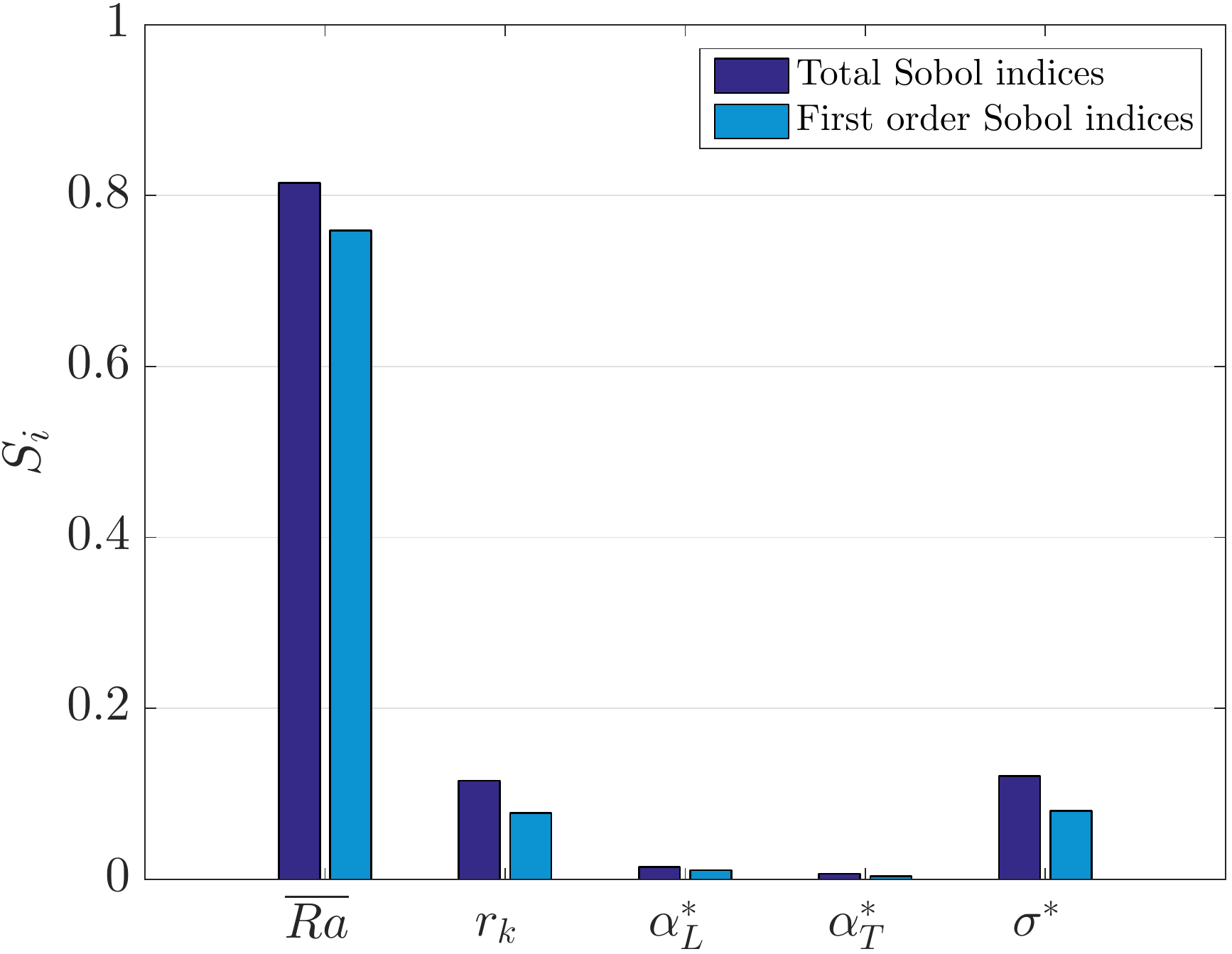}
		}
	\caption{Heterogeneous case - Total Sobol' indices for the model outputs}
	\label{fig:14}
\end{figure}

Fig. \ref{fig:14} shows bar-plots of the first order and total Sobol'
indices of the three model outputs ( $\overline{Nu}$, $u_{max}^\ast$ and
$v_{max}^\ast$). 

This figure allows drawing the following conclusions:
\begin{itemize}
\item The heterogeneity of the porous media does not affect the rank of
  the parameters regarding their influences on the model outputs.
  $\overline{Ra}$ and $\alpha_T^\ast$ remain the most influential
  parameters for $\overline{Nu}$, $\overline{Ra}$ and $r_k$ for
  $u_{max}^\ast$, and $\overline{Ra}$ for $v_{max}^\ast$.

	\item The uncertainty associated with the rate of change of the
          heterogeneity $\sigma^\ast$ has no effect on $\overline{Nu}$.
          This is coherent with the results obtained for the temperature
          distribution that shows that the effect of $\sigma^\ast$ is
          located relatively far from the hot wall in the slow-motion
          region.
	
	\item The influence of $\sigma^\ast$ is more important on
          $u_{max}^\ast$ and $v_{max}^\ast$, which is reasonable because
          the velocity is directly related to the permeability level of
          heterogeneity.
		
	\item One can also observe that the heterogeneity renders
          $\overline{Nu}$ and $v_{max}^\ast$ more sensitive to $r_k$. In
          fact, in vertically stratified heterogeneous domains (as it is
          the case here), the anisotropy can be at the origin of a
          vertical flow that can transmit the fluid from a certain layer
          to lower or upper ones with different permeability. Hence any
          change of $r_k$ may have a strong effect on the flow as
          different layers of heterogeneity can be involved.
\end{itemize}

\begin{figure}[!ht]
	\centering
	\subfigure
	[Univariate effects of $\sigma^\ast$ on $u_{max}^\ast$]
	{
		\includegraphics[width=.4\textwidth,clip = true, trim = 0 0 0 
		0]{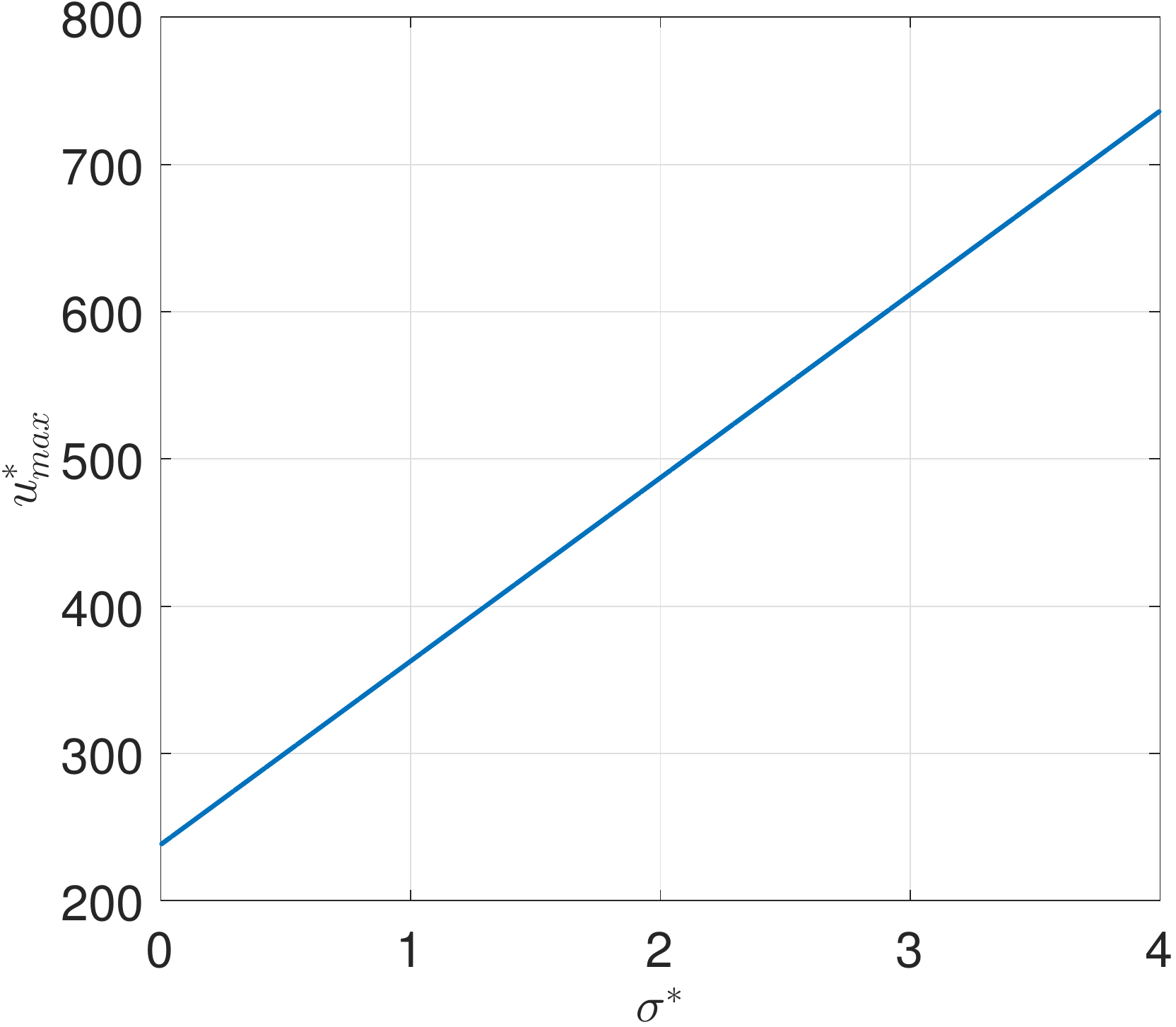}
	}
	\subfigure
	[Univariate effects of $\sigma^\ast$ on $v_{max}^\ast$]
	{\includegraphics[width=.4\textwidth,clip = true, trim = 0 0 0 
		0]{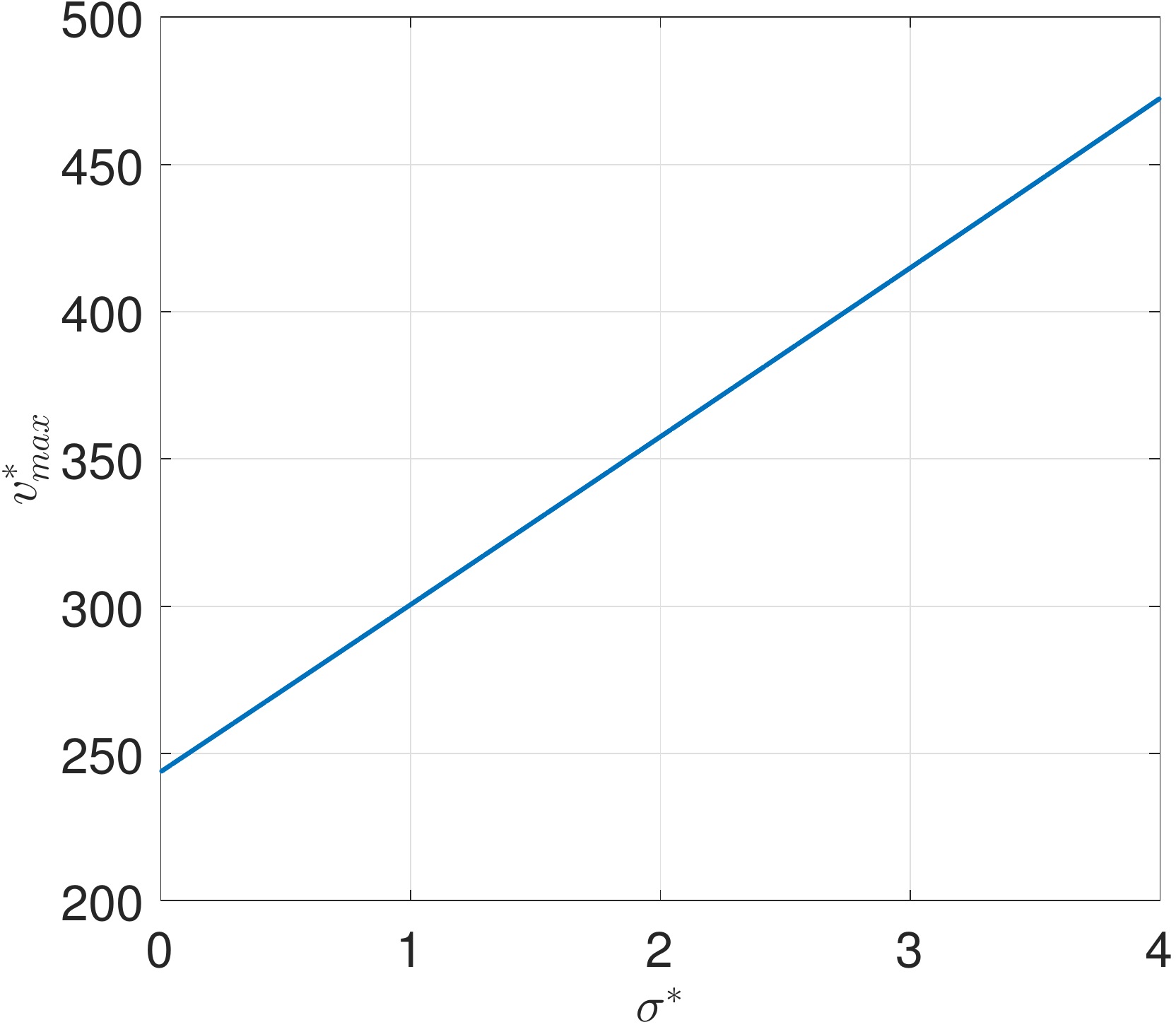}
	}

	\caption{Heterogeneous case - Univariate effects of $\sigma^\ast$ on $u_{max}^\ast$ and $v_{max}^\ast$.}
	\label{fig:15}
\end{figure}

The investigation of the marginal effects of the parameters on the model
outputs showed that the conclusions drawn for the homogeneous case still
hold for the heterogeneous porous media. This is why we hereby discuss
only the marginal effect of $\sigma^\ast$. The results show a slight
variation of $\overline{Nu}$ against $\sigma^\ast$. Indeed, the increase
of sigma leads to the attenuation of the local velocity at the lower
part of the hot wall and an intensification in the upper part. The
attenuation of the horizontal velocity around the bottom surface is
associated to a reduction of the thermal dispersion and by consequence a
decrease of the temperature gradient. Reverse process occurs at the
upper surface and leads to the increase of the thermal gradient. As a
consequence, the local Nusselt number decreases in the lower part of the
vertical wall and increases at the upper part. Upper and lower local
Nusselt numbers tend to balance out each other and lead to small
variation of the $\overline{Nu}$. Marginal effects of $\sigma^\ast$ on
$u_{max}^\ast$ and $v_{max}^\ast$ are given in Fig.~\ref{fig:15}. This
figure confirms the high sensitivity of $u_{max}^\ast$ and
$v_{max}^\ast$ to sigma. It indicates also an increasing variation of
$u_{max}^\ast$ and $v_{max}^\ast$ against $\sigma^\ast$ as a result of
the permeability increasing at the top surface of the cavity.

%\begin{figure}[!ht]
%	\centering
%	\includegraphics[width=.44\textwidth,clip = true, trim = 0 0 0 
%	0]{heterogene/vx_sigma_hete.pdf}
%	\caption{Variation of  $u_{max}^\ast$ as function of $\sigma$;  }
%	\label{fig:18}
%\end{figure}

%The same analysis is conducted for the maximum velocity $v_{max}^\ast$. Fig. \ref{fig:19}  shows bar-plots of the  total Sobol' indices of the maximum Velocity $v_{max}^\ast$. Results shows that the variability of $v_{max}^\ast$ is mainly due to main effects of the Rayleigh number $\overline{Ra}$. A small influence of $r_k$ and $\sigma^\ast$ are also observed. They explain $0.079\%$ and $0.08\%$ of the total variance, respectively. Interactions between parameters are negligible (less than $0.026$). We depict in Fig. \ref{fig:20} the marginal effect of $v_{max}^\ast$ with respect to $\sigma^\ast$. We note that an increase of $\sigma^\ast$ results in an increase of $v_{max}^\ast$. We note that an increasing of the heterogeneity level $\sigma_z$ results in an increasing of the maximum velocity $u_{max}^\ast$.
%
%\begin{figure}[!ht]
%	\centering
%	\includegraphics[width=.7\textwidth,clip = true, trim = 0 0 0 
%	0]{heterogene/vy_ST_hete.pdf}
%	\caption{Total Sobol' indices for  $v_{max}^\ast$;}
%	\label{fig:19}
%\end{figure}
%\begin{figure}[!ht]
%	\centering
%	\includegraphics[width=.7\textwidth,clip = true, trim = 0 0 0 
%	0]{heterogene/vy_sigma_hete.pdf}
%	\caption{Variation of  $v_{max}^\ast$ as function of $\sigma$;}
%	\label{fig:20}
%\end{figure}

\section{Summary and conclusions}  
The proposed study handled the topic of natural convection problem in heterogeneous porous media including velocity-dependent dispersion. The considered benchmark is the popular square cavity filled with a saturated porous medium and subject to differentially heated vertical walls and adiabatic horizontal surfaces. The simplicity of the geometry and the boundary conditions renders this problem especially suitable for testing numerical models but also useful to provide physical insights into the involving processes. It introduces however several uncertain parameters, namely: the average Rayleigh number ($\overline{Ra}$), the permeability anisotropy ratio ($r_k$), the non-dimensional dispersion coefficients ($\alpha_L^\ast$ and $\alpha_T^\ast$).\\

 The imperfect knowledge of the system parameters and their variability significantly affect the flow and heat transfer patterns. It is thus of utmost importance to properly account for the aforementioned uncertainties within the frame of uncertainty and sensitivity Analysis.  In this work, we analyze the impact of the uncertain parameters on the output quantities of interest (QoIs) allowing assessment of flow and heat transfer. To describe the flow process, we use the maximum dimensionless velocity components ($u^\ast_{max}$ and $v^\ast_{max}$). For the heat transfer process, the assessment is based on the spatial distribution of the dimensionless temperature ($T^\ast$) and the average Nusselt number $\overline{Nu}$ on the hot wall. The effect of heterogeneity was also investigated by considering a stratified heterogeneous porous media with an exponential distribution of the permeability as a function of depth. The rate of change of the permeability $\sigma^\ast$ is then considered uncertain.

Herein, we performed a comprehensive global sensitivity analysis and uncertainty quantification through performing the probability distributions by means of surrogate modeling. Sparse PCE are used for this purpose. Our results lead to the following major conclusions.

\begin{itemize}
	
	\item Sparse PCE proved particularly efficient in providing reliable results at a considerably low computational costs. All results derived for the homogeneous (resp. heterogeneous) case were obtained at the cost of only $150$ simulations of the computational model. Note that those runs could have been carried out in parallel on any distributed computing architecture. Due to the inexpensive-to-evaluate formulation of the sparse PCE meta-model, the probability density functions (PDF's) of the QoIs have been accurately estimated. An excellent agreement between sparse PCE and MC results is obtained.
	
	\item The Sobol' indices of the temperature distribution allow specifying the spatial zone of influence of each parameter. Results showed that the variability of the temperature distribution is largely influenced by the effect of $\alpha_L^\ast$ and $\alpha_T^\ast$. Nevertheless, the effect of $\alpha_L^\ast$ on the temperature distribution is more pronounced than that for $\alpha_T^\ast$ as its zone of influence is located in the region where the variance is maximum.  
	
	\item The variability of $\overline{Nu}$ is mainly due to main effects of $\overline{Ra}$ and $\alpha_T^\ast$. Indeed, the Rayleigh number dramatically influence the flow profile and heat transfer within the cavity, as well as the thermal boundary layer thickness. The variability of $u_{max}^\ast$ is mainly due to main effects of the ratio of anisotropy $r_k$, and  $\overline{Ra}$ while the variability of $v_{max}^\ast$ is mainly controlled by $\overline{Ra}$. 
	
	\item The effect of the heterogeneity results in a different distribution of the variance of the temperature while maintaining the same level of variability. The zone of largest temperature variance becomes located in the low permeable layer near the bottom surface of the cavity. In this case, the isotherms are more affected by the circulating flow than in the homogeneous case, especially in the high permeable zones, associated to a more intense convective flow. Results shows that the average Nusselt number is not sensitive to the heterogeneity rate while an increase of  $\sigma^\ast$ is associated to an increase of the maximum velocity $u_{max}^\ast$ and $v_{max}^\ast$.  
	
	\item Marginal effects of each parameters are also readily obtained from PCE. As opposed to classical "one-at-a-time" sensitivity analyses, where all parameters but one are frozen so as to study the effect of the remaining one, the univariate effect curves account for the uncertainties in the other parameters. Quantitative conclusions have been drawn, which confirm qualitative interpretations. 
\end{itemize}

This study represents a prototype to point out the benefit of GSA and UQ to understand how the complex system of natural convection in porous media behaves. This kind of study shall be useful for safe design and risk assessment of systems involving natural convection in porous enclosure.

%%%%%%%%%%%%%%%%%%

\bibliographystyle{chicago}
\bibliography{bib1}

\end{document}